\documentclass[aps,prb,twocolumn]{revtex4} %,preprint

\usepackage{graphicx}% Include figure files
\usepackage{dcolumn}% Align table columns on decimal point
\usepackage{bm}% bold math

\newcommand{\comment}[1]{}
\begin{document}
%\preprint{}   % Preprint number in upper right corner
\renewcommand{\theequation}{\arabic{section}.\arabic{equation}}

\title{Expansion for Quantum Statistical Mechanics
Based on Wave Function  Symmetrization}
%Quantum Statistical Mechanics. V.
%Wave Function  Symmetrization,
%Entropy Eigenfunctions,
%and the Classical Limit.

%\author{}
%\thanks{}

\author{Phil Attard}
%\affiliation{\protect\texttt{phil.attard1@gmail.com}}

\date{v2, May 11, 2016,\ phil.attard1@gmail.com}
%\\
%..  Begun: 1 Oct., 2015. Papers/Current/QSM/QSM5/QSM5.tex}

\begin{abstract}
An expansion for quantum statistical mechanics is derived
that gives classical statistical mechanics as the leading term.
Each quantum correction comes
from successively larger permutation loops,
which arise from the factorization of the symmetrization
of the wave function with respect to localized particle interchange.
Explicit application of the theory yields
the full fugacity expansion for the quantum ideal gas,
and the second fugacity coefficient for interacting quantum particles,
which agree with known results.
Compared to the  Lee-Yang virial cluster expansion,
the present expansion is expected to be more rapidly converging
and the individual terms appear to be simpler to evaluate.
The results obtained in this paper
are intended for practical computer simulation algorithms
for terrestrial  condensed matter quantum systems.
\end{abstract}

%The overlap factors in  addition enable non-orthogonal eigenfunctions
%to be used.
%It is argued that the entropy representation is the optimum
%one for a many-particle quantum system,
%and that wave packets are the most appropriate trial entropy eigenfunctions.
%It is shown that in the thermodynamic limit for interacting particles
%the width of the wave packets tends to zero, $\overline \xi \sim N^{-1/8}$,
%and they become exact  entropy eigenfunctions,
%with the relative root mean square energy expectation
%fluctuation scaling as $N^{-1/2}$.
%In this limit the entropy microstates become points in classical phase space.

%\keywords{}

\maketitle

%%%%%%%%%%%%%%%%%%%%%%%%%%%%%%%%%%%%%%%%%%%%%%%%%%%%%%%%%%%%%%%%%%%%%%%%%%
%
\section{Introduction}
\setcounter{equation}{0} \setcounter{subsubsection}{0}
%\renewcommand{\theequation}{\Alph{section}.\arabic{equation}}
%
%%%%%%%%%%%%%%%%%%%%%%%%%%%%%%%%%%%%%%%%%%%%%%%%%%%%%%%%%%%%%%%%%%%%%%%%%%

This paper addresses fundamental and practical questions concerning
the quantum mechanics of many particles.
The extant theme is practical:
How can one compute efficiently
the behavior of a many-particle quantum system?
But in answering this a more fundamental conceptual point is illuminated:
How does the observed classical world
arise from the underlying quantum mechanical equations?
The answers to both turn out to be intimately related.

There are three major challenges for the computation of the properties
of a many-particle quantum system:
the superposition of states,
the symmetrization of the wave function,
and the calculation of the eigenfunctions.

The superposition of states of a quantum system
poses insurmountable difficulties for computation
in that one has to store simultaneously all possible states of the system.
Whereas the calculation of a classical average grows linearly
with the number of states,
and can be accumulated one state at a time,
the calculation of a quantum expectation value grows quadratically
with the number of states,
and it depends upon the relative weights and phases of all the states
simultaneously.
Of course the problem is often avoided by restricting the computation
solely to the ground state,
since this excludes everything else that would
have had to be  superposed.
But this myopia fails if the ground state is degenerate,
which it is for large systems,
and it is of limited use for systems at non-zero temperatures.
The problem rapidly becomes prohibitive as the system size is increased,
because not only does the number of possible states increase,
but also the gaps between the accessible states decrease.

Symmetrization of the wave function upon particle interchange
likewise poses great computational challenges.
For example, the Hartree-Fock approximation
accounts for the symmetrization of a fermionic wave function
by invoking Slater determinants of one-particle orbitals.
Not only is the evaluation of a large determinant problematic,
(it grows factorially with particle number),
but in order  to account for correlations,
the number of determinants required grows exponentially with particle number.

It is not feasible to calculate
the full spectrum of eigenstates and eigenfunctions
of a realistic many-particle quantum system.
Again if the focus is restricted to the ground state,
there exist efficient computational techniques
for optimizing a trial wave function.
But even in this case the problem of degeneracy
creates the need to orthogonalize the set of ground state eigenfunctions,
which is a challenge for large systems.
It is often the case that the quality of the solution
is crucially dependent upon
an informed guess for the form of the trial wave function.
Again one has the problem that at non-zero temperatures
one requires the excited state eigenvalues and eigenfunctions,
and variational techniques in these cases are  more complicated,
or possibly even non-existent,
because the search has to be restricted to a sub-space orthogonal
to the full set of degenerate ground state and lower level eigenfunctions.
The set of trial wave functions has to be improved and orthogonalized
within each level,
the numbers of which increase,
and the gaps between which decrease,
with increasing system size.

These three problems do not exist for classical statistical mechanics
of large systems.
In the classical case there is apparently
no superposition of states,
no symmetrization of wave functions,
and indeed no need to guess,
or orthogonalize,
or indeed even obtain eigenfunctions.
Nowadays it is routine to use a computer
to characterize the physical properties
of a classical system of many thousands of particles.

Furthermore,
in the terrestrial sphere,
it is a matter of common observation
that most systems are dominated by classical behavior,
and that the quantum influence is either negligible
or no more than a small perturbation.
For example, in the case of liquid water at room temperature and pressure,
based on the value of $\rho \Lambda^3 $
($\Lambda$ is the thermal wave length and $\rho$ is the number density),
which characterizes spatial symmetrization effects,
the quantum correction to classical properties is approximately
one part in ten thousand.

These observations suggest that one should approach quantum mechanics
for many-particle systems
as a perturbation of  classical statistical mechanics.
That this is likely to be a fecund approach
is already hinted at by the fundamental difference
between quantum mechanics and quantum statistical mechanics.
The expectation value of an operator can be written
\cite{Messiah61,Merzbacher70}
\begin{equation}
%O(\psi) =
\langle \psi |\hat O | \psi \rangle
= \mbox{TR }  \hat \rho  \, \hat O ,
\end{equation}
where $\hat \rho \equiv | \psi\rangle \, \langle \psi |$
is the single wave function density operator.
The latter cannot be diagonalized
since it reflects a superposition of states.\cite{QSM1,QSM}
In contrast, the statistical average of a quantum operator is
\cite{Messiah61,Merzbacher70,Neumann27}
\begin{eqnarray}
\left< O \right>_\mathrm{stat}
& = &
\mbox{TR } \hat \wp  \, \hat O
\nonumber \\ & = &
\sum_{\bf n}\!'   \wp_{\bf n}   O_{\bf n n}^\mathrm{S} .
\end{eqnarray}
(The prime denotes the restriction to distinct states,
the formula for which is given in  \S \ref{Sec:Stat-Av} below.)
In this case the probability  operator can be diagonalized,
as in the second equality,
and the average becomes a weighted sum over a mixture of pure states.

Although it is widely known that the superposition of states
do not contribute to the quantum statistical average,
less  appreciated  is the mechanism by which superposition is suppressed,
and also the nature of the actual states that physically form the mixture.
The suppression of superposition states is tied to the collapse
of the wave function, and a deal of work has been done on this
in the context of open quantum systems.
\cite{Davies76,Breuer02,Weiss08}
A specific  example
is the quantum measurement process,
where so-called  environmental selection
has been modeled as  responsible for wave function collapse.\cite{Zeh01,Zurek03}
A more quantitative, thermodynamic treatment reveals
the nature of the collapsed states.
It has been shown that
due to interactions with a reservoir
(or heat bath, or environment),
the wave function of the sub-system (or open quantum system),
collapses into a mixture of entropy states.\cite{QSM1,QSM}
The second equality above invokes the entropy representation,
in which case the the probability operator is diagonal.
(It is fundamental to probability theory
that the probability is the exponential of the entropy,\cite{TDSM}
and so the operators are related by $ \hat \wp
=  e^{ \hat S_\mathrm{r}  /k_\mathrm{B} } /Z$.)
%= e^{ -\hat {\cal H}  /k_\mathrm{B}T }/Z$.

In this form the quantum statistical average is identical
to an average in classical probability theory:
it is a weighted sum over states.
In fact the result in classical statistical mechanics is entirely analogous,
\begin{equation}
\left< O \right>_\mathrm{cl}
=
\frac{1}{N! h^{3N} Z} \int \mathrm{d}{\bf \Gamma}\;
\wp({\bf \Gamma}) O({\bf \Gamma}) .
\end{equation}
Here a weighted integral over classical phase space appears.

Already at this introductory stage, one of the three challenges
posed by the quantum mechanics of many particle systems
has been solved:
use quantum statistical mechanics to eliminate the superposition of states.
There remain the problems of the symmetrization of the wave function
and of the identification, calculation, and orthogonalization
of the eigenstates and eigenfunctions.
But the direction of the solution to the two latter problems
is already indicated by the solution to the first problem.
First, since entropy states play a preferred role
in the collapse of the wave function,
one can anticipate that entropy eigenfunctions will provide
a unique basis in which to represent the quantum mechanics
of a many-particle system.
And second,
since one anticipates that classical statistical mechanics
will provide the leading term in the description
of a  many-particle system,
then one should also expect that each entropy microstate ${\bf n}$
can be identified with a point in classical  phase space ${\bf \Gamma}$.
In turn this means that the eigenvalues of the quantum many-particle system
must be the values of the  entropy function of classical phase space.

These arguments already identify the eigenstates
and eigenvalues in which one should cast quantum statistical mechanics
for its most efficient computation.
It remains of course to give the appropriate form
for the entropy eigenfunctions,
to symmetrize them, and to orthogonalise them.
Because one will eventually invoke the classical continuum,
rather than orthogonalise the entropy eigenfunctions,
it turns out to be better to account explicitly for their overlap
in the formulation of the partition function and the statistical average.
By good luck or good management,
the mathematical apparatus for dealing with non-orthogonality
will turn out to be identical
to that required for dealing with wave function symmetrization.
What will result below is a systematic expansion
for quantum statistical mechanics.
The first term is just classical statistical mechanics.
The second term distinguishes between bosons and fermions.

In identifying an entropy microstate with a point in classical phase space,
the theory developed below implicitly gives the probability
of simultaneous position and momentum states.
It should be stressed that this expression
is not directly related to the widely known expression
of Wigner\cite{Wigner32}
(see also Kirkwood,\cite{Kirkwood33}
and also Ch.~10 of Ref.~\onlinecite{Allen87}).
Wigner gave a quantum correction for thermodynamic equilibrium
that involved a function of simultaneous position and momentum
that shared many aspects with a phase space probability.
However, as Wigner himself recognized,\cite{Wigner32}
his probability-like function could take on negative values,
whereas the function given here is positive semi-definite.
Moreover, Wigner's function has no direct interpretation
as a quantity in classical statistical mechanics,
or as the quantum analogue of such a quantity,
whereas the eigenvalues of the entropy eigenfunction obtained here
have a direct classical interpretation.

The Wigner-Kirkwood expansion and the present theory
also disagree about the first order quantum correction
to the classical partition function and free energy.
This is derived in general in \S \ref{Sec:Zcl+Z1},
and explicitly for the ideal gas in \S \ref{Sec:IdealGas},
and for interacting particles in \S \ref{Sec:b2}.
%Following Allen and Tildesley, Ch.~10,\cite{Allen87}
The discrepancy between the present results
and that which arise from the Wigner-Kirkwood expansion
of the phase space distribution
function\cite{Wigner32,Kirkwood33,Allen87}
is due at least in part to the fact that the
Wigner-Kirkwood  analysis  does not distinguish between bosons and fermions
for the symmetry of the wave function,
whereas here this is taken into account.
Also, the  Wigner-Kirkwood  correction vanishes for the ideal gas,
whereas here the quantum corrections are non-zero.

More happily,
the present results agree
with that from the method of cluster expansions,
taken over to quantum system by Kahn and Uhlenbeck\cite{Kahn38}
and developed by Lee and Yang\cite{Lee59}
(at least for the second virial coefficient).
In contrast to Wigner and Kirkwood,
both the present and the Lee-Yang theories
give the same leading order correction
that distinguishes  bosons and fermions,
and that is also non-zero for the ideal gas.
From either the first term in the expansion
\cite{Uhlenbeck36,Uhlenbeck37}
or directly, the second virial coefficient can be obtained
(see Ch.~9 of Ref.~\onlinecite{Pathria72}).
This agrees with the low density limit of the present theory
for interacting quantum particles,
as is shown in \S \ref{Sec:b2}.
The present theory is also applied to the ideal quantum gas,
\S \ref{Sec:IdealGas}.
In \S \ref{Sec:Ideal-Packet-Fug},
the first two quantum corrections  are obtained explicitly
using wave packets in the large width limit.
In \S \ref{Sec:Ideal-Plane-Fug} using plane waves,
the full infinite series of quantum corrections is obtained explicitly.
Again these agree with the known results.

Although the present theory can be reduced to a fugacity expansion,
%in agreement with these  known results,
it is not actually an expansion in powers of density or fugacity.
Rather the expansion is in terms of increasing quantum permutation loop size,
which will be defined below.
This means that one can expect retaining even only the first
quantum correction will give accurate results for condensed matter systems
at terrestrial densities and temperatures.

A preliminary account of some of the material presented here can be found in
Ch.~5 and Appendix C of the author's book, Ref.~\onlinecite{QSM}.
This paper consolidates those results,
clarifies their meaning,
and extends them beyond the original presentation.
The justification for the general approach
and physical interpretation taken here,
and some of the specific results invoked in the derivations,
can be found elsewhere in the book
and are used here with only limited comment.
In particular, the collapse of the wave function
into entropy states will be taken directly from Ch.~1 of Ref.~\onlinecite{QSM}
without further justification.

%%%%%%%%%%%%%%%%%%%%%%%%%%%%%%%%%%%%%%%%%%%%%%%%%%%%%%%%%%%%%%%%%%%%%%%%%%
%
\section{Symmetrization of the Entropy Eigenfunction} \label{Sec:Symmetry}
\setcounter{equation}{0} \setcounter{subsubsection}{0}
%\renewcommand{\theequation}{\Alph{section}.\arabic{equation}}
%
%%%%%%%%%%%%%%%%%%%%%%%%%%%%%%%%%%%%%%%%%%%%%%%%%%%%%%%%%%%%%%%%%%%%%%%%%%

This section addresses two related issues:
the effect of wave function symmetrization
on the formulation of the partition function and statistical averages
in quantum statistical mechanics.
And the general formulation of quantum statistical mechanics
in the case that the chosen basis states are non-orthogonal.

It is of course fundamental to quantum mechanics that the wave function
must be either fully symmetric or fully anti-symmetric
with respect to the interchange of identical particles.
The consequences of this for enumerating quantum states
will be addressed in this section.
When one writes the trace as the sum over all microstates,
it will be shown that there is an additional factor due to symmetrization
that has to be included in the partition function and in statistical averages.

Although the set of eigenfunctions chosen as a basis in a given
quantum mechanical system are almost always chosen to be orthogonal,
there is no mathematical reason why a basis cannot include
sets of non-orthogonal degenerate eigenfunctions,
provided that the double counting that arises from this
is properly accounted for.
In many cases it may well be convenient to construct  an orthogonal set,
but there may be other cases where it is difficult or inefficient to do so.
One such case is the continuum limit,
where the distinction between individual states
ceases to have meaning,
as does the concept of the orthogonality of the eigenfunctions
that correspond to those states.
For this reason it is worthwhile to give a general formulation
of quantum statistical mechanics for non-orthogonal basis states
irrespective of the exact form
of the eigenfunctions corresponding to those states.
As will be shown in this section,
the treatment of this second issue
has the same functional form as the answer to the first problem.

%%%%%%%%%%%%%%%%%%%%%%%%%%%%%%%%%%%%%%%%%%%%%%%%%%%%%%%%%%%%%%%%%%%%%%%%%
\subsection{Orthogonal Entropy Eigenfunctions}

Consider an $N$ particle system, $j=1,2,\ldots,N$,
in three-dimensional space, $\alpha=x,y,z$.
Denote a normalized entropy eigenfunction by $\phi_{\bf m}({\bf r})$.
This is not yet symmetrized.
It is convenient to cast the following results in the position
representation:
${\bf r}= \{ {\bf r}_1,{\bf r}_2, \ldots, {\bf r}_N \}$
is the  $3N$-dimensional vector of particle positions.

The entropy microstates are labeled
by the integer vector ${\bf m}$.
In the simplest case,
which, for clarity and definiteness,
will sometimes be used as an example below,
the microstates consist of one-particle states,
${\bf m} = \{ {\bf m}_1,{\bf m}_2, \ldots, {\bf m}_N \}$.
In this picture each particle is in a definite state:
particle $j$ is in the state ${\bf m}_j$.
In the analysis of wave packets detailed in \S \ref{Sec:Entropy-Eigen} below,
the one-particle state labels ${\bf m}_j$ are  $6$-dimensional vector integers.
Wave packets are not essential for the general formalism,
and neither are one-particle states.

Messiah, Ch.~XIV, \S 2,\cite{Messiah61}
takes it as axiomatic that a complete set of dynamical variables
(position or momentum, spin) has a state space
that is the product of one-particle states.
Here instead the interest lies in entropy microstates,
and it will be assumed that there are configurations
where  some or all of the particles are clustered
into multi-particle states that cannot be decomposed
into a set of one-particle states.
Multi-particle states are not the same as multiple particles
in the same one-particle state.
An example is the entropy eigenfunction
for the collision of two particles,
which is often formulated as the product of a center of mass
wave function and a relative separation wave function
(cf.\ \S 9.5 of Pathria\cite{Pathria72}).
In the general multi-particle state case
there is a relationship between the order
of the elements of ${\bf m}$ and the order of the particle positions ${\bf r}$.
Unfortunately, it is difficult to be more precise
than this in the general case,
and so the multi-particle state formalism
makes for a less transparent notation
because neither the  precise make-up of the entropy microstate label ${\bf m}$,
nor the link between its components and
the individual particle positions in ${\bf r}$,
need be specified.

In the canonical equilibrium system,
which is the concern of this paper,
the entropy operator is proportional to the energy operator,
$\hat S_\mathrm{r} = - \hat{\cal H}/T$,
where $T$ is the temperature.
Therefore the entropy eigenfunctions are also energy eigenfunctions,
\begin{equation}
\hat{\cal H}({\bf r}) \, \phi_{\bf m}({\bf r})
= {\cal H}_{\bf m} \, \phi_{\bf m}({\bf r}) .
\end{equation}
Obviously  entropy states are also energy states,
and entropy eigenvalues are proportional to energy eigenvalues,
$S_{\mathrm{r},{\bf m}} =  -{\cal H}_{\bf m}/T$.

The entropy states are highly degenerate.
This means that entropy eigenfunctions with the same entropy eigenvalue
are not necessarily orthogonal.
However, one could orthogonalize them, by a Gram-Schmidt procedure or otherwise.
It will be assumed in the present sub-section
that the  entropy eigenfunctions for an orthonormal set,
\begin{equation}
\langle \phi_{\bf m'}\, | \, \phi_{\bf m} \rangle
= \delta({\bf m}'-{\bf m}) ,
\end{equation}
where the multi-dimensional Kronecker delta function
appears on the right hand side.
It will also be assumed that the set is complete,
$\sum_{\bf m}  \phi_{\bf m}({\bf r}) \, \phi_{\bf m}({\bf r}')
=  \delta({\bf r}-{\bf r}')$,
where the multi-dimensional Dirac delta function
appears on the right hand side.

%%%%%%%%%%%%%%%%%%%%%%%%%%%%%%%%%%%%%%%%%%%%%%%%%%%%%%%%%%%%%%%%%%%
\subsubsection{Symmetrization}

In quantum mechanics,
with respect to particle interchange,
the wave function is fully symmetric for identical bosons
and fully anti-symmetric for identical fermions.
Hence one must have
\cite{Messiah61,Merzbacher70}
\begin{eqnarray} \label{Eq:zeta^SA}
\phi_{{\bf m}}^\mathrm{S/A}({\bf r})
& \equiv &
\frac{1}{\sqrt{N!\chi_{\bf m}} }
\sum_{\hat{\mathrm P}} (\pm 1)^p
 \, \phi_{{\bf m}}(\hat{\mathrm P}{\bf r})
\nonumber \\ & \equiv &
\frac{1}{\sqrt{N!\chi_{\bf m}} }
\sum_{\hat{\mathrm P}} (\pm 1)^p
 \, \phi_{\hat{\mathrm P}{\bf m}}({\bf r}) .
\end{eqnarray}
The overlap factor,
which would be more precisely written $\chi^\pm(\phi_{\bf m})$,
is explained in the next sub-subsection.
The superscript  S signifies symmetric,
and it applies for bosons using $(+ 1)^p=1$ on the right hand side.
The superscript A signifies anti-symmetric,
and it applies for fermions using $(- 1)^p$ on the right hand side.
Here $\hat{\mathrm P}$ is the permutation operator,
and $p$ is its parity
(ie.\ the number of pair transpositions
that comprise the permutation).

The permutation operator changes the order
of the position arguments in the first equality,
which is to say that it interchanges particles.
There are of course $N!$ possible permutations of $N$ particles.
Since there must be a link between the order of the arguments
and the order of the elements in the microstate vector,
one can equivalently permute the latter,
as in the second equality.
(More precisely, a permutation of the arguments
is the same as the conjugate permutation of the one-particle states.)
In the case that the microstate consists of one-particle states,
for a given term on either right hand side,
one may have that the single particle state ${\bf m}_j$ applies
to the particle with position ${\bf r}_k$, $k \ne j$,
depending upon the particular permutation.
In the case that the microstates consist of multi-particle states,
the permutation operator may  be applied conceptually to the
elements of the microstate vector,
without being specific on the link between those elements and the particles.

%%%%%%%%%%%%%%%%%%%%%%%%%%%%%%%%%%%%%%%%%%%%%%%%%%%%%%%%%%%%%%%%%%%
\subsubsection{Overlap Factor}

The prefactor including $\chi_{\bf m}$ ensures the correct normalization,
\begin{equation}
\langle \phi^\mathrm{S/A}_{\bf m'}\, | \, \phi^\mathrm{S/A}_{\bf m} \rangle
= \delta({\bf m}'-{\bf m}) .
\end{equation}
This holds even if ${\bf m}'$ is a non-identical permutation of ${\bf m}$,
such that ${\bf m}' \equiv \hat{\mathrm P} {\bf m} = {\bf m}$,
$\hat{\mathrm P}\ne {\mathrm I}$.
This can occur if more than one particle is in the same one-particle state.

When more precision is needed,  the overlap factor will be denoted
$\chi_{\bf m} \equiv \chi^\pm(\phi_{\bf m}) $.
The normalization gives it as
\begin{eqnarray} \label{Eq:chi_m}
\chi_{\bf m}
& = &
\frac{1}{N!}
\sum_{\hat{\mathrm P},\hat{\mathrm P}'} (\pm 1)^{p+p'}
\langle \phi_{{\bf m}}(\hat {\mathrm P}{\bf r})
| \phi_{{\bf m}}(\hat {\mathrm P}'{\bf r}) \rangle
\nonumber \\  & = &
\sum_{\hat {\mathrm P}}
(\pm 1)^{p}
\langle \phi_{{\bf m}}(\hat {\mathrm P}{\bf r})
| \phi_{{\bf m}}({\bf r}) \rangle
\nonumber \\ & = &
\sum_{\hat{\mathrm P}} (\pm 1)^{p}
\int \mathrm{d}{\bf r}\;
\phi_{{\bf m}}(\hat{\mathrm P}{\bf r})^*
\phi_{{\bf m}}({\bf r}) .
\end{eqnarray}

The quantity $\chi_{\bf m}$ will be called
the overlap factor,
because it tells how much symmetrization
counts the same microstate multiple times.
In the present case a permutation of identical particle state labels
gives complete overlap of the entropy eigenfunctions,
and permutation of different labels no matter how close gives no overlap.
In the case treated in the next subsection,
permutation of nearby labels give partial overlap.

The overlap factor is most clearly explained in terms of one-particle states.
However, the formalism itself holds in general for multi-particle states.

For fermions,
no more than one particle may be in any one state,
and so any permutation of an allowed state
$\hat {\mathrm P} \ne \hat {\mathrm I}$
leads to $\hat {\mathrm P}{\bf m} \ne {\bf m}$ and
$\langle \phi_{{\bf m}} | \phi_{\hat {\mathrm P}{\bf m}} \rangle = 0$.
Hence for allowed fermion microstates, $\chi_{\bf m} = 1$.
Actually one can still formally use the formula above
for fermions even for microstates ${\bf m}$
that correspond to multiple occupancy of one or more single particle states.
For such forbidden states $\chi_{\bf m} = 0$,
and the symmetrized eigenfunction normalization factor,
$(\chi_{\bf m} N!)^{-1/2}$, diverges.
But this divergence goes as the square root,
whereas the symmetrized wave function itself vanishes linearly,
which means that
$ \phi_{ {\bf m}}^\mathrm{A}({\bf r})$ itself vanishes
for multiply-occupied fermion one-particle states.
Of course the overlap factor itself
is well-defined for such forbidden states, $\chi^-_{\bf m} = 0$,
and it is this that will occur in the formula given below.

For bosons,
more than one particle may occupy a given state.
Because of the orthogonality of the present entropy eigenfunctions,
the only non-zero contributions
are permutations that consist of
the transposition of identical one-particle states.
For example,
$\langle \phi_{\hat {\mathrm P}_{jk} {\bf m}} | \phi_{{\bf m}} \rangle =
\delta({\bf m}_j - {\bf m}_k)$.
(The permutation operator
for the transposition of particles $j$ and $k$
is denoted $\hat{\mathrm P}_{jk}$.)
Suppose
that the microstates are ordered, %$\hat O{\bf n} = {\bf n}$,
and that the first $M_1$ occupied  single particle states are the same
(ie.\ ${\bf m}_1 = {\bf m}_2 = \ldots = {\bf m}_{M_1}$),
the next $M_2$ particles are in the same state
(ie.\ ${\bf m}_{M_1+1} = {\bf m}_{M_1+2} = \ldots = {\bf m}_{M_1+M_2}$),
etc.,
with $\sum_j M_j = N$.
(It would be more precise to write $M_j({\bf m})$,
but for simplicity the microstate dependence of the occupancy numbers
is implicitly understood rather than explicitly stated.)
Clearly there are $\prod_j M_j!$ permutations $\hat {\mathrm P}$
for which  $\langle \phi_{\hat {\mathrm P}{\bf m}} | \phi_{{\bf m}} \rangle
= 1$, with the inner product vanishing for all other permutations.
Hence for bosons $\chi^+_{\bf m} = \prod_j M_j!$.

The formula for the overlap factor, Eq.~(\ref{Eq:chi_m}),
holds for multi-particle states,
as well as for when the entropy microstate consists of
one-particle states.
It will be shown in the next sub-section
that it also holds when the degenerate entropy eigenfunctions
are not orthogonal.

In general for identical particles the Hamiltonian operator
is unchanged by a permutation of the particles,
\begin{equation}
\hat {\cal H}({\bf r})
=
\hat {\cal H}(\hat{\mathrm P}{\bf r}).
\end{equation}
Obviously the same symmetry holds for the entropy operator.
This means that the symmetrized wave function remains
an eigenfunction of the entropy operator
with unchanged eigenvalue.

%%%%%%%%%%%%%%%%%%%%%%%%%%%%%%%%%%%%%%%%%%%%%%%%%%%%%%%%%%%%%%%%%%%
\subsubsection{Partition Function}

Because of symmetrization,
eigenfunctions that differ only by the permutation of the position arguments
are the same eigenfunction.
Similarly, for the case of one-particle states,
states that differ only by the permutation
of the single particle state indeces
are the same state.
One therefore needs some convention for ordering the entropy microstates.
This unique ordered arrangement may be denoted $\hat O {\bf m}$.
Obviously $  \hat O \hat{\mathrm P} {\bf m}  = \hat O {\bf m}  $.
An example of an ordering convention is
${\bf m}' = \hat O {\bf m} \Leftrightarrow
{\bf m}_1' \le {\bf m}_2' \le \ldots \le {\bf m}_N'$,
but this is not essential.
One has that
$\phi_{\bf m}^\mathrm{S/A}({\bf r})
= (\pm 1)^{o_{\bf m}} \phi_{\hat O{\bf m}}^\mathrm{S/A}({\bf r})$,
where $o_{\bf m}$ is the parity of the permutation that is required
to order ${\bf m}$.

This point holds also for multi-particle states.
The unique ordered arrangement, $\hat O {\bf m}$,
allows the distinct states of the system to be enumerated.

In general the free energy of a system
is minus the temperature times the total entropy
(ie.\ the entropy of the total system,
which is that of the sub-system
plus that of the reservoir with which it can exchange energy).
\cite{TDSM}
The  total entropy is the logarithm of the number of states,
which is the logarithm of the partition function.
This raises the questions: which states, and what is their weight?

It can be shown that the states are the entropy microstates
of the total system, and these have equal weight.\cite{QSM}
It also can be shown that  for the canonical equilibrium system
the number of reservoir microstates
for each sub-system entropy microstate is proportional
to the Boltzmann factor
(the exponential of the negative sub-system energy
divided by the temperature $T$ and Boltzmann's constant $k_\mathrm{B}$).
\cite{TDSM,QSM}
Hence the partition function is the sum over sub-system entropy microstates
weighted by the Boltzmann factor.

A crucial point for the present analysis is that
this sum must be over distinct states.
Obviously in counting the total number of states
one should not count the same state more than once.
Therefore  the partition function for the canonical equilibrium system is
\begin{eqnarray}
Z(N,V,T)
& = &
\mbox{TR } e^{-\beta \hat{\cal H}}
\nonumber \\ & \equiv &
\sum_{\hat O{\bf m}}  e^{-\beta {\cal H}_{\bf m}}
\nonumber \\ & = &
\frac{1}{N!} \sum_{\bf m} \chi_{\bf m} e^{-\beta {\cal H}_{\bf m}} .
\end{eqnarray}
Here and below $\beta \equiv 1/k_\mathrm{B}T$.
The second equality \emph{defines} that the trace means
the sum over distinct states.
Here the sum is over only the microstates
in which the single particle states are ordered.
(In the introduction a prime was used to denote this:
$\sum_{\hat O{\bf m}} \equiv \sum_{{\bf m}}'$.)
In the third equality, the sum is over all microstates,
with the factor $\chi_{\bf m}/N!$ correcting for multiple counting of
the same state.
The meaning of $ \chi_{\bf m}$ here is the same as in the normalization,
but it is worth repeating the argument
as it has a direct physical interpretation in terms of counting
the number of distinct microstates.
The discussion is given in terms of one-particle states.

If in the microstate ${\bf m}$ all the $N$ particles
are in different single particle states,
then each of the $N!$ permutations of these single particle states
are different and each is counted in the summation,
and the correction factor must be $1/N!$,
(ie.\ $\chi_{\bf m} = 1$).
If  in the microstate ${\bf m}$
there are $M_1$ particles in the first occupied single particle state,
$M_2$ in the second, etc.,
with $\sum_j M_j = N$,
then there are $N!/M_1!M_2!\ldots$ distinct permutations
of these single particle states,
and the correction factor is
$M_1!M_2!\ldots/N!$
(ie.\  $\chi_{\bf m} = \prod_j M_j!$).
For the case of fermions, $\chi_{\bf m}^- = 0 $
if any $M_j > 1$.

\comment{ %%%%%%%%%%%%%%%%%%%%%%%%%%%%%%%%%%%%%%%%%%%%%%%%%%%%%%%%%
It is straightforward to mathematically derive the above result.
One has
\begin{eqnarray}
Z(N,V,T)
& = &
\mbox{TR }  e^{-\beta \hat{\cal H}}
\nonumber \\ & = &
\sum_{\hat O{\bf m}}
\langle \phi_{\bf m}^\mathrm{S/A} |
 e^{-\beta \hat{\cal H}} | \phi_{\bf m}^\mathrm{S/A} \rangle
\nonumber \\ & = &
\sum_{\hat O{\bf m}}
e^{-\beta {\cal H}_{\bf m}}
\nonumber \\ & = &
\frac{1}{N!} \sum_{\bf m} \chi_{\bf m} e^{-\beta {\cal H}_{\bf m}}
\end{eqnarray}
} % end comment %%%%%%%%%%%%%%%%%%%%%%%%%%%%%%%%%%%%%%%%%%%%%%%%%%%%%

This point about only counting distinct states
has also been made by Messiah, Ch.~XIV, \S 6.\cite{Messiah61}
The symmetrized wave function normalization factor given by him
in the case of one-particle, orthogonal states
is equivalent to that given here.
Messiah does not explicitly use this factor
to convert the partition function or average to a sum over all states.
\cite{Messiah61}
Conversely, Pathria, Eq.~(9.6.2),\cite{Pathria72} gives the partition function
as the Boltzmann weighted sum over entropy (energy) states,
with the implication being that these are all states
(the issue of distinct states is not raised),
but no correction or overlap factor is exhibited.
The dearth of explicit treatments in the literature of the overlap factor
in the  sum over all states is perhaps a little surprising.
Perhaps the definition of trace is assumed implicitly
to restrict it to be the sum over distinct states
and it is not seen as necessary to explicitly reformulate this
as a sum over all states.
However for a number of practical reasons  it is actually
more convenient to invoke an unrestricted sum.
In this case
each entropy microstate has a specific number of non-zero permutations
associated with it, and since this number is microstate dependent,
the Boltzmann factor alone is not sufficient to correctly
weight each state in the unrestricted sum.

%There is some evidence that supports the first statement made
%in the penultimate paragraph.
It is clear that  $\chi_{\bf m}$ differs for fermions and bosons.
(It is more precise to denote it
$\chi^\pm(\phi_{\bf m})$.)
Hence the partition function and the free energy differ
for the two cases.
In contrast, the long-standing  Wigner-Kirkwood expansion
of the phase space distribution function
gives a first quantum correction for the Helmholtz free energy
that is the same for bosons as for fermions.\cite{Wigner32,Kirkwood33}
This result is repeated in contemporary texts.\cite{Allen87}
The problem stems from the fact  that
the Wigner-Kirkwood expansion does not take into account
the symmetrization of the wave function.
This again suggests that it is unfortunately common
to overlook the consequences of wave function symmetrization:
since permuted states are equivalent,
the partition function must be restricted to the sum over distinct states only,
and since bosons and fermions have different numbers of distinct states,
the partition function  and the free energy must differ
for bosons and for fermions.

%Similarly the otherwise reliable text by Pathria\cite{Pathria72}
%neglects such a factor in the partition function
%(although the explicit results given  by Pathria
%for the second virial coefficient do distinguish between bosons and fermions).

%%%%%%%%%%%%%%%%%%%%%%%%%%%%%%%%%%%
\subsubsection{Statistical Average} \label{Sec:Stat-Av}

The factor of $\chi_{\bf m}/N!$ is evidently the weight
that needs to be applied to the entropy microstates
(in addition to the reservoir weight,
which is the Maxwell-Boltzmann factor).
In consequence the statistical average of an operator must be
\begin{eqnarray}
\left< \hat O \right>_{N,V,T}
& = &
\mbox{TR } \hat \wp \, \hat O
\nonumber \\ & \equiv &
\sum_{\hat O{\bf m}}   \wp_{\bf m} \, O_{\bf mm}^\mathrm{S}
\nonumber \\ & = &
\sum_{\bf m}
\frac{\chi_{\bf m}}{N!} \, \wp_{\bf m} \, O_{\bf mm}^\mathrm{S}
\nonumber \\ & = &
\sum_{\bf m} \frac{\chi_{\bf m}}{N!}
\, \frac{e^{-{\cal H}_{\bf m}/k_\mathrm{B}T}}{Z(N,V,T)}
\, \langle {\bf m} | \hat O | {\bf m} \rangle .
\end{eqnarray}
(The superscript S here stands for the entropy representation,
not symmetric wave function.)
It seems best to keep the probability operator
in the traditional Maxwell-Boltzmann form
and to show the quantum symmetry weight factor explicitly.

%%%%%%%%%%%%%%%%%%%%%%%%%%%%%%%%%%%%%%%%%%%%%%%%%%%%%%%%%%%%%%%%%%%%%%%%%
\subsection{Non-Orthogonal Entropy  Eigenfunctions} \label{Sec:non-ortho-zeta}

The previous sub-section dealt with orthogonal entropy eigenfunctions.
These need to be distinguished from the eigenfunctions
that will be used in the analysis  in the present sub-section and below.
Although it will be argued that there is every reason to regard
the present eigenfunctions
just as legitimate as the conventional eigenfunctions
invoked above,
there are two aspects that call for closer scrutiny
in the present case:
the present eigenfunctions may be approximate entropy eigenfunctions,
and they need not form an orthogonal set.

The present eigenfunctions are denoted $\zeta_{\bf n}({\bf r})$.
Although it is simplest to assume that
the microstate label is an $N$-dimensional vector
of one-particle microstates,
${\bf n} = \{ {\bf n}_1, {\bf n}_2, \ldots, {\bf n}_N\}$,
this is not explicitly invoked,
and the following analysis applies as well to the multi-particle state case.
%(see \S \ref{Sec:local} below for more explicit analysis).
Permutations will usually be denoted in terms of states,
$\zeta_{\hat{\mathrm P}{\bf n}}({\bf r})$,
even though in the  multi-particle state case
this is most readily realized as the (conjugate) permutation
of the particle positions,
 $\zeta_{{\bf n}}(\hat{\mathrm P}^\dag{\bf r})$.

The eigenvalue equation is
\begin{equation}
\hat{\cal H} ({\bf r}) \, \zeta_{\bf n}({\bf r})
\approx {\cal H}_{\bf n} \, \zeta_{\bf n}({\bf r}).
\end{equation}
It is assumed that the eigenfunctions are such that
this can be systematically improved to make the inherent error in this
as small as necessary.
In the sections below it will be shown that
in the case that wave packets are used as the chosen  eigenfunctions,
then these become exact in the thermodynamic limit.

It will be assumed that the eigenfunctions obey a soft orthogonality,
\begin{equation}
\langle \zeta_{\bf n'} | \zeta_{\bf n} \rangle
= \delta_\xi({\bf n}'-{\bf n}).
\end{equation}
The function is a soft Kronecker-$\delta$,
with unit peak $\delta_\xi({\bf 0})= 1$
(ie.\ the eigenfunctions are normalized,
$\langle \zeta_{\bf n} | \zeta_{\bf n} \rangle =1$),
but it decays over a finite width characterized by $\xi$
(which means that the eigenfunctions are not orthogonal).
On physical grounds the most common realization of this would be a Gaussian.
The important point is that although the present eigenfunctions
do not form an orthogonal set,
they nevertheless enjoy an approximation to orthogonality.

In Appendix \ref{Sec:genoverlap},
based on the Gram-Schmidt orthogonalization procedure,
it is shown that the `true' entropy eigenfunctions
can be written as a linear combination of this complete set of non-orthogonal
approximate entropy eigenfunctions,
\begin{equation}
\phi_{\bf m}({\bf r})
=
\sum_{\bf n} c_{{\bf m} {\bf n}} \, \zeta_{\bf n}({\bf r}) .
\end{equation}
The coefficients are given by
\begin{eqnarray}
 c_{{\bf m} {\bf n}}
 & = &
\frac{1}{\eta} \left\{
\langle \zeta_{\bf n}|\phi_{\bf m}\rangle \,
- \sum_{\bf n'}\!\!^{(\ne {\bf n})}
\langle \zeta_{\bf n}|\zeta_{\bf n'}\rangle \,
\langle \zeta_{\bf n'}|\phi_{\bf m}\rangle
\right\}
\nonumber \\ & = &
\frac{1}{\eta}
\left\{
2 \langle \zeta_{\bf n}|\phi_{\bf m}\rangle \,
- \sum_{\bf n'}
\delta_{\xi}({\bf n}-{\bf n'}) \,
\langle \zeta_{\bf n'}|\phi_{\bf m}\rangle \right\}
\nonumber \\ & \approx &
\frac{1}{\eta}
\left\{
2 \langle \zeta_{\bf n}|\phi_{\bf m}\rangle \,
- \langle \zeta_{\bf n}|\phi_{\bf m}\rangle
\sum_{\bf n'} \delta_{\xi}({\bf n}-{\bf n'}) \,
 \right\}
\nonumber \\ & = &
\frac{2-\lambda}{\eta}
\langle \zeta_{\bf n}|\phi_{\bf m}\rangle .
\end{eqnarray}
Here
$\eta = (2-\lambda) \lambda$,
and
\begin{equation}
\lambda \equiv \sum_{\bf n'} \delta_{\xi}({\bf n}-{\bf n'})
\end{equation}
are constants that are assumed  independent of the microstate ${\bf n}$.
(In  Appendix \ref{Sec:genoverlap}, Eq.~(\ref{Eq:lambda(xi)}) shows that
for the case that the  entropy eigenfunction is a wave packet,
then  $ \delta_{\xi}({\bf n}-{\bf n'})$ is a Gaussian,
and $\lambda$ is indeed independent of the microstate ${\bf n}$,
and also of the wave packet width $\xi$.
Further, for wave packets and
the conventional volume element of classical statistical mechanics,
$\Delta_{\bf q}\Delta_{\bf p} = h$,
one has $\lambda = \left[{h}/{\Delta_{\bf q}\Delta_{\bf p}}\right]^{3N}
= 1$,
and hence $\eta = 1$.)
From the fact that
$c_{{\bf m} {\bf n}} \propto \langle \zeta_{\bf n}|\phi_{\bf m}\rangle $,
compared to the orthogonal case one sees that
the only thing that non-orthogonality introduces is a scale factor,
and for wave packets this is just unity.

Since the set $\zeta_{\bf n}$ is complete,
one can also show that
\begin{equation}
\frac{\lambda(2-\lambda)^2}{\eta^2}
\sum_{\bf n}
|\zeta_{\bf n}\rangle \,
\langle \zeta_{\bf n}|
= {\bf I} ,
\end{equation}
or
$ ({\lambda(2-\lambda)^2}/{\eta^2})
\sum_{\bf n}
 \zeta_{\bf n}({\bf r}')^* \,  \zeta_{\bf n}({\bf r})
= \delta( {\bf r}'-{\bf r})$.

With these results,
the partition function can be transformed
into a sum over the non-orthogonal approximate entropy eigenstates,
\begin{eqnarray}
\lefteqn{
Z(N,V,T)
} \nonumber \\
& = &
\sum_{\hat O{\bf m}}  e^{-\beta {\cal H}_{\bf m}}
\nonumber \\ & = &
\frac{1}{N!} \sum_{\bf m} \chi^\pm(\phi_{\bf m}) e^{-\beta {\cal H}_{\bf m}}
\nonumber \\ & = &
\frac{1}{N!} \sum_{\bf m}
\sum_{\hat P} (\pm 1)^{p}
\langle \phi_{\hat P{\bf m}} | \phi_{\bf m} \rangle
e^{-\beta {\cal H}_{\bf m}}
\nonumber \\ & = &
\frac{1}{N!} \sum_{\bf m}
\sum_{\hat P} (\pm 1)^{p}
\langle \phi_{\hat P{\bf m}} |  e^{-\beta \hat {\cal H}}\phi_{\bf m} \rangle
\nonumber \\ & = &
\frac{\lambda(2-\lambda)^2}{\eta^2 N!} \sum_{\bf n}
\sum_{\hat P} (\pm 1)^{p} \sum_{\bf m}
\langle \phi_{\hat P{\bf m}} | \zeta_{\bf n} \rangle \,
\langle \zeta_{\bf n} |  e^{-\beta \hat {\cal H}}\phi_{\bf m} \rangle
\nonumber \\ & = &
\frac{\lambda(2-\lambda)^2}{\eta^2 N!} \sum_{\bf n}
\sum_{\hat P} (\pm 1)^{p} \sum_{\bf m}
\langle \phi_{\bf m} | \zeta_{\hat P{\bf n}} \rangle \,
\langle e^{-\beta \hat {\cal H}} \zeta_{\bf n} | \phi_{\bf m} \rangle
\nonumber \\ & \approx &
\frac{\lambda(2-\lambda)^2}{\eta^2 N!} \sum_{\bf n}
\sum_{\hat P} (\pm 1)^{p} \sum_{\bf m}
\langle \zeta_{\bf n} | \phi_{\bf m} \rangle \,
\langle \phi_{\bf m} | \zeta_{\hat P{\bf n}} \rangle
e^{-\beta {\cal H}_{\bf n}}
\nonumber \\ & = &
\frac{\lambda(2-\lambda)^2}{\eta^2 N!}
\sum_{\bf n} \chi^\pm(\zeta_{\bf n}) e^{-\beta {\cal H}_{\bf n}} .
\end{eqnarray}
The third last equality invokes the Hermitian nature of the energy operator.
The approximation made in the penultimate equality
is that $\zeta_{\bf n}$ is an eigenfunction of $\hat {\cal H}$.
The overlap factor for non-orthogonal states
has the same functional form as for orthogonal states,
\begin{eqnarray} \label{Eq:chi-zeta}
\chi^\pm(\zeta_{\bf n})
& = &
\sum_{\hat P} (\pm 1)^{p}
\langle \zeta_{\hat P{\bf n}} | \zeta_{{\bf n}} \rangle
\nonumber \\ & = &
\sum_{\hat{\mathrm P}} (\pm 1)^{p}
\int \mathrm{d}{\bf r}\;
\zeta_{{\bf n}}(\hat{\mathrm P}{\bf r})^*
\zeta_{{\bf n}}({\bf r})
\nonumber \\ & = &
\sum_{\hat P} (\pm 1)^{p}
\delta_{\xi}({\bf n}-\hat P{\bf n}).
\end{eqnarray}

Since the prefactor $\lambda(2-\lambda)^2/\eta^2$ is a constant,
in general it can be discarded because it becomes an additive constant
for the free energy: only differences in thermodynamic potentials
have physical consequences.
For wave packets and
the conventional volume element of classical statistical mechanics,
$\Delta_{\bf q}\Delta_{\bf p} = h$,
one has $\lambda = \eta =1$.
(In \S \ref{Sec:olxi}  below the possibility is raised that
the optimum wave packet width $\xi$ varies with particle number $N$,
and possibly also with the microstate  ${\bf n}$.
But  Appendix \ref{Sec:genoverlap}, Eq.~(\ref{Eq:lambda(xi)}), shows
that for wave packets, $\lambda$ is independent of $\xi$,
and hence of $N$ and of ${\bf n}$.)
Hence the partition function for non-orthogonal entropy states
is formally the same as for orthogonal states,
\begin{equation} \label{Eq:Z-chi-zeta}
Z(N,V,T)
=
\frac{1}{N!}
\sum_{\bf n} \chi^\pm(\zeta_{\bf n}) e^{-\beta {\cal H}_{\bf n}} .
\end{equation}
All of the quantitative change to the weighting of the states
is carried by the overlap factor $\chi^\pm(\zeta_{\bf n})$.
Obviously an analogous result holds for the statistical average.

%%%%%%%%%%%%%%%%%%%%%%%%%%%%%%%%%%%
\subsubsection{Dimer Overlap Integral}

Before proceeding to the next section
on the formally exact expansion of the partition function,
it is worth seeking some physical insight into the overlap factor.
The discussion is cast in terms of single particle states,
which have the most intuitive physical interpretation.
(The same result in essence
is obtained more formally in the following section
for multi-particle states.)
One may take the soft Kronecker-$\delta$, $\delta_\xi({\bf n}'-{\bf n})$,
to characterize the proximity of the two states that are its argument.
This is the continuum analogue of state occupancy.

Suppose that in a given microstate ${\bf n}$ there is a single dimer,
which is to say just two particles whose respective single particle states are
in close proximity.
Without loss of generality, these can be taken to be particles 1 and 2.
The only permutations that contribute to the symmetrization
of the wave function are the identity and the transposition
of these two particles.
Hence in this dimer case one has
(writing the wave function as the product of one-particle wave functions
for simplicity,
and using the normality of the entropy eigenfunctions
to remove the non-permuted particles and indeces)
\begin{eqnarray} \label{Eq:chi(1,2)-tot}
\chi_{\bf n}
& = &
\int \mathrm{d}{\bf r}_1  \mathrm{d}{\bf r}_2 \;
\left\{
\zeta_{{\bf n}_1}({\bf r}_1)^*
\zeta_{{\bf n}_2}({\bf r}_2)^*
\zeta_{{\bf n}_1}({\bf r}_1) \zeta_{{\bf n}_2}({\bf r}_2)
\right. \nonumber \\ && \left. \mbox{ }
\pm
\zeta_{{\bf n}_2}({\bf r}_1)^*
\zeta_{{\bf n}_1}({\bf r}_2)^*
\zeta_{{\bf n}_1}({\bf r}_1)\zeta_{{\bf n}_2}({\bf r}_2)
\right\}
\nonumber \\ & = &
\left\{ 1 \pm
\int \mathrm{d}{\bf r}_1
\zeta_{{\bf n}_2}({\bf r}_1)^* \zeta_{{\bf n}_1}({\bf r}_1)
\right. \nonumber \\ && \left. \mbox{ } \times
\int \mathrm{d}{\bf r}_2 \;
\zeta_{{\bf n}_1}({\bf r}_2)^* \zeta_{{\bf n}_2}({\bf r}_2)
\right\}
\nonumber \\ & = &
1 \pm
\left| \int \mathrm{d}{\bf r}_1
\zeta_{{\bf n}_2}({\bf r}_1)^* \zeta_{{\bf n}_1}({\bf r}_1) \right|^2 .
\end{eqnarray}
This is real and positive.
The magnitude of the integral is less than or equal to unity.
Hence the right hand side is between zero and one for fermions,
and between one and two for bosons.
If ${\bf n}_1 ={\bf n}_2$, the integral would be unity,
and for bosons the right hand side would be two,
and for fermions it would be zero.
For non-orthogonal or continuous states
it would be intermediate between these two values.

%\newpage
%%%%%%%%%%%%%%%%%%%%%%%%%%%%%%%%%%%%%%%%%%%%%%%%%%%%%%%%%%%%%%%%%%%%%%%%%
%
\section{Expansion of the Grand Partition Function} \label{Sec:Zcl+Z1}
\setcounter{equation}{0} \setcounter{subsubsection}{0}
%
%%%%%%%%%%%%%%%%%%%%%%%%%%%%%%%%%%%%%%%%%%%%%%%%%%%%%%%%%%%%%%%%%%%%%%%%%

\subsection{Localization of Permuted States} \label{Sec:Local-n-r}

The  subsection that follows this one
gives an expansion for the partition function
that is based on a general series representation of the permutation operator,
the terms of which involve the overlap of a permutation of a set of particles.
The dimer overlap factor just discussed is an example.
Slightly more generally using multi-particle states,
for $N$ particles in the entropy microstate ${\bf n}$,
the transposition of particles $j$ and $k$
gives a dimer permutation overlap factor of
\begin{eqnarray}
\chi_{{\bf n};jk}^{(2)}
& = &
\pm
\langle \zeta_{\bf n}(\hat P_{jk}{\bf r}) | \zeta_{\bf n}({\bf r}) \rangle
 \\ & = &
\pm \int \mathrm{d}{\bf r} \;
\zeta_{\bf n}(\ldots  {\bf r}_k \ldots  {\bf r}_j  \ldots)^*
\zeta_{\bf n}(\ldots  {\bf r}_j \ldots  {\bf r}_k  \ldots) . \nonumber
\end{eqnarray}

The question addressed in the present sub-section is to what extent
the states ${\bf n}$ that give a non-zero overlap factor
correspond to a localization of the permuted particles' states.
For example, does the inner product of the non-permuted
particles cancel them out leaving just a two-particle state,
\begin{equation}
\chi_{{\bf n};jk}^{(2)}
=
\pm \int \mathrm{d}{\bf r}_j \,  \mathrm{d}{\bf r}_k\;
\zeta_{{\bf n}_{jk}}( {\bf r}_k ,  {\bf r}_j )^*
\zeta_{{\bf n}_{jk}}( {\bf r}_j ,  {\bf r}_k ) ?
\end{equation}
Obviously if the eigenfunction were the product of one-particle eigenfunctions,
as in the example above, this would hold.
Perhaps less drastically,
can one  identify a localized cluster $c< N$ of particles
in a multi-particle state ${\bf n}_c \in  {\bf n}$
that contain the permuted particles,
\begin{equation}
\chi_{{\bf n};jk}^{(2)}
=
\pm \int \mathrm{d}{\bf r}^c
\zeta_{{\bf n}_{c}}( \hat P_{jk}{\bf r}^c)^*
\zeta_{{\bf n}_{c}}({\bf r}^c) ?
\end{equation}

The reason for discussing the possibility of localized clusters
is that the product of closed permutation loops will occur in the expansion
below.
If each permutation loop can be considered to be contained within a
localized cluster, then their states are independent
and the overlap of the product factorizes into the product
of the individual permutation loop overlap factors.
This considerably simplifies the subsequent analysis.

The concept of localization applies to the microstate ${\bf n}$
rather than to the particles' positions ${\bf r}$ (but see below).
The permutation of far-separated particle positions does not necessarily
give zero overlap in every state.
For example the wave function for two real particles tends to the product
of two free particle wave functions in the limit of large separations,
$\zeta_{{\bf n}_{12}}({\bf r}_1,{\bf r}_2) \rightarrow
V^{-1} e^{i {\bf k}_1 \cdot {\bf r}_1 } e^{i {\bf k}_2 \cdot {\bf r}_2 }$,
$ r_{12} \rightarrow \infty$.
The dimer overlap factor for two such ideal particles is
\begin{eqnarray}
\chi_{{\bf n}_{12}}^{(2)}
& = &
\frac{\pm 1}{V^2}
\int \mathrm{d}{\bf r}_1 \,  \mathrm{d}{\bf r}_2\;
e^{i {\bf k}_{12} \cdot {\bf r}_1 }
e^{i {\bf k}_{21} \cdot {\bf r}_2 }
\nonumber \\ & = &
\pm \delta({\bf k}_{1}-{\bf k}_{2}) ,
\end{eqnarray}
where a Kronecker-$\delta$ appears.
This is only non-zero if the two particles are in the same
one-particle state, which is as close together as one can get.
Even though the product of the original
and the permuted wave functions
have non-zero overlap,
$e^{i {\bf k}_{12} \cdot {\bf r}_{12} }  \ne 0$,
for any positions ${\bf r}_1$ and ${\bf r}_2$
and any states ${\bf k}_1$ and ${\bf k}_2$,
upon integration only neighboring, in fact identical, single particle
states give a non-zero overlap factor.

One can instead look at localization
in the position representation.
For the case of a two particle ideal system,
Pathria,  Eq.~(5.5.24),\cite{Pathria72}
gives the probability of a particular configuration
as proportional to
\begin{eqnarray}
\wp({\bf r}_1,{\bf r}_2)
& \propto &
\langle {\bf r}_1,{\bf r}_2
| e^{- \beta \hat{\cal H}}
| {\bf r}_1,{\bf r}_2\rangle
%\nonumber \\ & = &
%\frac{V^2 \Lambda^{-6}}{2!}
%\left\{ 1 \pm
%f({\bf r}_1-{\bf r}_2) f({\bf r}_2-{\bf r}_1) \right\}
\nonumber \\ & = &
\frac{V^2 \Lambda^{-6}}{2!}
\left\{ 1 \pm
 e^{ -2 \pi r_{12}^2/\Lambda^2}  \right\} ,
\end{eqnarray}
where $\Lambda = \sqrt{ 2 \pi \hbar^2/m k_\mathrm{B}T }$
is the thermal wave length.
(This is derived and further discussed in Appendix~\ref{Sec:Local-Posn}.)
This says that the quantum correction due to symmetrization
decays rapidly with separation.
The conclusion is that quantum permutation
is only significant when the particles are
in the same locality in position space.

It is worth mentioning that a similar point
about localization is made by  Messiah, Ch.~XIV, \S 8.\cite{Messiah61}
He shows that if the particles are represented by wave packets,
then the wave function need not be symmetrized with respect to
interchange of particles in non-overlapping wave packets.
This is another way of viewing the clusters discussed above.

From these examples
it may be concluded that  quantum symmetrization effects are significant
only if the permuted particles are in the same locality.
This holds in entropy microstate space  or in position space.
It is henceforth assumed that this concept of localization holds in general.

%%%%%%%%%%%%%%%%%%%%%%%%%%%%%%%%%%%%%%%%%%%%%%%%%%%%%%%%%%%%%%%%%%%%%%%%%
\subsection{General Formalism} \label{Sec:Zcl+Z1-0}

Now an expansion is formulated for the quantum partition function.
The expansion is based upon the symmetrization of the wave function,
which, it will be recalled from
Eqs.~(\ref{Eq:chi-zeta}) and (\ref{Eq:Z-chi-zeta}),
requires the sum over all permutations of the particle positions.
The present expansion is in contrast to
the well-known virial expansion based upon quantum cluster integrals,
which is derived in Appendix \ref{Sec:Zb}.

Any permutation can be cast as the product of disconnected loops.
A loop is the cyclic permutation of a set of particles.
This can be written as a sequence of connected pair transpositions,
which is to say that  the end label of one transposition
is the start label of the next transposition.
Because the partition function is the sum over all entropy microstates,
the nodes of the loops can be re-labeled as convenient.
A monomer is a one-particle loop
(the identity permutation).
A dimer is a two-particle loop (a single transposition),
for example, $ 1 \rightarrow 2 \rightarrow 1$,
which is equivalent to the  single transposition $\hat{\mathrm P}_{21}$.
The trimer, or three-particle loop,
$ 1 \rightarrow 2 \rightarrow 3 \rightarrow 1$,
is equivalent to  the double transposition
$\hat{\mathrm P}_{32} \hat{\mathrm P}_{21} $.
This permutation can also be achieved as
$\hat{\mathrm P}_{12}\hat{\mathrm P}_{13} $
and as $\hat{\mathrm P}_{13}\hat{\mathrm P}_{12} $.
The tetramer
$ 1 \rightarrow 2 \rightarrow 3 \rightarrow 4 \rightarrow 1$,
is equivalent to the triple transposition
$\hat{\mathrm P}_{43} \hat{\mathrm P}_{32} \hat{\mathrm P}_{21}  $,
and others besides.
In general an $l$-mer is an $l$-particle loop.
An $l$-mer of particles $1, 2, \ldots, l$ in order
can be written as the application of $l-1$ successive transpositions,
$\hat{\mathrm P}^{l-1} \equiv \hat{\mathrm P}_{l,l-1}
\ldots \hat{\mathrm P}_{32} \hat{\mathrm P}_{21}$.

Evidently the parity of a loop is the parity of the number of nodes minus one.
That is, an $l$-mer has parity $l-1$.

Henceforth the word loop will be restricted to dimers and above.
Monomers will not be considered to be a loop and will
instead be named explicitly.

The permutation operator breaks up into loops
\begin{eqnarray}
\sum_{\hat{\mathrm P} } (\pm1)^p\; \hat{\mathrm P}
& = &
\hat {\mathrm I}
\pm \sum_{i,j} \!' \; \hat{\mathrm P}_{ij}
+ \sum_{i,j,k} \!' \; \hat{\mathrm P}_{ij} \hat{\mathrm P}_{jk}
\nonumber \\ & & \mbox{ }
+ \sum_{i,j,k,l} \!\!' \; \hat{\mathrm P}_{ij} \hat{\mathrm P}_{kl}
\pm \ldots
\end{eqnarray}
The prime on the sums restrict them to unique loops,
with each index being different.
The first term is just the identity.
The second term is a dimer loop,
the third term is a trimer loop,
and the fourth term is the product of two different dimers.

The overlap factor,
$\chi_{\bf n}$ $= \sum_{\hat{\mathrm P} }(\pm1)^p$
$ \langle \zeta_{\bf n}( \hat{\mathrm P}{\bf r})
| \zeta_{\bf n}({\bf r}) \rangle$,
is the sum of the expectation values of these loops.
The monomer overlap factor
is unity,
\begin{equation}
\chi^{(1)}_{\bf n} = \langle \zeta_{\bf n}({\bf r})
| \zeta_{\bf n}({\bf r}) \rangle = 1.
\end{equation}

The dimer overlap factor in the microstate ${\bf n}$
for particles $j$ and $k$ is
\begin{eqnarray}
\chi^{(2)}_{{\bf n};jk}
& = &
\pm \langle \zeta_{\bf n}( \hat{\mathrm P}_{jk}{\bf r})
| \zeta_{\bf n}({\bf r}) \rangle
\nonumber \\ & = &
\langle \zeta_{{\bf n}_{jk}}( \hat{\mathrm P}_{jk}{\bf r}^c)
| \zeta_{{\bf n}_{jk}}({\bf r}^c) \rangle .
\end{eqnarray}
This dimer overlap factor for entropy microstate ${\bf n}$
is non-zero if particles $j$ and $k$ belong
to the same localized multi-particle state,
as was discussed in the preceding sub-section.
This can be signified by the
multi-particle state label ${\bf n}_{jk}$,
which is a sub-label of the entropy microstate ${\bf n}$.
It is important to note that this multi-particle state
can, and most usually does, include other monomers
that are in the vicinity of the dimer permutation loop
consisting of particles $j$ and $k$,
as determined by the microstate  ${\bf n}$.
The attached monomers' state is intrinsic to the multi-particle state
label ${\bf n}_{jk}$.
Particles that are not part of the same cluster as the dimer
do not contribute to the expectation value
(nor do their state),
as is indicated by the transition to the second equality,
where ${\bf r}^c$ signifies the reduced configuration
that consists of the coordinates of the dimer
and neighboring monomer particles.

Obviously the trimer overlap factor is
\begin{eqnarray}
\chi^{(3)}_{{\bf n};jkl}
& = &
\langle \zeta_{\bf n}( \hat{\mathrm P}_{jk}\hat{\mathrm P}_{kl}{\bf r})
| \zeta_{\bf n}({\bf r}) \rangle
\nonumber \\ & = &
\langle
\zeta_{{\bf n}_{jkl}}( \hat{\mathrm P}_{jk}\hat{\mathrm P}_{kl}{\bf r}^c)
| \zeta_{{\bf n}_{jkl}}({\bf r}^c) \rangle .
\end{eqnarray}
Again the multi-particle state label ${\bf n}_{jkl}$
localizes the three loop particles and the attached monomers
that belong to the same cluster, as implied by the microstate ${\bf n}$.

It will be assumed that the loops are so dilute that
each of them can be taken to belong to separate clusters
(ie.\ localities).
They are therefore independent of each other.
Hence the overlap factor for the product of dimer loops shown explicitly above
reduces to the product of dimer overlap factors,
\begin{eqnarray}
\chi^{(2,2)}_{{\bf n};ij,kl}
& = &
\langle \zeta_{\bf n}( \hat{\mathrm P}_{ij}\hat{\mathrm P}_{kl}{\bf r})
| \zeta_{\bf n}({\bf r}) \rangle
\nonumber \\ & \approx &
\langle \zeta_{\bf n}( \hat{\mathrm P}_{ij}{\bf r})
| \zeta_{\bf n}({\bf r}) \rangle \,
\langle \zeta_{\bf n}(\hat{\mathrm P}_{kl}{\bf r})
| \zeta_{\bf n}({\bf r}) \rangle
\nonumber \\ & = &
\chi^{(2)}_{{\bf n};ij} \, \chi^{(2)}_{{\bf n};kl}.
\end{eqnarray}
By definition of distinct permutations,
the $i,\, j,\, k, $ and $l$ must all be different.
The justification for this factorization
is  that there are many more microstates ${\bf n}$
that consist of many small clusters than consist of a few large clusters.
Hence the sum over microstates will be dominated
by terms in which  pairs of dimers are far apart and independent of each other
than there terms in which the two
are close together and influencing each other in a single cluster.
The numbers of the two types of terms scales as $V^2$ versus $V$,
and so the correction to the leading term given here is negligible
in the thermodynamic limit.
By this same argument,
a similar factorization holds for the overlap factors of all products
of permutation loops.

Because the overlap factor $\chi_{\bf n}$ is the sum over all permutations,
it can be rewritten as the sum over all possible monomers and loops.
This gives the loop expansion for the overlap factor
for use in the partition function as
\begin{eqnarray} \label{Eq:chi_n}
\chi_{\bf n}
& = &
1 + \sum_{ij}\!' \chi_{{\bf n};ij}^{(2)}
+ \sum_{ijk}\!' \chi_{{\bf n};ijk}^{(3)}
\nonumber \\ && \mbox{ }
+ \sum_{ijkl}\!'  \chi_{{\bf n};ij}^{(2)} \chi_{{\bf n};kl}^{(2)}
+ \ldots
\end{eqnarray}
Note that the parity factor for fermions and bosons, $(\pm 1)^{l-1}$,
has been incorporated into the definition of the $\chi_{{\bf n}}^{(l)}$.

It is convenient to work in the grand canonical system.
The grand canonical partition function is
\begin{equation}
\Xi(\mu,V,T)
=
\sum_{N=0}^\infty \frac{z^N}{N!}
\sum_{\bf n} \chi_{\bf n}
 e^{-\beta {\cal H}_{\bf n}} ,
\end{equation}
where the fugacity is $z=e^{\beta \mu}$, $\mu$ being the chemical potential.

One can insert into this
the loop expansion of the overlap factor
and evaluate the terms one at a time.
The monomer term is
\begin{equation}
\Xi_1
=
\sum_{N=0}^\infty \frac{z^N}{N!}  \sum_{\bf n}
 e^{-\beta {\cal H}_{\bf n}} .
\end{equation}
As discussed below,
one can use wave packets for the entropy eigenfunction,
and in the thermodynamic limit they have zero width,
in which case this is just
the classical grand canonical equilibrium partition function
evaluated as an integral over phase space.
It is a moot point whether or not it is better to write
$ \Xi_\mu$ or $ \Xi_\mathrm{cl}$  instead of   $ \Xi_1$.

The single dimer term is
\begin{eqnarray}
\Xi_2
& = &
\sum_{N=2}^\infty \frac{z^N}{N!}  \sum_{\bf n}
 e^{-\beta {\cal H}_{\bf n}}
\sum_{ij}\!' \chi_{{\bf n};ij}^{(2)}
\nonumber \\ & = &
\sum_{N=2}^\infty  \frac{z^NN(N-1)}{2N!} \sum_{\bf n}
 e^{-\beta {\cal H}_{\bf n}}
 \chi_{{\bf n};12}^{(2)} .
\end{eqnarray}
The second equality follows because the sum over all states
makes all dimer pairs equivalent.
Because of this the subscript 12 on  $\chi_{{\bf n};12}^{(2)}$
is redundant and will be dropped for this and the following overlap factors.
Recall that the entropy microstates ${\bf n}$ are for the dimer particles
and the $N-2$ monomer particles.
It is convenient to divide the dimer term by the monomer term,
in which case it becomes an average,
\begin{eqnarray}
\frac{\Xi_2}{\Xi_1}
& = &
\frac{ \displaystyle
\sum_{N=2}^\infty  \frac{z^N}{N!} \frac{N(N-1)}{2} \sum_{\bf n}
 e^{-\beta {\cal H}_{\bf n}}
 \chi_{{\bf n}}^{(2)}
}{ \displaystyle
\sum_{N=0}^\infty  \frac{z^N}{N!} \sum_{\bf n}
 e^{-\beta {\cal H}_{\bf n}}
}
\nonumber \\ & = &
 \left< \frac{N(N-1)}{2} \chi^{(2)} \right>_\mu
\nonumber \\ & \approx &
\frac{1}{2!} \left< N^2 \chi^{(2)} \right>_\mu .
\end{eqnarray}
(It is not essential to replace $N(N-1)$ by $N^2$ here and below,
but it does save space,
and it gives a necessary factorization below.)
This scales with volume because the overlap factor is only non-zero
when the two particles are close together
(ie.\ $\Xi_1 \sim V^{\overline N}$, $\Xi_2 \sim V^{\overline N-1}$,
and $\overline N  \sim V$).
Here and below, the $N$ that appears explicitly in the grand canonical average
is the total number of particles in each term,
which is to say that there are $l$ loop particles and $N-l$ monomers,
all able to interact  depending upon their proximity in each microstate.

The single trimer term is
\begin{eqnarray}
\Xi_3
& = &
\sum_{N=3}^\infty  \frac{z^N}{N!} \sum_{\bf n}
 e^{-\beta {\cal H}_{\bf n}}
\sum_{jkl}\!' \chi_{{\bf n};jkl}^{(3)}
\nonumber \\ & = &
\sum_{N=3}^\infty
\frac{2!z^N}{3!(N-3)!}\sum_{\bf n}
 e^{-\beta {\cal H}_{\bf n}}
 \chi_{{\bf n};123}^{(3)} .
\end{eqnarray}
The combinatorial pre-factor comes from two contributions.
In general there are $N!/l!(N-l)!$ ways of choosing $l$ different particles
from $N$ particles.
And there are $(l-1)!$ distinct ways of arranging these
in a loop of $l$ particles.
Obviously  $l=3$ here.
The subscript 123 is redundant and can be dropped.
The ratio of partition functions is
\begin{equation}
\frac{\Xi_3}{\Xi_1}
=
\frac{2!}{3!} \left< N^3  \chi^{(3)} \right>_\mu .
\end{equation}
For brevity $N(N-1)(N-2)$ has been replaced by $N^3$,
but again this is not essential.
This average also scales with the volume.

The double dimer product term is
\begin{eqnarray}
\Xi_{22}
& = &
\sum_{N=4}^\infty  \frac{z^N}{N!} \sum_{\bf n}
 e^{-\beta {\cal H}_{\bf n}}
\sum_{ijkl}\!' \chi^{(2,2)}_{{\bf n};ij,kl}
 \\ & \approx &
\sum_{N=4}^\infty  \frac{z^N}{N!} \sum_{\bf n}
 e^{-\beta {\cal H}_{\bf n}}
\sum_{ijkl}\!' \chi_{{\bf n};ij}^{(2)} \chi_{{\bf n};kl}^{(2)}
\nonumber \\ & = &
\sum_{N=4}^\infty  \frac{z^N}{2^3(N-4)!}\sum_{\bf n}
 e^{-\beta {\cal H}_{\bf n}}
 \chi_{{\bf n};12}^{(2)}  \chi_{{\bf n};34}^{(2)} .\nonumber
\end{eqnarray}
Hence
\begin{eqnarray}
\frac{\Xi_{22}}{\Xi_1}
& = &
\frac{1}{2^3  }
\left<  N^4 \chi_{12}^{(2)}  \chi_{34}^{(2)} \right>_\mu
\nonumber \\ & \approx &
\frac{1}{2!}
\left[ \frac{1}{2!} \left< N^2  \chi^{(2)}  \right>_\mu \right]^2 .
\end{eqnarray}
In the second equality the average of the product has been written
as the product of the averages,
which is valid if the dimer loops are dilute.
Of course this factorization of the average is consistent
with the factorization of the expectation value:
there are many more microstates in which the two dimers are far apart
and independent than there are those in which they are close together
and influencing each other.
As mentioned, the contribution from two independent dimers scales with $V^2$,
whereas that from two interacting dimers scales with $V$.
%The right hand side of the above ratio scales with $V^2$.

Continuing in this fashion it is clear that
\begin{eqnarray}
\lefteqn{
\Xi(\mu,V,T)
} \nonumber \\
& = &
\Xi_1
\left\{ 1 +
\frac{1}{2!} \left< N^2 \chi^{(2)} \right>_\mu
+ \frac{2!}{3!} \left< N^3  \chi^{(3)} \right>_\mu
\right. \nonumber \\ && \left. \mbox{ }
+ \frac{1}{2!}
\left[ \frac{1}{2!} \left< N^2  \chi^{(2)}  \right>_\mu \right]^2
+ \ldots
\right\}
\nonumber \\ & = &
\Xi_1 \sum_{\{m_l\}} \prod_{l=2}^\infty \frac{1}{m_l!}
\left[ \frac{(l-1)!}{l!}
\left< \frac{N!}{(N-l)!}  \chi^{(l)}  \right>_\mu \right]^{m_l}
\nonumber \\ & = &
\Xi_1 \prod_{l=2}^\infty \sum_{m_l=0}^\infty
\frac{1}{m_l!}
\left[ \frac{(l-1)!}{l!}
\left< \frac{N!}{(N-l)!} \chi^{(l)}  \right>_\mu \right]^{m_l}
\nonumber \\ & = &
\Xi_1 \prod_{l=2}^\infty \exp
\left[ \left< \frac{(l-1)!N!}{l!(N-l)!} \chi^{(l)}  \right>_\mu \right] .
\end{eqnarray}
Here $m_l$ is the number of loops of $l$ particles.
Here  $N!/(N-l)!$ has been written in place of $N^l$,
although in truth either is justified.

The grand potential is
\begin{eqnarray}
\Omega(\mu,V,T) & = &
- k_\mathrm{B}T \ln \Xi(\mu,V,T)
\nonumber \\ & \equiv &
\sum_{l=1}^\infty \Omega_l .
\end{eqnarray}
Here the monomer grand potential is $\Omega_1 = - k_\mathrm{B}T \ln \Xi_1$,
and the grand potential for an $l$-loop is
\begin{equation} \label{Eq:Omega_l=<chi^l>}
\Omega_l
=
- k_\mathrm{B}T \left< \frac{(l-1)!N!}{l!(N-l)!}   \chi^{(l)}  \right>_\mu
, \;\; l \ge 2.
\end{equation}
%Each of these coefficients scale with volume,
%so that the grand potential itself is extensive.
Recall that the average $l$-loop overlap factor is given by
\begin{eqnarray}
\lefteqn{
\left< \frac{(l-1)!N!}{l!(N-l)!}   \chi^{(l)}  \right>_\mu
} \nonumber \\
& = &
\frac{1}{\Xi_1}
\sum_{N=l}^\infty  \frac{z^N}{N!} \frac{(l-1)!N!}{l!(N-l)!} \sum_{\bf n}
 e^{-\beta {\cal H}_{\bf n}}
 \chi_{{\bf n}}^{(l)}   .
\end{eqnarray}
Recall also that the symmetry factor for bosons and fermions,
$(\pm 1)^{l-1}$, is included in the definition of the loop overlap factor
$ \chi_{{\bf n}}^{(l)}$.
This average is discussed in the sub-subsection
at the end of the present subsection.

Each loop grand potential $\Omega_l$
is a sum  beginning at $N=l$, with the first term multiplied by $z^l$.
Although the expansion of the grand potential
has the appearance of a fugacity expansion,
this is is not the whole story.
The other terms in each
grand canonical sum $N > l$ are multiplied by $z^N$,
and the denominator for the average also depends upon the fugacity.

The loop potentials are extensive,
which is to say that they scale with the volume of the system
(assuming no externally applied potential).
Hence one can define the loop grand potential density,
\begin{equation}
\omega_l(\mu,T) \equiv \frac{\Omega_l(\mu,V,T)}{V}.
\end{equation}
The reason that the loop potentials are extensive
is that the center of mass of the $l$-mer is free to roam throughout
the homogeneous  volume $V$.
(The sum over the loop states ${\bf n}_l$ contains
amongst other things equivalent integrals of the particles' position states
over the volume $V$.)
But the particles of the loop must remain in the vicinity
of the center of mass because of the connectivity of the loop.
This is most easily seen when the loop entropy eigenfunction is
taken as the product of wave packets of finite width.
In this case the expectation value
of the original and permuted eigenfunction
is only non-zero if the successive wave packets overlap,
In other words, the loop states ${\bf n}_l$
that contribute to the loop potential
are those in which the particles are close enough
for their wave packets to overlap.
It is essential that the loop potentials are extensive
because the grand potential of the quantum system
has to be extensive.

As mentioned,
the monomer term corresponds to classical statistical mechanics,
$\Omega_1(\mu,V,T) = \Omega_\mathrm{cl}(\mu,V,T)$.
The monomer partition function can be written,
\begin{eqnarray}
\Xi_1(\mu,V,T)
& = &
\sum_{N=0}^\infty
\frac{z^N}{N!}
\sum_{{\bf n}}
e^{-\beta {\cal H}({\bf n}) }
 \nonumber \\ & = &
\sum_{N=0}^\infty
\frac{z^N}{h^{3N} N!}
\int \mathrm{d}{\bm \Gamma}^N e^{-\beta {\cal H}({\bm \Gamma}^N) } .
\end{eqnarray}
It will be shown below that in the thermodynamic limit
$N \rightarrow \infty$
the entropy eigenfunctions become wave packets of zero width,
and the corresponding entropy microstates become a point
in classical phase space.
As is traditional, the volume per entropy microstate
has been taken to equal Planck's constant, $\Delta_p\Delta_q =h$,
but this is not essential.

In the traditional thermodynamic limit,
the fugacity is fixed,
as is  the most likely number density $\overline \rho(z)$,
and the most likely  number and volume become infinite,
$\overline N = V \overline \rho(z) \rightarrow \infty$.
The sum over $N$ is dominated by the terms $N \approx \overline N$.
In this case zero width wave packets
are the exact entropy eigenfunctions (for interacting particles).
In a fugacity expansion,
typically the number is fixed, say $N=2$ for the second term,
and the limit is taken that
$z  \rightarrow 0 $ and $V \rightarrow \infty$.
In this case one cannot invoke zero width wave packets
as entropy eigenfunctions,
which point is relevant for the analysis of the second fugacity coefficient
in \S \ref{Sec:b2} below.

For a statistical average of an operator,
one can also use the expansion for the overlap factor, Eq.~(\ref{Eq:chi_n}).

Two approximations were made in the derivation of this
expansion for the grand potential.
The first was
that the expectation value of the product of permutation loops
is equal to the product of the expectation values
of the individual loops,
which is to say that the overlap factor for a product of loops
is equal to the product of overlap factors.
The second was that the average of the product of overlap factors
is equal to the product of the average of each overlap factor.
These two approximations are obviously related to each other
and it is expected that both are exact in the thermodynamic limit.
On physical grounds one can see
that the approximations will be accurate when the loops do not interact,
which is the case for an ideal gas,
or for real particles at low densities or high temperatures,
when there are many accessible microstates.
The loop grand potentials are of course  the successive quantum
corrections to classical statistical mechanics.

%\newpage
%%%%%%%%%%%%%%%%%%%%%%%%%%%%%%%%%%%%%%%%%%%%%%%%%
\subsubsection{Approximations to the Loop Grand Potential} \label{Sec:<chi_n>}

The quantum corrections to the grand potential
involve the average of the total loop overlap factor,
the evaluation of which appears challenging.
One requires the entropy eigenstates and eigenfunctions
for the $l$ loop particles and $N-l$ monomer particles,
and thence the inner product of the permuted and the original eigenfunction,
and the statistical average over the eigenstates,
and this must be done for each $N$ in the grand canonical sum.
It is worth discussing some practical approaches.

The simplest approximation is to neglect the monomers altogether.
That is, one need only average the `bare' loop overlap factor
over the $l$-particle eigenstates,
$l \ge 2 $,
\begin{eqnarray}
-\beta \Omega_l
& = &
\left< \frac{(l-1)!N!}{l!(N-l)!}   \chi^{(l)}  \right>_\mu
\nonumber \\ & \approx &
\frac{z^l(l-1)!}{l!} \sum_{{\bf n}_l}
 e^{-\beta {\cal H}_{{\bf n}_l}}
 \chi_{{\bf n}_l}^{(l)} .
\end{eqnarray}
The entropy eigenstates and the eigenfunctions for the overlap factor
here are for a system of  $l  $ particles only.

A more sophisticated approach includes the monomers as follows.
As will be shown in \S \ref{Sec:olxi},
in the thermodynamic limit, $N \rightarrow \infty$,
the entropy eigenfunction can be cast as
the product of wave packets of zero width.
The problem with this is that these correspond
to  single-particle, orthogonal states that have zero overlap.
This suggests that one ought to avoid the thermodynamic limit
in evaluating the loop overlap average,
and instead focus upon the terms $N \agt l$
in which case finite width overlapping wave packets
could be used as entropy eigenfunctions.
However this in turn would be inconsistent with the denominator $\Xi_1$,
which is in the thermodynamic limit,
with zero width wave packets, and which reduces to
the classical grand partition function.

One way to reconcile these two competing requirements
is to take the thermodynamic limit for the monomers
but not for the loop particles.
In this hybrid picture the wave function of the system
factorizes into a multi-particle microstate wave function
for the $l$ loop particles,
$\zeta^{(l)}_{{\bf n}_l}$,
and zero width wave packets for the $N-l$ monomer particles,
$\prod_{j=l+1}^{N} \zeta_{{\bf n}_j}({\bf r}_j)$.
Hence the monomers  occupy a point in classical phase space state,
${\bf \Gamma}^{N-l} = \{ {\bf q}^{N-l} , {\bf p}^{N-l} \}$.
The loop entropy eigenfunction depends upon the positions of the monomers,
$\zeta^{(l)}_{{\bf n}_l}({\bf r}^{l};{\bf q}^{N-l})$,
since these represent an external potential for the Hamiltonian operator,
$\hat{\cal H}({\bf r}^{l};{\bf q}^{N-l})
= \hat{\cal K}({\bf r}^{l})
+ \hat{\cal U}({\bf r}^{l})
+ \hat{\cal U}^\mathrm{ext}({\bf r}^{l};{\bf q}^{N-l})$.
The eigenvalue equation to be solved is
\begin{equation}
\hat{\cal H}({\bf r}^{l};{\bf q}^{N-l}) \,
\zeta^{(l)}_{{\bf n}_l}({\bf r}^{l};{\bf q}^{N-l})
=
{\cal H}_{{\bf n}_l}({\bf q}^{N-l} ) \,
\zeta^{(l)}_{{\bf n}_l}({\bf r}^{l};{\bf q}^{N-l}) .
\end{equation}

With these the average $l$-loop overlap factor for $l \ge 2$
is given by
\begin{eqnarray} \label{Eq:Omega_l=<chi^l_tot>}
\lefteqn{
-\beta \Omega_l
} \nonumber \\
& = &
\left< \frac{(l-1)!N!}{l!(N-l)!}   \chi^{(l)}  \right>_\mu
\nonumber \\ & = &
\frac{1}{\Xi_1}
\sum_{N=l}^\infty  \frac{z^N}{N!} \frac{(l-1)!N!}{l!(N-l)!} \sum_{\bf n}
 e^{-\beta {\cal H}_{\bf n}}
 \chi_{{\bf n}}^{(l)}
 \nonumber \\ & \approx &
\frac{1}{\Xi_1}
\sum_{N=l}^\infty   \frac{z^N(l-1)!}{l!(N-l)!h^{3(N-l)}}
 \int \mathrm{d} {\bf \Gamma}^{N-l}\;
 e^{-\beta {\cal H}({\bf \Gamma}^{N-l})}
 \nonumber \\ & & \mbox{ } \times
 \sum_{{\bf n}_l}
  e^{-\beta {\cal H}_{{\bf n}_l}({\bf q}^{N-l} )}
 %\nonumber \\ & & \mbox{ } \times
 \chi^{(l)}_{{\bf n}_l}({\bf q}^{N-l})
 \nonumber \\ & = &
\frac{(l-1)!z^l}{l!}   \left< \sum_{{\bf n}_l}
  e^{-\beta {\cal H}_{{\bf n}_l}({\bf q}^{N-l} )}
 %\nonumber \\ & & \mbox{ } \times
 \chi^{(l)}_{{\bf n}_l}({\bf q}^{N-l})
 \right>_\mu
\nonumber \\ & \equiv &
(l-1)! z^l
\left< \chi^{(l)}_\mathrm{tot}({\bf q}^{N-l})  \right>_\mu  .
\end{eqnarray}
Here the total weighted overlap factor is
\begin{equation}
\chi^{(l)}_\mathrm{tot}({\bf q}^{N-l})
\equiv
\frac{1}{l!}  \sum_{{\bf n}_l}
  e^{-\beta {\cal H}_{{\bf n}_l}({\bf q}^{N-l} )}
 %\nonumber \\ & & \mbox{ } \times
 \chi^{(l)}_{{\bf n}_l}({\bf q}^{N-l}) ,
\end{equation}
and the $l$-loop overlap factor is
\begin{eqnarray}
\lefteqn{
 \chi^{(l)}_{{\bf n}_l}({\bf q}^{N-l})
}  \\
& = &
(\pm 1)^{l-1}
\int \mathrm{d}{\bf r}^{l} \;
 \zeta^{(l)}_{{\bf n}_l}(\hat{\mathrm P}^{l-1}{\bf r}^{l};{\bf q}^{N-l})^* \;
 \zeta^{(l)}_{{\bf n}_l}({\bf r}^{l};{\bf q}^{N-l}) ,
 \nonumber
\end{eqnarray}
where the loop permutator is
$\hat{\mathrm P}^{l-1} \equiv \hat{\mathrm P}_{l,l-1}
\ldots \hat{\mathrm P}_{32} \hat{\mathrm P}_{21}$.
The monomer grand partition function is just the classical grand
partition function given above.
The final average, $\langle \ldots \rangle_\mu$,
is a classical grand canonical average over the monomers.

A different approximation is to define some small volume
$V' \ll V$ that encompasses the loop
and to evaluate the grand canonical average for $\Omega_l$
over the monomers.
One would need to evaluate
the entropy eigenfunctions for the loop and monomer particles
$N' \ll \overline N(\mu,V,T)$ explicitly.
Because the loop grand potential is extensive in the volume,
one can scale this up as required,
but the computational burden is much reduced compared to using the
macroscopic volume $V$ and number $\overline N(\mu,V,T)$.

In either case it is still a challenging problem to obtain
all the eigenstates, eigenvalues, and eigenfunctions of the loop particles
for every configuration of the monomers.
From the computational point of view however,
one only has to focus upon the likely configurations of the monomers
(ie.\ low energy),
the monomers in close vicinity to the loop,
the low energy or ground states of the loop,
and the states  of the loop corresponding
to mutually overlapping particles such that their loop permutation
gives a non-zero inner product.
In these circumstances an approximate solution to the eigenvalue
equation suffices.

One practical point that can be made
is that for typical terrestrial condensed matter,
only small loops need to be evaluated, and $l \Lambda^3 \ll \rho^{-1}$
(ie.\ the volume per loop is much smaller than the volume per monomer).
This means that the monomers effectively see the loop
as a point particle of strength $l$,
perhaps with multipole corrections for each state ${\bf n}_l$,
and the classical average can be calculated on that basis.
The loop eigenstates can be calculated in the monomer mean field.
No doubt these and the other attributes can be exploited to make the problem
computationally feasible.

%\newpage
%%%%%%%%%%%%%%%%%%%%%%%%%%%%%%%%%%%%%%%%%%%%%%%%%
\subsection{Zeroth Order Entropy Eigenfunction} \label{Sec:zeta(0)}

In \S \ref{Sec:Entropy-Eigen}, which follows,
a detailed analysis is given of
the entropy eigenfunction  cast as a wave packet,
\begin{eqnarray}
\zeta_{{\bf n}}({\bf r})
& \equiv &
 C^{-1}
\exp \left\{ \frac{-\left({\bf r}-{\bf q}_{{\bf n}} \right)
\cdot \left({\bf r}-{\bf q}_{{\bf n}} \right) }{4 \xi^2}
\right. \nonumber \\ && \mbox{ } \left.
- \frac{1}{i\hbar} {\bf p}_{{\bf n}}
\cdot ({\bf r}-{\bf q}_{{\bf n}})
 \right\} ,
\end{eqnarray}
with $ C^{2} = ( 2\pi \xi^2)^{3N/2}$.
Ideal particles,
to which the analysis is applied at the end of this subsection,
have an exact entropy eigenfunction that corresponds to a
wave packet of infinite width, $\xi \rightarrow \infty$.

For the present minimum uncertainty wave packet,
and neglecting monomers,
the dimer overlap factor  is
\begin{eqnarray} \label{Eq:chi(2)12}
\chi_{12}^{(2)}
& \equiv &
\pm
\langle  \zeta_{{\bf n}_2{\bf n}_1} | \zeta_{{\bf n}_1{\bf n}_2}\rangle
\nonumber \\ & = &
\pm \int \mathrm{d}{\bf r}_1  \mathrm{d}{\bf r}_2 \;
\zeta_{{\bf n}_2}({\bf r}_1)^*
\zeta_{{\bf n}_1}({\bf r}_2)^*
\zeta_{{\bf n}_1}({\bf r}_1) \zeta_{{\bf n}_2}({\bf r}_2)
\nonumber \\ & = &
\pm
\int \mathrm{d}{\bf r}_1
\zeta_{{\bf n}_2}({\bf r}_1)^* \zeta_{{\bf n}_1}({\bf r}_1)
%\nonumber \\ &&  \mbox{ } \times
\int \mathrm{d}{\bf r}_2 \;
\zeta_{{\bf n}_1}({\bf r}_2)^* \zeta_{{\bf n}_2}({\bf r}_2)
\nonumber \\ & = &
\pm
\left| \int \mathrm{d}{\bf r}_1
\zeta_{{\bf n}_2}({\bf r}_1)^* \zeta_{{\bf n}_1}({\bf r}_1) \right|^2
\nonumber \\ & = &
\frac{\pm 1}{( 2\pi \xi^2)^{3}}
\left| \int \mathrm{d}{\bf r}_1
\right. \nonumber \\ && \mbox{ } \left. \times
e^{ -\left({\bf r}_1-{\bf q}_{{\bf n}_2} \right)^2/4 \xi^2 }
%\right. \nonumber \\ && \mbox{ } \left.
e^{ {\bf p}_{{\bf n}_2} \cdot ({\bf r}_1-{\bf q}_{{\bf n}_2}) /i\hbar  }
\right. \nonumber \\ && \mbox{ } \left. \times
e^{ -\left({\bf r}_1-{\bf q}_{{\bf n}_1} \right)^2/4 \xi^2 }
%\right. \nonumber \\ && \mbox{ } \left.
e^{ -{\bf p}_{{\bf n}_1} \cdot ({\bf r}_1-{\bf q}_{{\bf n}_1}) /i\hbar }
%\right. \nonumber \\ && \mbox{ } \left.
\right|^2
\nonumber \\ & = &
\pm \exp \left\{
\frac{-1}{4\xi^2}
\left( {\bf q}_{{\bf n}_2}-{\bf q}_{{\bf n}_1}\right)^2
- \frac{\xi^2}{\hbar^2} \left({\bf p}_{{\bf n}_2}-{\bf p}_{{\bf n}_1}\right)^2
\right\} .
\nonumber \\
\end{eqnarray}
%Several steps have been skipped in obtaining the final equality.
This is evidently an un-normalized Gaussian in position and momentum
that ties the two particles together.
It is implicitly assumed here that the two particles comprising the dimer
can be treated independently of the remaining monomers,
which don't therefore contribute to the scalar product.

The entropy microstate ${\bf n}$ embodied in the wave packet
is just a point in two-particle phase space,
 ${\bf \Gamma} = \{{\bf \Gamma}_1, {\bf \Gamma}_2\} $,
with ${\bf \Gamma}_j = \{ {\bf q}_j, {\bf p}_j \}$, $j =1,2$.
Hence one simply replaces
${\bf q}_{{\bf n}_1} \Rightarrow  {\bf q}_{1}$, etc.
The classical canonical equilibrium total weighted dimer overlap factor is
\begin{eqnarray}
\lefteqn{
\chi^{(2)}_\mathrm{tot}
% \left< \chi^{(2)} \right>_{\mathrm{cl},N}
} \nonumber \\
& = &
\frac{1}{ 2 h^{6}}
\int \mathrm{d}{\bm \Gamma}_1 \, \mathrm{d}{\bf \Gamma}_2 \;
e^{- \beta{\cal H}({\bm \Gamma}_{1},{\bm \Gamma}_{2})}
\chi^{(2)}({\bm \Gamma}_{1},{\bm \Gamma}_{2})
\nonumber \\ & = &
\frac{\pm 1}{2 h^{6}}
\int \mathrm{d}{\bm \Gamma}_1 \, \mathrm{d}{\bf \Gamma}_2 \;
e^{- \beta{\cal H}({\bm \Gamma}_{1},{\bm \Gamma}_{2})}
\nonumber \\ &&  \mbox{ } \times
 e^{ -\left( {\bf q}_{1}-{\bf q}_{2}\right)^2 /{4\xi^2}}
%\nonumber \\ &&  \mbox{ } \times
e^{ - \xi^2 \left({\bf p}_{1}-{\bf p}_{2}\right)^2
/{\hbar^2} } .
\end{eqnarray}
Again this neglects the monomers
since the Hamiltonian only depends upon the two dimer particles.
Consistent with this is that the grand canonical average over the monomers
may be neglected, and the first quantum correction
for the grand potential may be equated to this total weighted dimer
overlap factor,
$-\beta \Omega_2 = z^2 \chi^{(2)}_\mathrm{tot}$.

Assume that the Hamiltonian is of the form
\begin{equation}
{\cal H}({\bm \Gamma}_{1},{\bm \Gamma}_{2}) =
\frac{1}{2m} [ p_1^2 + p_2^2 ]
+ U({\bf q}_1,{\bf q}_2).
\end{equation}
For the momentum integral one has
%\begin{equation}
%\frac{1}{h^{6}} \int \mathrm{d}{\bf p}_1 \, \mathrm{d}{\bf p}_2 \;
%e^{-\beta [ p_1^2 + p_2^2 ] /2m}
%=\Lambda^{-6} ,
%\end{equation}
%and
\begin{eqnarray}
\frac{1}{h^{6}}  \int \mathrm{d}{\bf p}_1 \, \mathrm{d}{\bf p}_2 \;
\lefteqn{
e^{-\beta [ p_1^2 + p_2^2 ]/2m}
e^{ - \xi^2 \left({\bf p}_{1}-{\bf p}_{2}\right)^2/{\hbar^2} }
} \nonumber \\
& = &
%\Lambda^{-6}  \left[ 1  + \frac{4 m \xi^2}{\beta\hbar^2}  \right]^{-3/2}
% \nonumber \\ & = &
\Lambda^{-6}  \left[ 1  + \frac{8 \pi\xi^2}{\Lambda^2}  \right]^{-3/2} .
\end{eqnarray}
where the thermal wave-length is
$\Lambda \equiv \sqrt{ 2 \pi \hbar^2 /mk_\mathrm{B}T}$.

\comment{ %%%%%%%%%%%%%%%%%%%%%%%%%%%%%%%%%%%%%%%%%%%%%%%%%%%%%%%%%
In detail this is
\begin{eqnarray}
\lefteqn{
\frac{1}{\hbar^{3N}} \int \mathrm{d}{\bf p} \,
e^{-\beta {\bf p}\cdot{\bf p}/2m}
e^{ - \xi^2 \left({\bf p}_{1}-{\bf p}_{2}\right)^2/{\hbar^2} }
} \nonumber \\
& = &
\frac{ (2\pi mk_\mathrm{B}T )^{-3} }{ \Lambda^{3N} }
\int \mathrm{d}{\bf p}_1 \,\mathrm{d}{\bf p}_2 \;
\nonumber \\ && \mbox{ } \times
e^{-\beta [ p_1^2 + p_2^2 ]/2m}
e^{ - \xi^2 [ p_1^2 + p_2^2 - 2 {\bf p}_{1}\cdot{\bf p}_{2}]/{\hbar^2} }
\nonumber \\ & = &
\frac{ (2\pi mk_\mathrm{B}T )^{-3} }{ \Lambda^{3N} }
\int \mathrm{d}{\bf p}_1 \,\mathrm{d}{\bf p}_2 \;
\nonumber \\ && \mbox{ } \times
\exp\left\{
- \left[ \beta' - \frac{ \xi^4}{ \hbar^4\beta'} \right] p_1^2 \right\}
\nonumber \\ && \mbox{ } \times
\exp\left\{ - \beta'
\left[  {\bf p}_2 - \frac{ \xi^2}{ \hbar^2\beta'} {\bf p}_1 \right]^2 \right\}
\nonumber \\ & = &
\frac{ (2\pi mk_\mathrm{B}T )^{-3} }{ \Lambda^{3N} }
( 2 \pi)^3 2^{-3} \left[ \beta'^2 - \frac{ \xi^4}{ \hbar^4} \right]^{-3/2}
\nonumber \\ & = &
\Lambda^{-3N} \left( \frac{\beta}{2m}\right)^3
 \left[ \left( \frac{\beta}{2m}\right)^2
 + 2\frac{\beta \xi^2}{2m \hbar^2}  \right]^{-3/2}
 \nonumber \\ & = &
\Lambda^{-3N}
 \left[ 1  + \frac{4 m \xi^2}{\beta\hbar^2}  \right]^{-3/2} ,
\end{eqnarray}
where $\beta' \equiv \beta/2m + \xi^2/\hbar^2$.
} % end comment %%%%%%%%%%%%%%%%%%%%%%%%%%%%%%%%%%%%%%%%%%%%%%%%%%%%

Hence the weighted total dimer overlap factor can be written
\begin{eqnarray}
\lefteqn{
\chi^{(2)}_\mathrm{tot}
}  \\
& = &
\frac{ \pm Q(2,V,T)}{\Lambda^{6}}
\left[ 1  + \frac{8 \pi\xi^2}{\Lambda^2}  \right]^{-3/2}
\left<
e^{ -\left( {\bf q}_{1}-{\bf q}_{2}\right)^2 /{4\xi^2}}
\right>_\mathrm{cl}
\nonumber \\ & = &
\frac{ \pm Q(2,V,T)}{\Lambda^{6}V}
\left[ 1  + \frac{8 \pi\xi^2}{\Lambda^2}  \right]^{-3/2}
\int \mathrm{d}{\bf r} \; g(r)
e^{ -r^2 /{4\xi^2}} , \nonumber
\end{eqnarray}
where $g(r)$ is the radial distribution function,
a spherically symmetric pair potential having been assumed,
and $Q(2,V,T)$
is the canonical equilibrium configuration integral for two particles.

For the case of an ideal gas,
$U({\bf q}) = 0$, $g(r) = 1$, and $Q(2,V,T)=V^2/2$
this gives
\begin{eqnarray} \label{Eq:chi(2)^id}
\chi^{(2)}_\mathrm{tot,id}
& = &
\frac{ \pm V}{2\Lambda^6}
\left[ 1  + \frac{8 \pi\xi^2}{\Lambda^2}  \right]^{-3/2}
(4 \pi \xi^2)^{3/2} .
 \nonumber \\ & \sim &
 \frac{ \pm  V}{2^{5/2}\Lambda^3} , \; \xi \rightarrow \infty.
\end{eqnarray}
The wave function for a free particle
corresponds to the wave packet with
$\xi \rightarrow \infty$.
This  ideal gas limit also emerges from a general optimization procedure
in \S \ref{Sec:olxi}.
In this case the dimer loop average overlap factor
is independent of the width of the wave packet.
(For the case of interacting particles,
$U({\bf q}) \ne 0$,
the optimized result is that $\xi \rightarrow 0$
in the thermodynamic limit.
Of course $N=2$ is some way from the thermodynamic limit.)

This result gives the first quantum correction to the grand partition function
for the ideal gas as
\begin{equation}
-\beta \Omega_2 =  z^2   \chi^{(2)}_\mathrm{tot,id}
=  \frac{ \pm  z^2 V}{2^{5/2}\Lambda^3} .
\end{equation}
This agrees with the result known by other methods.\cite{Pathria72}

%The higher order quantum corrections
%to the classical grand partition function of the ideal gas
%are discussed in the following section.
%The strength of the quantum correction is determined
%by the dimensionless parameter $\rho \Lambda^3$.
%For liquid water at standard temperature and pressure,
%this parameter is $4 \times 10^{-4}$.

It is worth noting that the quantum correction for the ideal gas
given here and below is non-zero,
and it differs for bosons and fermions.
In contrast the quantum correction given by Wigner
\cite{Wigner32,Kirkwood33,Allen87}
vanishes if the potential energy is zero
(ie.\ if the system is ideal),
and it is the same for bosons and for fermions.
The latter problem can be explained by the fact that Wigner
explicitly neglects wave function symmetrization.
The former problem is harder to explain;
one possible interpretation is that the Wigner-Kirkwood formulation is unsound.

%%%%%%%%%%%%%%%%%%%%%%%%%%%%%%%%%%%%%%%%%%%%%%%%%%%%%%%%%%%%%%%%%%%%%%%%%%%
%
\section{Wave Packets for Entropy Eigenfunctions} \label{Sec:Entropy-Eigen}
\setcounter{equation}{0} \setcounter{subsubsection}{0}
%
%%%%%%%%%%%%%%%%%%%%%%%%%%%%%%%%%%%%%%%%%%%%%%%%%%%%%%%%%%%%%%%%%%%%%%%%%%%

This section explores a particular form for the entropy
eigenfunction that is both approximate and non-orthogonal.
The focus is on the way in which classical phase space
corresponds to the entropy microstates. This is made clear by
choosing the minimum uncertainty wave packet as the reference
entropy wave eigenfunction. In \S \ref{Sec:olxi}, the bare wave
packet is used, and its width is chosen to minimize the fluctuation
in the energy eigenvalue.
In \S\S \ref{Sec:f-zeta} and \ref{Sec:ctsU}
a modifier function is used to give a series of
systematic improvements of the minimum uncertainty wave packet as
the entropy eigenfunction.
It is shown that in the thermodynamic limit, $N\rightarrow \infty$,
the modifying terms vanish,
and the wave packet width goes to zero,
which means that for macroscopic systems
the exact entropy eigenfunction consists of
the product of  bare wave packets of zero width.

Although the entropy eigenfunction
that is based on the minimum uncertainty wave packet
is \emph{not} a position or momentum eigenfunction
(except in the thermodynamic limit),
it is shown that the entropy microstates  that follow from it
have the interpretation of simultaneous position and momentum states.
As such the final result is rather close to the phase space formulation
of classical statistical mechanics.

%%%%%%%%%%%%%%%%%%%%%%%%%%%%%%%%%%%%%%%%%%%%%%%%%%%%%%%%%%%%%%%%%%%%%%%%%%%
\subsection{Entropy Microstates and Phase Space}

Since most systems in the terrestrial world  appear classical,
it makes sense to treat quantum statistical mechanics
as a perturbation of classical statistical mechanics.
Since the latter is situated in the phase space of particle
positions and momenta, in the first place one should formulate
the entropy microstate labels as a vector of single particle sates,
\begin{equation}
{\bf n} = \{ {\bf n}_1, {\bf n}_2, \ldots , {\bf n}_N \} .
\end{equation}
And in the second place one should associate
the single particle states with a point in position and momentum space,
\begin{equation}
{\bf n}_j = \{ n_{qjx}, n_{qjy}, n_{qjz}, n_{pjx}, n_{pjy}, n_{pjz}\} .
\end{equation}
The subscript $q$ is associated with the positions,
and the subscript $p$ is associated with the momenta.

The entropy state label ${\bf n} = \{ {\bf n}_q , {\bf n}_p \}$
maps to the phase space point
${\bf \Gamma}_{\bf n} = \{ {\bf q}_{{\bf n}_q}, {\bf p}_{{\bf n}_p} \}$.
The entropy states are most simply taken to
form a uniform grid in phase space,
\begin{equation} \label{Eq:q=nDq,p=nDp}
q^\alpha_{n_{qj\alpha}}
= n_{qj\alpha} \Delta_{q} ,
\mbox{ and }
p^\alpha_{n_{pj\alpha}}
= n_{pj\alpha} \Delta_{p} ,
\end{equation}
where $\alpha = x, y, z$.
It is is not essential to specify the grid width,
but it is traditional to take $\Delta_{q}\Delta_{p} = h$,
where $h$ is Planck's constant.
Choosing this value is conventional, but it has no physical consequences.
What is important is that
the grid spacing is small enough
that the microstates form a continuum.
In this case there are many more available states
than there are particles,
so that there is never more than one particle in a one-particle state
(ie.\ all the ${\bf n}_j$ in a given microstate ${\bf n}$ are different).
However, the corollary of this is that the entropy eigenfunctions
necessarily overlap and form a non-orthogonal set.

Let the operators and the wave functions be represented
in configuration or position space,
\begin{equation}
{\bf r} = \{ {\bf r}_1, {\bf r}_2, \ldots , {\bf r}_N \} ,
\;\;
{\bf r}_{j} = \{ r_{jx}, r_{jy}, r_{jz} \}.
\end{equation}
In the position representation
the position operator for the whole system is
$\hat{\bf q} = {\bf r}$
and the momentum operator is
$\hat{\bf p}  = -i \hbar \partial /\partial {\bf r}
\equiv  -i \hbar \nabla_{\bf r}$.

It would be straightforward to extend the present analysis to include
spin variables.
But since the present interest lies in resolving the fundamental
conceptual issues with the least possible distraction,
this is deferred until another day.

%%%%%%%%%%%%%%%%%%%%%%%%%%%%%%%%%%%%%%%%%%%%%%%%%%%%%%%%%%%%%%%%%%%%%%%%%%%
\subsection{Minimum Uncertainty Wave Packet}

The wave function
that is most readily interpreted
in terms of the position and momentum
to an individual particle is a wave packet.
Accordingly, at the first level of approximation
consider the entropy eigenfunction to be product of single particle
minimum uncertainty wave packets,
\begin{eqnarray}
\lefteqn{
\zeta_{{\bf n}}({\bf r})
}  \\
& \equiv &
\frac{1}{C}
\exp \left\{ \frac{-\left({\bf r}-{\bf q}_{{\bf n}} \right)
\cdot \left({\bf r}-{\bf q}_{{\bf n}} \right) }{4 \xi^2}
%\right. \nonumber \\ && \mbox{ } \left.
- \frac{1}{i\hbar} {\bf p}_{{\bf n}}
\cdot ({\bf r}-{\bf q}_{{\bf n}})
 \right\}
\nonumber \\ & = &
\frac{1}{C} \prod_{j=1}^N \prod_{\alpha=}^{x,y,z}
e^{ -\left({r}_{j\alpha}-{q}^\alpha_{n_{qj\alpha}} \right)^2 /4 \xi^2}
%\nonumber \\ && \mbox{ }
e^{ -  {p}^\alpha_{n_{pj\alpha}}
\left({r}_{j\alpha}-{q}^\alpha_{n_{qj\alpha}} \right) /i\hbar } .
\nonumber
\end{eqnarray}
Here and below the scalar product is a sum over
the $j=1,2,\ldots,N$ particle labels and the three coordinates
$\alpha =x,y,z$,
for example
${\bf p}_{{\bf n}}\cdot {\bf r}
= \sum_{j,\alpha}
p^\alpha_{n_{pj\alpha}} r_{j\alpha}$.
The normalizing factor is given by $C^2 =(2\pi\xi^2)^{3N/2}$.

This product of wave packets is most compactly written as
\begin{eqnarray}  \label{Eq:zeta^0_n}
 \zeta_{{\bf n}}({\bf r})
& \equiv &
\frac{1}{ C}
e^{  -{\bm \varepsilon}_{{\bf n}}({\bf r})^2/ 4 \xi^2 }
%\right. \nonumber \\ && \mbox{ } \left.
e^{- {\bf p}_{{\bf n}} \cdot {\bm \varepsilon}_{{\bf n}}({\bf r})/{i\hbar} },
\end{eqnarray}
where ${\bm \varepsilon}_{{\bf n}}({\bf r})
\equiv {\bf r}-{\bf q}_{{\bf n}} $.
It is assumed that the Gaussian form for the reference wave packet
is sufficient to keep this small
(but, it will turn out,
not in the case of the ideal gas).
This provides the basis for a systematic expansion
for the entropy eigenfunction
(at least in the case of a continuous potential).

The probability amplitude,
$ \zeta_{{\bf n}}^* \,  \zeta_{{\bf n}}$,
is a Gaussian of width $\xi$ in position space
and $\hbar/2\xi$ in momentum space
(per particle, per direction).
To within an error of these magnitudes,
the minimum uncertainty wave function
is approximately a simultaneous eigenfunction
of the position and momentum operators,
\begin{equation}
\hat {\bf q} |  \zeta_{\bf n}({\bf r}) \rangle
= {\bf r} |  \zeta_{\bf n}({\bf r}) \rangle
\approx {\bf q}_{\bf n} |  \zeta_{\bf n}({\bf r}) \rangle,
\end{equation}
and
\begin{eqnarray}
\hat {\bf p} |  \zeta_{\bf n}({\bf r}) \rangle
& = &
\left[ \frac{\hbar}{ 2 i \xi^2} \left({\bf q}_{\bf n}-{\bf r} \right)
+ {\bf p}_{\bf n} \right]  |  \zeta_{\bf n}({\bf r}) \rangle
\nonumber \\ & \approx &
{\bf p}_{\bf n} |  \zeta_{\bf n}({\bf r}) \rangle .
\end{eqnarray}
These follow from the sharply peaked nature of the wave packet,
which means that any prefactor that is a slowly-varying function of ${\bf r}$
can be evaluated at ${\bf q}_{\bf n}$.
Because the wave packet is approximately
a simultaneous position-momentum eigenfunction,
it is also an energy and hence an entropy eigenfunction.
It therefore forms a suitable reference wave function
that can be systematically corrected
to form a true entropy eigenfunction.

As will be shown in detail in Eq.~(\ref{Eq:hat H_zeta}) below,
the energy operator acting on  the minimum uncertainty wave packet yields
\begin{eqnarray}
\lefteqn{
\hat{\cal H}({\bf r}) \zeta_{\bf n}({\bf r})
} \nonumber  \\
& = &
\left\{ U({\bf r})
+ \frac{1}{2m} {\bf p}_{{\bf n}}\cdot{\bf p}_{{\bf n}}
+\frac{3N\hbar^2}{4m\xi^2}
\right. \nonumber \\ &&  \left. \mbox{ }
 - \frac{\hbar^2}{8m\xi^4}
{\bm \varepsilon}_{{\bf n}}({\bf r}) \cdot  {\bm \varepsilon}_{{\bf n}}({\bf r})
 - \frac{\hbar^2}{2mi\hbar\xi^2}
{\bm \varepsilon}_{{\bf n}}({\bf r}) \cdot  {\bf p}_{{\bf n}}
\right\} \zeta_{\bf n}({\bf r})
\nonumber \\ & \approx &
\left\{ {\cal H}({\bf q}_{\bf n},{\bf p}_{\bf n}) +\frac{3N\hbar^2}{4m\xi^2}
\right\} \zeta_{\bf n}({\bf r}) .
\end{eqnarray}
The second equality assumes that the wave packet is sharply peaked so that
${\bm \varepsilon}_{{\bf n}}({\bf r})$ and
variations in the potential can be neglected.
This gives the eigenvalue,
with the first term being the classical Hamiltonian function
of the nominal positions and momenta.
The second term is an immaterial constant.
Hence $\zeta_{\bf n}$ is approximately
an  entropy eigenfunction.

\comment{ %%%%%%%%%%%%%%%%%%%%%%%%%%%%%%%%%%%%%%%%%%%%%%%%
For the bare minimum uncertainty wave packet,
the expectation values of the position and momentum operators
are ${\bf q}_{\bf n}$ and ${\bf p}_{\bf n}$, respectively.
Below, systematic modifications of the wave packet
that progressively improve it as an entropy eigenfunction
will be explored.
In this case the expectation values do not coincide exactly
with ${\bf q}_{\bf n}$ and ${\bf p}_{\bf n}$.
For this reason the ${\bf q}_{\bf n}$ and ${\bf p}_{\bf n}$
will be called the nominal positions and momenta of the particles
in the entropy microstate ${\bf n}$.
%(The set ${\bf r}$ will be called the actual configuration,
%or the actual particle positions.)
} % end comment %%%%%%%%%%%%%%%%%%%%%%%%%%%%%%%%%%%%%%%%%%%%%%%%

The nominal positions and momenta,
${\bf q}_{{\bf n}}$ and $ {\bf p}_{{\bf n}}$,
occur naturally
in the Hamiltonian function of classical mechanics,
and they therefore play precisely the same role
as the classical  positions and momenta.
Therefore it is these rather than the expectation values,
the eigenvalues of the position and momentum operators,
or the operators themselves
that ought to be regarded as the analogues
of the classical  positions and momenta.
Just like their classical counterparts,
the nominal positions and momenta
are not restricted by
any non-commutative behavior or uncertainty relations.
Ultimately it is this interpretation of the eigenvalues of the entropy operator
as the nominal positions and momenta,
${\bf q}_{{\bf n}}$ and $ {\bf p}_{{\bf n}}$,
that explains how classical mechanics arises from quantum mechanics.

As mentioned above,
the reference wave packets are not necessarily
orthogonal to each other.
If the spacing between states is large,
$\Delta_q \gg \xi$ and $\Delta_p \gg \hbar/2\xi$,
then there is no overlap between
the wave packets of different states,
and in this case they form an orthogonal set:
interchanging particles in different one-particle states
creates a wave function orthogonal to the original
as the one-particle states do not overlap.
However, the larger the states,
the smaller is the ratio of accessible states to particles,
and so the more likely it is for the states
to be occupied by more than one particle.
As just mentioned,
the classical continuum corresponds
to the small grid width limit.
In this case the wave packets do not form an orthogonal set
(unless some specific orthogonalization scheme is invoked).
In what follows it will be assumed that the set
of entropy eigenfunctions is not orthogonal.

%It may be desirable to replace the wave packet length scale
%by a state dependent variance,
%$\xi^2 \Rightarrow {\bm \sigma}_{\bf n}$,
%in which case this would no longer be the product of single particle
%wave functions.

%%%%%%%%%%%%%%%%%%%%%%%%%%%%%%%%%%%%%%%%%%%%%%%%%%%%%%%%%%%%%%%%%%%%%%%%%%%
\subsection{Optimized Wave Packet Width for the Entropy Eigenfunction}
\label{Sec:olxi}

The minimal uncertainty wave packet just described
is approximately an entropy eigenfunction, as the last result above shows.
There are at least two ways that the  approximation inherent
in this can be reduced:
by optimizing the wave packet width $\xi$,
which is explored here,
and by introducing a pre-factor function that modifies the wave packet,
which is explored
in \S\S \ref{Sec:f-zeta} and \ref{Sec:ctsU}.

In principle, it ought to be straightforward
to construct approximate entropy eigenfunctions.
Since the entropy microstates are highly degenerate,
one can form the subspace spanned by all true entropy eigenfunctions,
$\{\phi \}$, of energy ${\cal H}_{\bf m}$.
One can project the approximate entropy eigenfunction,
$\zeta_{\bf n}({\bf r})$, onto this subspace,
\begin{equation}
\zeta_{\bf n}^\perp({\bf r})
=
\sum_{\bf m} \!^{({\cal H}_{\bf m}= E_{\bf n})} \;
\langle \phi_{\bf m}| \zeta_{\bf n}\rangle \, \phi_{\bf m}({\bf r}).
\end{equation}
The function $\zeta_{\bf n}^\perp$
is an exact entropy eigenfunction with eigenvalue $-E_{\bf n}/T$:
$\hat {\cal H} \, \zeta_{\bf n}^\perp({\bf r})
= E_{\bf n} \zeta_{\bf n}^\perp({\bf r})$.

Obviously, $\zeta_{\bf n}({\bf r})$ is an approximation to
$\zeta^\perp_{\bf n}({\bf r})$.
Given that the degeneracy of the true entropy eigenfunctions increases
with the size of the system,
one expects that
$\zeta^\perp_{\bf n}({\bf r}) \rightarrow \zeta_{\bf n}({\bf r})$
in the thermodynamic limit, $N \rightarrow \infty$.
That is, the increasing size of the degenerate  sub-space allows
increasing flexibility in fitting an arbitrary function.

With enough adjustable parameters in the approximate entropy eigenfunction,
it can be made more and more exact.
For example, symbolize the energy expectation value
of the approximate entropy eigenfunction as
\begin{equation}
E_{\bf n} \equiv
\langle \zeta_{\bf n} | \hat {\cal H}  | \zeta_{\bf n} \rangle .
\end{equation}
Then one criteria for improving
the eigenfunction is to minimize the fluctuation in the energy expectation,
\begin{eqnarray}
\Delta E_{\bf n}^2 & \equiv &
\left\langle \zeta_{\bf n} \left| \left\{ \hat {\cal H} - E_{\bf n} \right\}^2
\right| \zeta_{\bf n} \right\rangle
\nonumber  \\ & = &
\left\langle \zeta_{\bf n} \left|  \hat {\cal H}^2
\right| \zeta_{\bf n} \right\rangle
-E_{\bf n}^2 .
\end{eqnarray}
This is of course non-negative.
The vanishing of this is a necessary but not
a sufficient condition for $\zeta_{\bf n}({\bf r})$
to be an entropy eigenfunction.

This section is concerned with using
the minimum uncertainty wave packet itself
as the approximate entropy eigenfunction.
The aim is to determine the width of the wave packet
by minimizing the energy fluctuation.
Recall that the minimum uncertainty wave function is
\begin{eqnarray}
 \zeta_{{\bf n}}({\bf r})
& \equiv &
\frac{1}{ C}
e^{  -{\bm \varepsilon}_{{\bf n}}({\bf r})^2/ 4 \xi^2 }
%\right. \nonumber \\ && \mbox{ } \left.
e^{- {\bf p}_{{\bf n}} \cdot {\bm \varepsilon}_{{\bf n}}({\bf r})/{i\hbar} },
\end{eqnarray}
where ${\bm \varepsilon}_{{\bf n}}({\bf r})
\equiv {\bf r}-{\bf q}_{{\bf n}} $.
%The tilde has been dropped
%as this is the only entropy eigenfunction
%that will be analyzed in this section.

For the minimum uncertainty wave packet,
\begin{eqnarray}
\lefteqn{
\hat{\cal H}({\bf r}) \zeta_{\bf n}({\bf r})
}   \\
& = &
\left\{ U({\bf r})
+ \frac{1}{2m} {\bf p}_{{\bf n}}\cdot{\bf p}_{{\bf n}}
+\frac{3N\hbar^2}{4m\xi^2}
\right. \nonumber \\ &&  \left. \mbox{ }
 - \frac{\hbar^2}{8m\xi^4}
{\bm \varepsilon}_{{\bf n}}({\bf r}) \cdot  {\bm \varepsilon}_{{\bf n}}({\bf r})
 - \frac{\hbar^2}{2mi\hbar\xi^2}
{\bm \varepsilon}_{{\bf n}}({\bf r}) \cdot  {\bf p}_{{\bf n}}
\right\} \zeta_{\bf n}({\bf r}) .
\nonumber
\end{eqnarray}
(This is derived in Eq.~(\ref{Eq:hat H_zeta}) below.)
Expanding the potential energy to second order about ${\bf q}_{\bf n}$,
this gives the expectation value of the energy as
\begin{eqnarray}
E_{\bf n}
& = &
 U({\bf q}_{\bf n})
+ \frac{\xi^2}{2} \mbox{TR }{\bf U}_{\bf n}''
+ \frac{1}{2m} {\bf p}_{{\bf n}}\cdot{\bf p}_{{\bf n}}
+\frac{3N\hbar^2}{4m\xi^2}
\nonumber \\ &&  \mbox{ }
 - \frac{\hbar^2}{2m} \frac{1}{4\xi^4} 3 N \xi^2
\nonumber \\ & = &
 {\cal H}({\bf q}_{\bf n},{\bf p}_{\bf n})
+ \frac{\xi^2}{2} \mbox{TR }{\bf U}_{\bf n}''
+\frac{3N\hbar^2}{8m\xi^2} .
\end{eqnarray}
(This does not hold for a discontinuous potential.)
Here and below the expansion of the potential energy
is terminated at the quadratic term.
The expected energy is just the classical Hamiltonian of the nominal positions
and momenta, plus a constant,
plus the term ${\xi^2} \mbox{TR }{\bf U}_{\bf n}''/{2}$,
which is also a constant in the thermodynamic limit
(and it will turn out to be  ${\cal O}(N^{3/4})$,
which is relatively negligible  in the thermodynamic limit).

Note that here and everywhere analogous below,
the trace is the ordinary matrix trace,
which is the sum over the $6N$--one-particle states in the microstate ${\bf n}$.
It is not the quantum trace of an operator,
which is the sum over unique microstates.

The expectation value of the square of the Hamiltonian operator
is derived in detail in Appendix \ref{Sec:<H2>n}.
The result is
\begin{eqnarray}
\lefteqn{
\langle \zeta_{\bf n} | \hat{\cal H}^2 | \zeta_{\bf n}\rangle
} \nonumber  \\
& = &
\left(  U_{\bf n} + {\cal K}_{\bf n} \right)^2
+
\frac{3N\hbar^2}{4m\xi^2 }
\left(  U_{\bf n} + {\cal K}_{\bf n} \right)
+ \frac{ 9N^2 \hbar^4}{2^6 m^2   \xi^4 }
\nonumber  \\ &  &  \mbox{ }
+ \left\{
\xi^2{\cal K}_{\bf n}
+ \xi^2  U_{\bf n}
+ \frac{3N\hbar^2}{8m}
\right\}
\mbox{TR }{\bf U}_{\bf n}''
\nonumber \\ && \mbox{ }
+ \frac{\xi^4}{4} ( \mbox{TR }{\bf U}_{\bf n}'')^2
+ \frac{ \xi^4}{2} {\bf U}_{\bf n}'' : {\bf U}_{\bf n}''
\nonumber \\ && \mbox{ }   %%%%%%%%%%%%%%%%%%%%%%  O(N):
- \frac{i\hbar}{2m} {\bf U}_{\bf n}' \cdot {\bf p}_{\bf n}
+ \frac{  \hbar^2}{2m} \mbox{TR }{\bf U}_{\bf n}''
+  \frac{6 \xi^4}{4} \sum_{j,\alpha} ( U''_{j\alpha,j\alpha})^2
\nonumber  \\ &  &  \mbox{ }
+ \xi^2 {\bf U}_{\bf n}' \cdot {\bf U}_{\bf n}'
%\nonumber  \\ &  &  \mbox{ }
+ \frac{\hbar^2}{m\xi^2}  {\cal K}_{\bf n}
+ \frac{ N \hbar^4}{8m^2\xi^4 } .
\end{eqnarray}
The first six terms are ${\cal O}(N^2)$,
and the final five terms are ${\cal O}(N)$.
The factor $ {\bf U}_{\bf n}' \cdot {\bf p}_{\bf n}$
is ${\cal O}(N^{1/2})$ and can be neglected.
Note that the expansion of the potential energy has been
terminated at the second derivative.

Now the square of the expectation value of the energy is
\begin{eqnarray}
E_{\bf n}^2
& = &
\left( U_{\bf n} + {\cal K}_{\bf n} + \frac{3N\hbar^2}{8m\xi^2} \right)^2
+ \frac{\xi^4}{4} ( \mbox{TR }{\bf U}_{\bf n}'' )^2
\nonumber \\ && \mbox{ }
+
\xi^2 \left( U_{\bf n} + {\cal K}_{\bf n} + \frac{3N\hbar^2}{8m\xi^2} \right)
\mbox{TR }{\bf U}_{\bf n}'' .
\end{eqnarray}
This cancels with all but the final three terms that are ${\cal O}(N^2)$
in the expectation value of the square of the energy operator.
This leaves the energy fluctuation as
\begin{eqnarray} \label{Eq:DE_n^2}
%\lefteqn{
\Delta E_{\bf n}^2
%} \nonumber  \\
& = &
\langle \zeta_{\bf n} | \hat{\cal H}^2 | \zeta_{\bf n}\rangle
-
\langle \zeta_{\bf n} | \hat{\cal H} | \zeta_{\bf n}\rangle^2
\nonumber \\ &=&
\frac{\xi^4}{4} \left[ ( \mbox{TR }{\bf U}_{\bf n}'')^2
+ 2 {\bf U}_{\bf n}'' : {\bf U}_{\bf n}''\right]
\nonumber \\ && \mbox{ }   %%%%%%%%%%%%%%%%%%%%%%  O(N):
+\frac{  \hbar^2}{2m} \mbox{TR }{\bf U}_{\bf n}''
+  \frac{6 \xi^4}{4} \sum_{j,\alpha} ( U''_{j\alpha,j\alpha})^2
\nonumber  \\ &  &  \mbox{ }
+ \xi^2 {\bf U}_{\bf n}' \cdot {\bf U}_{\bf n}'
%\nonumber  \\ &  &  \mbox{ }
+ \frac{\hbar^2}{m\xi^2}  {\cal K}_{\bf n}
+ \frac{ N \hbar^4}{8m^2\xi^4 } .
\end{eqnarray}
Evidently every term here is  non-negative.
This means that this expression for the energy fluctuation
is positive semi-definite,
which is of course a necessary condition.

The derivative of the energy fluctuation with respect to
the square of the wave packet width is
\begin{eqnarray}
\frac{\partial (\Delta E_{\bf n}^2)}{\partial (\xi^2)}
& = &
\frac{\xi^2}{2} \left[ ( \mbox{TR }{\bf U}_{\bf n}'')^2
+ 2 {\bf U}_{\bf n}'' : {\bf U}_{\bf n}''\right]
\nonumber \\ && \mbox{ }   %%%%%%%%%%%%%%%%%%%%%%  O(N):
+  \frac{6 \xi^2}{2} \sum_{j,\alpha} ( U''_{j\alpha,j\alpha})^2
\nonumber  \\ &  &  \mbox{ }
+ {\bf U}_{\bf n}' \cdot {\bf U}_{\bf n}'
%\nonumber  \\ &  &  \mbox{ }
- \frac{\hbar^2}{m\xi^4}  {\cal K}_{\bf n}
- \frac{ 2  N \hbar^4}{8m^2\xi^6 } .
\end{eqnarray}
This is negative at small $\xi^2$ and positive at large $\xi^2$.
This order of the slopes is the same order as
that in which they occur for a parabolic minimum.
Hence  $\Delta E_{\bf n}^2$ certainly has at least one minimum
at some intermediate value of  $\xi^2$.

Because this derivative is the sum of positive terms,
that scale with $N^2$ and $\xi^2$,
(and $N$ and $\xi^2$, and $N$ and $\xi^0$),
and negative terms that scale with $N$ and $\xi^{-6}$
(and $N$ and $\xi^{-4}$),
for the two slopes to remain comparable $\xi$ must decrease
with increasing $N$.
In the limit $N\rightarrow\infty$, the first and the final term
on the right hand side of the derivative dominate all the others,
as is clear by inspection.
Keeping only these two terms,
the derivative vanishes when
\begin{eqnarray}
0
& = &
\frac{\overline \xi_{\bf n}\!\!^2}{2} \left[ ( \mbox{TR }{\bf U}_{\bf n}'')^2
+ 2 {\bf U}_{\bf n}'' : {\bf U}_{\bf n}''\right]
- \frac{  N \hbar^4}{4m^2 \overline \xi_{\bf n}^6 } ,
\end{eqnarray}
or
\begin{equation}
\overline \xi_{\bf n}\!\!^8
=
\frac{ 2 {N \hbar^4}/{4m^2}
}{
( \mbox{TR }{\bf U}_{\bf n}'')^2
+ 2 {\bf U}_{\bf n}'' : {\bf U}_{\bf n}''   } .
\end{equation}
Since the denominator is ${\cal O}(N^2)$,
this means that $\overline \xi \propto N^{-1/8}$,
$N\rightarrow\infty$.

This result holds in the thermodynamic limit;
for finite $N$,
one should minimize the full expression for the energy fluctuation,
Eq.~(\ref{Eq:DE_n^2}),
to obtain $\overline \xi_{\bf n}(N)$.
In this case, the $N$-dependence of the wave packet
has to be accounted for in evaluating the individual terms
in the grand canonical partition function and average,
particularly in the case of a low density expansion.
This has implications for the analysis in \S \ref{Sec:non-ortho-zeta}.

For an ideal gas, ${\bf U}=0$,
this shows that $\overline \xi\,\!^\mathrm{id} \rightarrow \infty$.
The same conclusion is reached by minimizing directly the full expression
for $\Delta E_{\bf n}^2$, Eq.~(\ref{Eq:DE_n^2}).
This result for the ideal gas was used in the discussion of
Eq.~(\ref{Eq:chi(2)^id}) above,
and will be used in the analysis of the quantum ideal gas
in \S \ref{Sec:IdealGas} below.
(Of course the energy eigenfunction of an ideal particle
is known by elementary methods to be of the form
$\zeta_{\bf k}({\bf r}) \propto e^{i {\bf k}\cdot {\bf r}}$,
which is indeed a wave packet with $\xi = \infty$,
but it is consoling to derive this result directly within the
present formalism.)

With this optimum wave packet width,
the relative root mean square energy expectation fluctuation scales as
\begin{equation}
\frac{1}{E_{\bf n}} \sqrt{ \Delta E_{\bf n}^2}
\sim \frac{1}{N^{5/4}}  \sqrt{N^{3/2}}
\sim N^{-1/2} .
\end{equation}
(This includes the scaling of $\overline \xi$ with $N$.)
This is a measure of the relative accuracy
of the minimum uncertainty wave packet
as an entropy eigenfunction.
Clearly in the thermodynamic limit
the wave packet is an exact entropy eigenfunction.

Note that
\begin{equation}
\mbox{TR }{\bf U}_{\bf n}''
= \langle \mbox{TR }{\bf U}'' \rangle_\mathrm{stat}
+ {\cal O}(N^{1/2}) ,
\end{equation}
and
\begin{equation}
{\bf U}_{\bf n}'' : {\bf U}_{\bf n}''
= \langle {\bf U}'' : {\bf U}'' \rangle_\mathrm{stat}
+ {\cal O}(N^{3/2}) .
\end{equation}
These mean that the optimum wave packet width
can be taken to be independent of the entropy microstate
to leading order:  $\overline \xi_{\bf n} =\overline \xi $.

The fact that for  real particles, $U \ne 0$,
the optimum wave packet width vanishes in the thermodynamic limit
poses difficulties for the calculation of the loop overlap factors.
Since zero width wave packets can't overlap,
it is evident that  one must obtain
the loop overlap factors before taking the thermodynamic limit.
One solution, suggested in \S \ref{Sec:<chi_n>},
is that the thermodynamic limit be invoked for the monomer particles,
in which case their entropy eigenfunctions are zero width wave packets,
and their entropy eigenstates are points in classical phase space.
The $l$ particles that comprise the $l$-loop are not in the thermodynamic limit,
and their entropy eigenfunctions and eigenstates can be calculated
as a function of the monomer configuration,
either exactly, or approximately as wave packets of finite, optimized, width.
This enables the $l$-loop overlap factor to be obtained.

%%%%%%%%%%%%%%%%%%%%%%%%%%%%%%%%%%%%%%%%%%%%%%%%%%%%%%%%%
\subsection{Modification of the Wave Packet}  \label{Sec:f-zeta}

In this section an alternative way of systematically improving
the wave packet as an entropy eigenfunction is explored.
Denoting the bare or reference wave packet, Eq.~(\ref{Eq:zeta^0_n}),
by a tilde,
one can formally write the entropy eigenfunction as
\begin{equation} \label{Eq:zetaSn(q)}
\zeta_{\bf n}({\bf r})
=
\frac{\tilde C_{\bf n}}{C_{\bf n}}
f_{\bf n}({\bf r}) \tilde \zeta_{\bf n}({\bf r}) ,
\end{equation}
with the new normalization factor given by
\begin{equation}
C_{\bf n}^2
= \tilde C_{\bf n}^2 \;
\langle \, f_{\bf n}({\bf r}) \tilde \zeta_{\bf n}({\bf r})
 \,| \,
f_{\bf n}({\bf r}) \tilde \zeta_{\bf n}({\bf r}) \, \rangle ,
\end{equation}
which ensures that
$\langle \zeta_{\bf n} | \zeta_{\bf n} \rangle = 1$.

The idea behind this formulation
is that the mixture of pure entropy states
is sufficiently complete.
(Sufficient because although
it does not include the superposition of entropy states,
these are not required for an open sub-system
that can exchange with a reservoir.)
An entropy state ${\bf n} = \{ {\bf n}_{q},{\bf n}_{p}\}$
corresponds to a point in classical phase space,
which the grid points are chosen to cover completely.
Conversely, any phase space point corresponds to an entropy eigenstate.
The corresponding entropy eigenfunction
$\zeta_{{\bf n}}$
must be sharply peaked
in configuration space, $ {\bf r} \approx {\bf q}_{{\bf n}}$,
and in momentum space, $ {\bf p} \approx {\bf p}_{{\bf n}}$,
for this to have any meaning.
The function ${\bm \varepsilon}_{\bf n}({\bf r})$
measures the departure of the actual configuration
from the nominal locality of the state,
and so it provides a suitable ordering quantity
to give the systematic corrections
to the entropy eigenfunction.

With the momentum operator for the system being
$\hat {\bf p} = - i \hbar \nabla_{\bf r} $,
the Hamiltonian operator is
\begin{eqnarray}
\hat{\cal H}({\bf r})
& = &
U({\bf r}) - \frac{\hbar^2}{2m}  \nabla_{\bf r}^2
\nonumber \\ & = &
U({\bf r})
- \frac{\hbar^2}{2m}
\sum_{j=1}^N \sum_{\alpha=x,y,z}
\frac{\partial^2}{\partial {r}^{2}_{j\alpha} }.
\end{eqnarray}

The derivatives of the reference wave function are
\begin{equation}
\nabla_{\bf r} \tilde \zeta_{\bf n}({\bf r})
=
\left[
\frac{-1}{2\xi^2} {\bm \varepsilon}_{{\bf n}}({\bf r})
- \frac{1}{i\hbar} {\bf p}_{{\bf n}}
\right]
\tilde \zeta_{\bf n}({\bf r}) ,
\end{equation}
and
\begin{eqnarray}
\nabla_{\bf r}^2 \tilde \zeta_{\bf n}({\bf r})
& = &
\left\{
\left[ \frac{-1}{2\xi^2} {\bm \varepsilon}_{{\bf n}}({\bf r})
- \frac{1}{i\hbar} {\bf p}_{{\bf n}} \right]^2
-\frac{3N}{2\xi^2}
\right\} \tilde \zeta_{\bf n}({\bf r}) .
\nonumber \\
\end{eqnarray}

Since in the present canonical equilibrium case,
the entropy operator is proportional to the energy operator,
$\hat S_{\mathrm{r}} = -\hat{\cal H}/T$,
the entropy eigenfunctions can be obtained
by exploring the effects of the Hamiltonian operator
on the wave function.
One has
\begin{eqnarray} \label{Eq:hat H_zeta}
\lefteqn{
\hat{\cal H}({\bf r}) \zeta_{\bf n}({\bf r})
} \nonumber  \\
& = &
U({\bf r}) \zeta_{\bf n}({\bf r})
- \frac{\hbar^2}{2m}
\frac{\tilde C_{\bf n}}{C_{\bf n}}
\left\{
 \tilde \zeta_{\bf n}({\bf r}) \nabla_{\bf r}^2 f_{\bf n}({\bf r})
\right. \nonumber \\ && \left. \mbox{ }
+ 2
\nabla_{\bf r} f_{\bf n}({\bf r}) \cdot
\nabla_{\bf r}\tilde \zeta_{\bf n}({\bf r})
+
f_{\bf n}({\bf r}) \nabla_{\bf r}^2\tilde \zeta_{\bf n}({\bf r})
\right\}
\nonumber  \\ & = &
U({\bf r}) \zeta_{\bf n}({\bf r})
- \frac{\hbar^2}{2m}
\left\{
\frac{1}{f_{\bf n}({\bf r})} \nabla_{\bf r}^2 f_{\bf n}({\bf r})
\right. \nonumber \\ && \left. \mbox{ }
+
\frac{2}{f_{\bf n}({\bf r})}
\nabla_{\bf r} f_{\bf n}({\bf r}) \cdot
\left[
\frac{-1}{2\xi^2} {\bm \varepsilon}_{{\bf n}}({\bf r})
- \frac{1}{i\hbar} {\bf p}_{{\bf n}}
\right]
\right. \nonumber \\ && \left. \mbox{ }
+
\left[ \frac{-1}{2\xi^2} {\bm \varepsilon}_{{\bf n}}({\bf r})
- \frac{1}{i\hbar} {\bf p}_{{\bf n}} \right]^2
-\frac{3N}{2\xi^2}
\right\} \zeta_{\bf n}({\bf r})
\nonumber \\ & = &
\left\{ U({\bf q}_{\bf n})
+ \frac{1}{2m} {\bf p}_{{\bf n}}\cdot{\bf p}_{{\bf n}}
+\frac{3N\hbar^2}{4m\xi^2}
\right\} \zeta_{\bf n}({\bf r})
 \nonumber \\ &&  \mbox{ }
 +
 \left\{  U({\bf r}) -  U({\bf q}_{\bf n}) \right\}
 \zeta_{\bf n}({\bf r})
 \nonumber \\ &&  \mbox{ }
 - \frac{\hbar^2}{2m}
\left\{
\frac{1}{f_{\bf n}({\bf r})} \nabla_{\bf r}^2 f_{\bf n}({\bf r})
\right. \nonumber \\ && \left. \mbox{ }
-
\frac{2}{f_{\bf n}({\bf r})}
\nabla_{\bf r} f_{\bf n}({\bf r}) \cdot
\left[
\frac{1}{2\xi^2} {\bm \varepsilon}_{{\bf n}}({\bf r})
+ \frac{1}{i\hbar} {\bf p}_{{\bf n}}
\right]
\right. \nonumber \\ && \left. \mbox{ }
+
\frac{1}{4\xi^4}
{\bm \varepsilon}_{{\bf n}}({\bf r}) \cdot  {\bm \varepsilon}_{{\bf n}}({\bf r})
+ \frac{1}{i\hbar\xi^2}
{\bm \varepsilon}_{{\bf n}}({\bf r}) \cdot  {\bf p}_{{\bf n}}
\right\} \zeta_{\bf n}({\bf r}) .
\nonumber \\
\end{eqnarray}

The first bracketed term is the eigenvalue
\begin{eqnarray}
{\cal H}_{\bf n}
& = &
U({\bf q}_{\bf n})
+ \frac{1}{2m} {\bf p}_{{\bf n}}\cdot{\bf p}_{{\bf n}}
+\frac{3N\hbar^2}{4m\xi^2}
\nonumber \\ & \equiv &
{\cal H}({\bf q}_{\bf n},{\bf p}_{\bf n}) +\frac{3N\hbar^2}{4m\xi^2}.
\end{eqnarray}
The classical Hamiltonian function of phase space
appears on the right hand side.
It is this that justifies interpreting the entropy eigenstates
as position and momentum points in classical phase space.
In addition to the Hamiltonian energy,
the eigenvalue has an added constant,
${3N\hbar^2}/{4m\xi^2}$.
This is immaterial as it has no physical effect.
In classical mechanics energy is defined only in relative
rather than absolute terms.
It would be possible cancel this constant with an addition to the second order
term in the modifying function,
but this appears to create problems elsewhere.

For this to be a true entropy eigenfunction,
the remaining terms must vanish
\begin{eqnarray}
%\lefteqn{
0
%} \nonumber  \\
& = &
 U({\bf r}) -  U({\bf q}_{\bf n})
 \nonumber \\ &&  \mbox{ }
 - \frac{\hbar^2}{2m}
\left\{
\frac{1}{f_{\bf n}({\bf r})} \nabla_{\bf r}^2 f_{\bf n}({\bf r})
\right. \nonumber \\ && \left. \mbox{ }
-
\frac{2}{f_{\bf n}({\bf r})}
\nabla_{\bf r} f_{\bf n}({\bf r}) \cdot
\left[
\frac{1}{2\xi^2} {\bm \varepsilon}_{{\bf n}}({\bf r})
+ \frac{1}{i\hbar} {\bf p}_{{\bf n}}
\right]
\right. \nonumber \\ && \left. \mbox{ }
+
\frac{1}{4\xi^4}
{\bm \varepsilon}_{{\bf n}}({\bf r}) \cdot  {\bm \varepsilon}_{{\bf n}}({\bf r})
+ \frac{1}{i\hbar\xi^2}
{\bm \varepsilon}_{{\bf n}}({\bf r}) \cdot  {\bf p}_{{\bf n}}
\right\}.
\end{eqnarray}
This is a second order differential equation for the modifier function
of the entropy eigenfunction.
For a given potential $U({\bf r})$
and a given state ${\bf n}$,
which fixes the nominal momenta ${\bf p}_{{\bf n}} = {\bf n}_p \Delta_p$
and positions ${\bf q}_{{\bf n}} = {\bf n}_q \Delta_q$,
this has to be solved for $f_{\bf n}({\bf r})$.
%Suitable boundary conditions have to be imposed.

Obviously this is satisfied if $f_{\bf n}({\bf r})=1$
and $ U({\bf r}) -  U({\bf q}_{\bf n})
= {\cal O}({\bm \varepsilon}_{{\bf n}}({\bf r}))$
is negligible.

%%%%%%%%%%%%%%%%%%%%%%%%%%%%%%%%%%%%%%%%%%%%%%%%%%%%%%%%%
\subsection{Expansion For Continuous Potentials} \label{Sec:ctsU}

This sub-section gives an ansatz for the entropy eigenfunction
to third order, based on solving
the under-determined zeroth and first order eigenvalue equations.
The eigenvalue equation is satisfied to first order.
In Appendix \ref{Sec:2Oeigen},
the zeroth, first, and second order eigenvalue equations are given,
which are under-determined for the entropy eigenfunction to fourth order.
That analysis gives non-zero expressions for the first three coefficients
of the eigenfunction expansion,
which ensure that the eigenvalue equation is satisfied to second order.
Although different to the present expressions
(since both result from an under-determined system),
they agree that the coefficients are ${\cal O}(N^{-1})$
and are therefore negligible in the thermodynamic limit.

As mentioned above,
one expects the entropy eigenfunctions to be sharply peaked
about the points in classical phase space.
In this case the wave packet
that is a Gaussian about these points
is a suitable reference function.
The function ${\bm \varepsilon}_{\bf n}({\bf r})$
measures the departure of the actual configuration
from the nominal locality of the state,
and so it provides a suitable ordering quantity
to give the systematic corrections
to the entropy eigenfunction.

In the case of a continuous potential,
one has the obvious expansion  about the nominal position,
\begin{eqnarray}
\lefteqn{
U({\bf r}) -  U({\bf q}_{\bf n})
} \nonumber \\
& = &
{\bf U}'({\bf q}_{\bf n}) \cdot {\bm \varepsilon}_{{\bf n}}({\bf r})
 + \frac{1}{2} {\bf U}''({\bf q}_{\bf n}) :
 {\bm \varepsilon}_{{\bf n}}({\bf r}) \,  {\bm \varepsilon}_{{\bf n}}({\bf r})
 \nonumber \\ && \mbox{ }
 + \ldots .
\end{eqnarray}
The gradient of the potential that appears here, ${\bf U}_{{\bf n}}'$,
in component form is
\begin{equation}
\left\{ {\bf U}_{{\bf n}}'  \right\}_{j\alpha}
=
\left.
\frac{\partial U({\bf r})}{\partial r_{j\alpha}}
\right|_{{\bf r} = {\bf q}_{{\bf n}}}.
\end{equation}

The modifying function has the expansion
\begin{eqnarray} \label{Eq:fn(q)}
f_{\bf n}({\bf r})
& = &
1
+ {\bf f}^{\bf n} \cdot {\bm \varepsilon}_{{\bf n}}({\bf r})
+ \frac{1}{2} {\bf f}^{\bf nn}
: {\bm \varepsilon}_{{\bf n}}({\bf r})
{\bm \varepsilon}_{{\bf n}}({\bf r})
+ \ldots
\nonumber \\ & = &
1
+ \sum_{j,\alpha} {f}^{\bf n}_{j\alpha}
[ r_{j\alpha} - q^\alpha_{n_{qj\alpha}}]
\nonumber \\ && \mbox{ }
+ \frac{1}{2} \sum_{j\alpha,k\beta}
{\bf f}^{\bf nn}_{j\alpha,k\beta}
[ r_{j\alpha}  - q^\alpha_{n_{qj\alpha}}]
[ r_{k\beta} - q^\beta_{n_{qk\beta}}]
\nonumber \\ && \mbox{ }
+ \ldots
\end{eqnarray}
The nature of the functional form chosen for the entropy eigenfunction
of the system
means that it is no longer the product
of single particle wave functions,
 $\zeta_{\bf n}({\bf r})
 \ne \prod_j  \zeta_{{\bf n}_j}({\bf r}_j)$.
As mentioned above, the modifier may shift the expected position,
$\langle \zeta_{\bf n}|
\hat {\bf q}
| \zeta_{\bf n} \rangle
\ne {\bf q}_{\bf n}$,
and the expected momentum,
$\langle \zeta_{\bf n}|
\hat {\bf p}
| \zeta_{\bf n} \rangle
\ne {\bf p}_{\bf n}$,
from the nominal ones of the state, but this is not significant.

The idea is to obtain the coefficients
in the expression for $f_{\bf n}({\bf r})$
by requiring that the eigenvalue equation for the entropy operator
be satisfied up to a certain order in the expansion parameter.
This means that the eigenvalue must be independent of
${\bm \varepsilon}_{\bf n}({\bf r})$ up to that order.
In the  case  now analyzed
the eigenvalue equation will be expanded to linear order.
This will require keeping the terms to cubic
order in this expansion for $f_{\bf n}({\bf r})$.

Expanding the `eigenvalue' to linear order in $\varepsilon$
one obtains
\begin{eqnarray} \label{Eq:Hzeta}
\lefteqn{
\hat{\cal H}({\bf r}) \zeta_{\bf n}({\bf r})
} \nonumber  \\
& = &
U({\bf r}) \zeta_{\bf n}({\bf r})
- \frac{\hbar^2}{2m}
\frac{\tilde C_{\bf n}}{C_{\bf n}}
\left[
 \tilde \zeta_{\bf n}({\bf r}) \nabla_{\bf r}^2 f_{\bf n}({\bf r})
\right. \nonumber \\ && \left. \mbox{ }
+ 2
\nabla_{\bf r} f_{\bf n}({\bf r}) \cdot
\nabla_{\bf r}\tilde \zeta_{\bf n}({\bf r})
+
f_{\bf n}({\bf r}) \nabla_{\bf r}^2\tilde \zeta_{\bf n}({\bf r})
\right]
\nonumber \\ & = &
\left[ U({\bf q}_{{\bf n}} )
+ {\bf U}_{{\bf n}}' \cdot {\bm \varepsilon}_{{\bf n}}({\bf r})
+ {\cal O}({\varepsilon}^2) \right]
\zeta_{\bf n}({\bf r})
\nonumber \\ &  & \mbox{ }
- \frac{\hbar^2}{2m}
\left[
\frac{ \mbox{TR }{\bf f}^{\bf nn}
+ \mbox{TR }{\bf f}^{\bf nnn} \cdot {\bm \varepsilon}_{{\bf n}}({\bf r})
}{ f_{\bf n}({\bf r}) }
\right. \nonumber \\ &  & \left. \mbox{ }
- \frac{1}{\hbar^2}{\bf p}_{{\bf n}} \cdot {\bf p}_{{\bf n}}
-\frac{3N}{2\xi^2}
+ \frac{1}{i\hbar \xi^2}
{\bf p}_{{\bf n}} \cdot {\bm \varepsilon}_{{\bf n}}({\bf r})
\right. \nonumber \\ &  & \left. \mbox{ }
- \frac{2}{f_{\bf n}({\bf r})}
\left\{
\frac{{\bm \varepsilon}_{{\bf n}}({\bf r})}{2\xi^2}
+ \frac{{\bf p}_{{\bf n}}}{i\hbar}
\right\}
\right. \nonumber \\ && \left. \mbox{ }
 \! \cdot \!
\left\{ {\bf f}^{\bf n}
+
\frac{1}{2} {\bf f}^{\bf nn} \cdot {\bm \varepsilon}_{{\bf n}}({\bf r})
+
\frac{1}{2}  {\bm \varepsilon}_{{\bf n}}({\bf r})\cdot{\bf f}^{\bf nn}
\right\}
\right]
\zeta_{\bf n}({\bf r})
%%%%%%%%%%%%%%%%%%%%%%%%%%%%%%%%%%%
\nonumber \\ & = &
\left[ U({\bf q}_{{\bf n}} )
+ \frac{1}{2m}   {\bf p}_{{\bf n}} \cdot   {\bf p}_{{\bf n}}
+ \frac{3N\hbar^2}{4m\xi^2}
\right. \nonumber \\ &  & \left. \mbox{ }
- \frac{\hbar^2}{2m} \mbox{TR }{\bf f}^{\bf nn}
- \frac{i \hbar}{m} {\bf p}_{{\bf n}} \cdot {\bf f}^{\bf n}
\right] \zeta_{\bf n}({\bf r})
\nonumber \\ &  & \mbox{ }
+ \left\{ {\bf U}_{{\bf n}}'
+ \frac{\hbar^2}{2m} \mbox{TR}({\bf f}^{\bf nn}) {\bf f}^{\bf n}
- \frac{\hbar^2}{2m} \mbox{TR}({\bf f}^{\bf nnn})
\right. \nonumber \\ &  & \left. \mbox{ }
+ \frac{i\hbar}{2m\xi^2} {\bf p}_{{\bf n}}
- \frac{i\hbar}{m}  {\bf p}_{{\bf n}}
\cdot {\bf f}^{\bf nn}
+ \frac{i\hbar}{m}
{\bf p}_{{\bf n}} \cdot {\bf f}^{\bf n}
{\bf f}^{\bf n}
\right. \nonumber \\ &  & \left. \mbox{ }
+ \frac{\hbar^2}{2m\xi^2}{\bf f}^{\bf n}
\right\}
\cdot {\bm \varepsilon}_{{\bf n}}({\bf r}) \;
%\nonumber \\ &  &  \mbox{ } \times
\zeta_{\bf n}({\bf r})
%\nonumber \\ &  & \mbox{ }
+ {\cal O}({\bm \varepsilon}^2) .
\end{eqnarray}

%%%%%%%%%%%%%%%%%%%%%%%%%%%%%%%%%%%%%%%%%%%%%%%%%%%%%%%%%%%%%%%%%%%%%%%%%%%
%\subsection{The Entropy Eigenvalue}

The eigenvalue is the coefficient of $\varepsilon^0$,
which is the term within the first bracket,
\begin{eqnarray}
{\cal H}_{\bf n} & = &
U({\bf q}_{{\bf n}} )
+ \frac{1}{2m}   {\bf p}_{{\bf n}} \cdot   {\bf p}_{{\bf n}}
+ \frac{3N\hbar^2}{4m\xi^2}
\nonumber \\ &  &  \mbox{ }
- \frac{\hbar^2}{2m} \mbox{TR }{\bf f}^{\bf nn}
- \frac{i \hbar}{m} {\bf p}_{{\bf n}} \cdot {\bf f}^{\bf n} .
\end{eqnarray}
As a scalar equation,
this is obviously insufficient to determine
the vector ${\bf f}^{\bf n} $ and the matrix ${\bf f}^{\bf nn} $ coefficients.
Nevertheless, guided by the structure of the equation,
one can make a guess at a reasonable ansatz.

Since the Hamiltonian operator is an Hermitian operator,
the eigenvalue has to be real.
By inspection,
it is clear that the imaginary part of this and the other extraneous terms
can be canceled by invoking the ansatz
\begin{eqnarray}  \label{Eq:fnn}
{\bf f}^{\bf nn} & = &
\frac{1}{i\hbar} \left[
{\bf p}_{\bf n} {\bf f}^{\bf n} +  {\bf f}^{\bf n} {\bf p}_{\bf n}
\right] .
\end{eqnarray}
Note that dyadic products appear on the right hand side,
and that these are symmetrized.
%(Actually there is reason to doubt that such a dyadic ansatz
%can accommodate the higher order eigenvalue conditions.)
One could add to this a term ${\bf I}/2\xi^2$
to cancel the constant ${3N\hbar^2}/{4m\xi^2}$
in the eigenvalue, but this is unnecessary
and it creates problems further on in the analysis.
In component form this ansatz is
\begin{equation}
{\bf f}^{{\bf nn}}_{j \alpha ; k\beta }
=
\frac{1}{i\hbar}
\left\{
{p}^\alpha_{n_{pj\alpha}}
{f}^{{\bf n}}_{k \beta}
+
 {f}^{{\bf n}}_{j \alpha} {p}^\beta_{n_{pk\beta}}
\right\} .
\end{equation}

This ansatz inserted into
the zeroth order term gives the eigenvalue
\begin{eqnarray}
{\cal H}_{\bf n}
& = &
U({\bf q}_{{\bf n}} )
+ \frac{1}{2m}   {\bf p}_{{\bf n}} \cdot   {\bf p}_{{\bf n}}
+ \frac{3N\hbar^2}{4m\xi^2}
\nonumber \\ & \equiv &
{\cal H}({\bf q}_{{\bf n}},{\bf p}_{{\bf n}})
+ \frac{3N\hbar^2}{4m\xi^2} .
\end{eqnarray}
The second equality is the classical Hamiltonian
function of the positions and the momenta,
plus an immaterial constant.
Because entropy eigenfunctions are the same as energy eigenfunctions
in the present equilibrium case,
the entropy eigenvalue is just $S_{\mathrm{r},{\bf n}} = - {\cal H}_{\bf n}/T$.

The fact that the parameters of the entropy state ${\bf n}$,
the nominal positions ${\bf q}_{{\bf n}}$
and nominal momenta ${\bf p}_{{\bf n}}$,
appear in the classical Hamiltonian function
as the classical positions and momenta of the particles
justifies interpreting them as the positions and the momenta of the particle
 in the entropy state ${\bf n}$.
Conversely, and speaking strictly,  the expectation values,
in the entropy state ${\bf n}$,
$\langle \zeta_{\bf n}| \hat {\bf q} | \zeta_{\bf n} \rangle
\equiv {\bf q}_{\bf nn}^\mathrm{S}
\ne {\bf q}_{\bf n}$,
and
$\langle \zeta_{\bf n}| \hat {\bf p} | \zeta_{\bf n} \rangle
\equiv {\bf p}_{\bf nn}^\mathrm{S}
\ne {\bf p}_{\bf n}$,
cannot be interpreted as the positions and the momenta
of the particles in the entropy state ${\bf n}$.
Of course to within an error of the order of
$\xi$ in position space and $\hbar/2\xi$ in momentum space,
per particle per direction,
the expectation values do equal the nominal values
and one need not distinguish them.

%%%%%%%%%%%%%%%%%%%%%%%%%%%%%%%%%%%%%%%%%%%%%%%%%%%%%%%%%%%%%%%%%%%%%%%%%%%
%\subsection{Second Order Eigenfunction} \label{Sec:2Oeigen-sum}

In the eigenvalue equation the  term
linear in ${\bm \varepsilon}_{{\bf n}}({\bf r})$
depends upon the first, second and third order terms in the expansion
for the entropy eigenfunction.
For this linear term to vanish,
which it must if the modified wave packet is to be a true eigenfunction,
the coefficient in braces must be zero,
\begin{eqnarray} \label{Eq:0-times-eps^1}
{\bf 0} & = &
{\bf U}'_{\bf n}
+ \frac{ i \hbar}{2m\xi^2} {\bf p}_{\bf n}
+ \frac{ \hbar^2}{2m}
\mbox{TR}\{ {\bf f}^{\bf n  n} \}
{\bf f}^{\bf n}
\nonumber \\ && \mbox{ }
+ \frac{ i \hbar}{m} {\bf p}_{\bf n} \cdot {\bf f}^{\bf n} {\bf f}^{\bf n}
- \frac{ i\hbar}{m} {\bf p}_{\bf n} \cdot {\bf f}^{\bf nn}
%\nonumber \\ && \mbox{ }
+ \frac{\hbar^2}{2m\xi^2} {\bf f}^{\bf n}
\nonumber \\ && \mbox{ }
- \frac{ \hbar^2}{2m}
\mbox{TR}^{(1)}  \{ {\bf f}^{\bf n  n  n} \}
\nonumber \\ & = &
{\bf U}'_{\bf n}
+ \frac{ i \hbar}{2m\xi^2} {\bf p}_{\bf n}
- \frac{ i\hbar}{m} {\bf p}_{\bf n} \cdot {\bf f}^{\bf n}
{\bf f}^{\bf n}
\nonumber \\ && \mbox{ }
+ \frac{ i \hbar}{m} {\bf p}_{\bf n} \cdot {\bf f}^{\bf n} {\bf f}^{\bf n}
%\nonumber \\ && \mbox{ }
- \frac{ 1}{m} {\bf p}_{\bf n} \cdot {\bf p}_{\bf n} {\bf f}^{\bf n}
- \frac{ 1}{m} {\bf p}_{\bf n} \cdot {\bf f}^{\bf n}  {\bf p}_{\bf n}
\nonumber \\ && \mbox{ }
+ \frac{\hbar^2}{2m\xi^2} {\bf f}^{\bf n}
%\nonumber \\ && \mbox{ }
- \frac{ \hbar^2}{2m}
\mbox{TR}^{(1)}  \{ {\bf f}^{\bf n  n  n} \}
\nonumber \\ & = &
{\bf U}'_{\bf n}
%\nonumber \\ && \mbox{ }
- \frac{ 1}{m} {\bf p}_{\bf n} \cdot {\bf p}_{\bf n} {\bf f}^{\bf n}
- \frac{ 1}{m} {\bf p}_{\bf n} \cdot {\bf f}^{\bf n}  {\bf p}_{\bf n}
\nonumber \\ && \mbox{ }
%\nonumber \\ && \mbox{ }
- \frac{ \hbar^2}{2m}
\mbox{TR}^{(1)}  \{ {\bf f}^{\bf n  n  n} \} .
\end{eqnarray}
The ansatz for ${\bf f}^{\bf nn}$ has been inserted into the second equality.
In the final equality,
the term ${\hbar^2} {\bf f}^{\bf n} /{2m\xi^2}$,
which is ${\cal O}(N^{-1})$ has been neglected compared to
the rest, which are ${\cal O}(N^{0})$,
 in the thermodynamic limit.

In view of the fact that ${\bf f}^{\bf nn}$ was linear on ${\bf f}^{\bf n}$,
a similar ansatz can be invoked for ${\bf f}^{\bf nnn}$,
\begin{equation}
{\bf f}^{\bf nnn}
=
\frac{-1}{\hbar^2} \left[ {\bf p}_{\bf n} {\bf p}_{\bf n}  {\bf f}^{\bf n}
+ {\bf p}_{\bf n} {\bf f}^{\bf n} {\bf p}_{\bf n}
+ {\bf f}^{\bf n} {\bf p}_{\bf n}{\bf p}_{\bf n} \right] .
\end{equation}
In both cases
the coefficient is linear in ${\bf f}^{\bf n}$
and a factor of $1/i\hbar$
is associated with each ${\bf p}_{\bf n}$ in the ansatz.
There is some reason to doubt that such a triadic ansatz is
more generally viable
based on the analysis of
the second order eigenvalue equation in Appendix \ref{Sec:2Oeigen}.

With this ansatz the coefficient of the linear term in the
the eigenvalue equation becomes
\begin{equation}
{\bf 0}
=
{\bf U}'_{\bf n}
%\nonumber \\ && \mbox{ }
- \frac{ 1}{2m} {\bf p}_{\bf n} \cdot {\bf p}_{\bf n} {\bf f}^{\bf n}   .
\end{equation}
This has immediate solution
\begin{equation}
{\bf f}^{\bf n} =
\frac{1}{K_{\bf n}} {\bf U}'_{\bf n}.
\end{equation}
Here $ K_{\bf n} = {\bf p}_{\bf n} \cdot {\bf p}_{\bf n} /2m$
is the kinetic energy of the entropy microstate
and $\{ {\bf U}'_{\bf n} \}_{j\alpha} =  \nabla_{j\alpha}  U({\bf q}_{\bf n})$
is the negative force vector of the entropy microstate.
Clearly ${\bf f}^{\bf n} ={\cal O}(N^{-1})$.

With this result,
the entropy eigenfunction has been determined to third order
in ${\bm \varepsilon}$
and the entropy eigenvalue has been determined to first order in
${\bm \varepsilon}$.
The explicit result is dependent on the assumed ansatz
for ${\bf f}^{\bf nn}$ and for ${\bf f}^{\bf nnn}$.

Choosing instead ${\bf f}^{\bf nnn}= 0$ gives
\begin{equation} %\label{Eq:fn=}
{\bf f}^{{\bf n}}
=
\frac{1}{ 2K_{{\bf n}}  }
\left[ {\bf U}_{{\bf n}}' - \frac{\phi_{\bf n}}{m} {\bf p}_{{\bf n}} \right] ,
\end{equation}
where
\begin{equation} %\label{Eq:phi_n}
\phi_{\bf n} \equiv {\bf f}^{{\bf n}} \cdot {\bf p}_{\bf n}
=
\frac{  {\bf U}_{{\bf n}}'  \cdot {\bf p}_{\bf n} }{ 4 K_{\bf n} }  .
\end{equation}
%Again, going to the second order in the eigenvalue
%as in Appendix \ref{Sec:2Oeigen} appears to rule out this ansatz.

For such an undetermined set of equations,
there are many forms of solutions,
including the present ones and the one explored in Appendix \ref{Sec:2Oeigen}.
These will no doubt differ in their internal consistency,
and their reliability and accuracy.
The main point is
that it is possible to generate the physical eigenvalue
and to develop a genuine entropy eigenfunction
based upon the above expansion.

The two ansatz explored in this section both have ${\bf f}^{\bf n}$,
${\bf f}^{\bf nn}$, and ${\bf f}^{\bf nnn}$
being ${\cal O}(N^{-1})$.
A similar conclusion is reached in Appendix \ref{Sec:2Oeigen}
where the eigenvalue equation is extended to second order.
In the thermodynamic limit these can be neglected
compared to the leading term of unity,
\begin{equation}
f({\bf r}) =
1 + {\cal O}(N^{-1}) .
\end{equation}
From this it follows that in the thermodynamic limit,
 the entropy eigenfunction is the reference wave packet itself,
\begin{equation}
\zeta_{\bf n}({\bf r})
=
%\frac{\tilde C_{\bf n}}{C_{\bf n}}
\tilde \zeta_{\bf n}({\bf r}) .
\end{equation}

For the first ansatz above, ${\bf f}^{\bf n} \propto {\bf U}'_{\bf n}$,
and all the higher order coefficients were linearly proportional to this.
Hence in the case of an ideal gas, ${\bf U}_{\bf n} = 0$,
all the coefficients beyond the zeroth rigorously vanish,
and  $f_{\bf n}^\mathrm{id}({\bf r})=1$.
In this case $\zeta_{\bf n}({\bf r}) = \tilde \zeta_{\bf n}({\bf r}) $
irrespective of the thermodynamic limit.
Hence the minimum uncertainty wave packet itself
is the entropy eigenfunction of the ideal gas.
(It will turn out that this is true if $\xi \gg \Lambda$;
see \S\S \ref{Sec:olxi} and \ref{Sec:IdealGas}  for details.)

%%%%%%%%%%%%%%%%%%%%%%%%%%%%%%%%%%%%%%%%%%%%%%%%%%%%%%%%%%%%%%%%%%%%%%%%%
%
\section{Quantum Ideal  Gas} \label{Sec:IdealGas}
\setcounter{equation}{0} \setcounter{subsubsection}{0}
%\renewcommand{\theequation}{\Alph{section}.\arabic{equation}}
%
%%%%%%%%%%%%%%%%%%%%%%%%%%%%%%%%%%%%%%%%%%%%%%%%%%%%%%%%%%%%%%%%%%%%%%%%%%

This section derives the fugacity expansion for the quantum ideal gas.
The purpose is to illustrate
and to verify the present permutation loop expansion.
Two derivations are given.
The first is based upon wave packets of width $\xi$,
and it confirms explicitly that in the limit $\xi \rightarrow \infty$
the correct result is obtained,
at least for the first three coefficients.
That derivation is rather lengthy and a little messy.
The second derivation, in \S \ref{Sec:Ideal-Plane-Fug},
is based upon plane waves.
It is rather shorter and yields all the fugacity coefficients explicitly.

%%%%%%%%%%%%%%%%%%%%%%%%%%%%%%%%%%%%%%%%%%%%%%%%%%%%%%%%%%%%%%%%%%%%%%%%%%
\subsection{Wave Packet Derivation}
\label{Sec:Ideal-Packet-Fug}

The terms in the expansion for the grand partition function
and the grand potential, \S \ref{Sec:Zcl+Z1},
involve the grand canonical average of the loop overlap factor.
This is
\begin{eqnarray}
%\lefteqn{
-\beta \Omega_l
%} \nonumber \\
& = &
\left< \frac{(l-1)!(N)!}{l!(N-l)!} \chi^{(l)}({\bm \Gamma}^l)
\right>_{\mu}
\nonumber \\ & = &
\frac{1}{\Xi_\mu}
\frac{(l-1)!}{l!}
\sum_{N=l}^\infty
\frac{z^N }{(N-l)!h^{3N} }
\nonumber \\ && \mbox{ } \times
\int \mathrm{d}{\bm \Gamma}^{N} \;
e^{-\beta {\cal H}({\bm \Gamma}^{N})}
\chi^{(l)}({\bm \Gamma}^l) .
\end{eqnarray}
Here  $\Xi_\mathrm{\mu} = \Xi_1$ is the same classical partition function
as in \S \ref{Sec:Zcl+Z1}.
The entropy eigenstate denoted by ${\bf n}$ in \S \ref{Sec:Zcl+Z1}
is here denoted by a point in classical phase space,
${\bm \Gamma}^N$.
The $l$-mer overlap factor, $ \chi^{(l)}_{\bf n}
\Rightarrow \chi^{(l)}({\bm \Gamma}^{N})
\Rightarrow \chi^{(l)}({\bm \Gamma}^{l})$
does not depend on the monomers in the present case of an ideal gas.

For the ideal gas,
the integrals involving the monomer ${\bm \Gamma}^{N-l}$
are independent of each other and of the ${\bm \Gamma}^l$.
Hence they can be performed to give $\Xi_\mathrm{\mu,id}$,
which cancels with the same term in the denominator leaving
\begin{eqnarray}
%\lefteqn{
-\beta \Omega_l^\mathrm{id}
%} \nonumber \\
& = &
\frac{z^l (l-1)!}{h^{3l} l!}
%\nonumber \\ && \mbox{ } \times
\int \mathrm{d}{\bm \Gamma}^{l} \;
e^{-\beta {\cal H}({\bm \Gamma}^{l})}
\chi^{(l)}({\bm \Gamma}^l) .
\end{eqnarray}

If for the loop with $l$ nodes
one defines $j' \equiv j+1 \mbox{ mod }l$, $j=1,2,\ldots , l$,
and $l' =1$,
the overlap factor is
\begin{eqnarray}
\chi_{12\ldots l}^{(l)}
& \equiv &
(\pm 1)^{l-1}
\langle  \zeta_{1'2'\ldots l'} | \zeta_{12\ldots l}\rangle
\nonumber \\ & = &
(\pm 1)^{l-1} \prod_{j=1}^l (2\pi\xi^2)^{-3/2}
\int \mathrm{d}{\bf r}_j \;
\nonumber \\ && \mbox{ } \times
e^{ -\left({\bf r}_j-{\bf q}_{j} \right)^2/4 \xi^2 }
%\right. \nonumber \\ && \mbox{ } \left.
e^{ -{\bf p}_{j} \cdot ({\bf r}_j-{\bf q}_j) /i\hbar  }
\nonumber \\ && \mbox{ }  \times
e^{ -\left({\bf r}_j-{\bf q}_{j'} \right)^2/4 \xi^2 }
%\right. \nonumber \\ && \mbox{ } \left.
e^{ {\bf p}_{j'} \cdot ({\bf r}_j-{\bf q}_{j'}) /i\hbar }
\nonumber \\ & = &
(\pm 1)^{l-1} \prod_{j=1}^l (2\pi\xi^2)^{-3/2}
\int \mathrm{d}{\bf r}_j \;
\exp \left\{ \rule{0cm}{0.5cm}
\right. \nonumber \\ && \left.
\frac{-1}{2\xi^2} \left( {\bf r}_j
- \frac{1}{2} \left[ {\bf q}_{j} + {\bf q}_{j'} \right]
+ \frac{\xi^2}{i\hbar} \left[ {\bf p}_{j} - {\bf p}_{j'} \right]
\right)^2
\right. \nonumber \\ && \left.
- \frac{1}{4\xi^2} \left[ {\bf q}_{j}^2 + {\bf q}_{j'}^2 \right]
+ \frac{1}{i\hbar} \left[ {\bf p}_{j}\cdot {\bf q}_{j}
- {\bf p}_{j'}\cdot{\bf q}_{j'} \right]
\right. \nonumber \\ && \left.
+ \frac{1}{8\xi^2} \left[ {\bf q}_{j} + {\bf q}_{j'} \right]^2
- \frac{\xi^2}{2\hbar^2} \left[ {\bf p}_{j} - {\bf p}_{j'} \right]^2
\right. \nonumber \\ && \left.
-  \frac{1}{2i\hbar} \left[ {\bf q}_{j} + {\bf q}_{j'} \right]
\cdot \left[ {\bf p}_{j} - {\bf p}_{j'} \right]
\right\}
\nonumber \\ & = &
(\pm 1)^{l-1} \exp \sum_{j=1}^l  \left\{ %\rule{0cm}{0.5cm}
\frac{-1}{8\xi^2} \left[ {\bf q}_{j} - {\bf q}_{j'} \right]^2
\right. \nonumber \\ && \left.
- \frac{\xi^2}{2\hbar^2} \left[ {\bf p}_{j} - {\bf p}_{j'} \right]^2
%\right. \nonumber \\ && \left.
+  \frac{1}{2i\hbar} \left[ {\bf q}_{j} \cdot {\bf p}_{j'}
- {\bf q}_{j'} \cdot {\bf p}_{j} \right]
\right\} .
\nonumber \\
\end{eqnarray}
The final equality follows because  many of the sums telescope.
For the case of a dimer loop, the final term vanishes,
but not more generally.

The Hamiltonian for $N \ge l$ ideal particles is
\begin{equation}
{\cal H}^\mathrm{id}({\bm \Gamma}^N)
=
\frac{1}{2m} \sum_{j=1}^N {\bf p}_j \cdot {\bf p}_j.
\end{equation}

For the  ideal gas,
because the monomers do not interact with the loop particles,
their integrals factorize out and cancel with the partition function.
Hence in the following explicit results for the momentum integrals
one can take $N=l$.

Adding the exponent of the Boltzmann factor
%with $U({\bf q}^N) = 0$
to the exponent of the Gaussian overlap factor,
the exponent for the average overlap factor is
\begin{eqnarray}
\mbox{exp}
& = &
\frac{-\beta}{2m} \sum_{j=1}^N {\bf p}_j \cdot {\bf p}_j
%\nonumber \\ &&
+ \sum_{j=1}^l  \left\{ %\rule{0cm}{0.5cm}
\frac{-\left[ {\bf q}_{j} - {\bf q}_{j'} \right]^2}{8\xi^2}
\right. \nonumber \\ && \left.
- \frac{\xi^2}{2\hbar^2} \left[ {\bf p}_{j} - {\bf p}_{j'} \right]^2
%\right. \nonumber \\ && \left.
+  \frac{1}{2i\hbar} \left[ {\bf q}_{j} \cdot {\bf p}_{j'}
- {\bf q}_{j'} \cdot {\bf p}_{j} \right]
\right\}
\nonumber \\ & = &
- \sum_{j=l+1}^N {\bf p}_j \cdot {\bf p}_j
\nonumber \\ && \mbox{ }
+ \sum_{\alpha=x,y,z}
\left\{ \frac{-1}{2} \underline{\underline A} : \underline P \, \underline P
+ \underline B \cdot \underline P
+ C \right\}_\alpha .
\end{eqnarray}
The braces deal with a single component of direction at a time.
Since these all give the same answer it is not necessary to specify which one.
The determinant of the full matrix is the cube of the
determinant of the one-component matrix.

The elements of the momentum vector are
$\{\underline P\}_j = {p}_j$, $j=1,2,\ldots , l$.
The elements of the almost tri-diagonal matrix
$\underline{\underline A} $ are
\begin{equation}
A_{jk} =
\left\{ \begin{array}{ll}
\displaystyle
\frac{\beta}{m} + \frac{2 \xi^2}{\hbar^2} , & j= k ,
\vspace{0.2cm} \\
\displaystyle
\frac{-\xi^2}{\hbar^2}(1+\delta_{l2}) ,
& j = k \pm 1 \mbox{ mod }l, \vspace{0.2cm} \\
0 , & \mbox{otherwise} .
\end{array}
\right.
\end{equation}
Note  that this holds for $l \ge 2$.
Note that $A_{1l} = A_{l1} = {-\xi^2}/{\hbar^2}$, $l \ge 3$.

In view of this define
\begin{equation}
a \equiv \frac{\beta}{m} + \frac{2 \xi^2}{\hbar^2}
, \mbox{ and }
b \equiv \frac{-\xi^2}{\hbar^2}.
\end{equation}
These simplify the analysis below.

The elements of the vector  $\underline B$ for $j=1,2, \ldots, l$
are
\begin{equation}
B_j = \frac{1}{2 i \hbar} \left[ {q}_{j-1} - {q}_{j+1} \right].
\end{equation}
The indeces are counted mod $l$, such that $0$ is the same as $l$
and $l+1$ is the same as $1$.

The configuration contribution is homogeneous,
which is to say that only the difference in positions of the particles
of the loop matter.
(This is also true if the potential is non-zero
provided that no external potential acts on the system.)
One can therefore define $q_j' = q_j - q_l$, $j=1,2,\ldots, l-1$.
The integration over  ${\bf q}_l$ just gives a factor of $V$
(if $U({\bf q}^N) = 0$).
With this
\begin{eqnarray}
B_j
& = &
\frac{1}{2 i \hbar} \left[ {q}_{j-1}' - {q}_{j+1}' \right]
\nonumber \\ & = &
\frac{1}{2 i \hbar}
\left\{
\begin{array}{ll}
- {q}_{2}' , & j = 1 \vspace{.2cm} \\
\left[ {q}_{j-1}' - {q}_{j+1}' \right]
, & 2 \le j \le l-2 \vspace{.2cm}  \\
 {q}_{l-2}' , & j = l-1 \vspace{.2cm}  \\
 \left[ {q}_{l-1}' - {q}_{1}' \right]
, & j = l .
\end{array} \right.
\end{eqnarray}

This vector may be written
\begin{equation}
\underline B = \underline{\underline D} \, \underline Q' ,
\end{equation}
where $\underline Q' = \{ q_1',q_2', \ldots , q_{l-1}'\}$
is an $(l-1)$-component vector
and $\underline{\underline D}$ is an $l \times (l-1)$  matrix
with entries
\begin{equation}
D_{jk}
=
\frac{1}{2i\hbar}
\left\{ \begin{array}{ll}
-1 , & k = j+ 1  \mbox{ mod }l,  \\
1 , & k = j- 1 ,  \\
0 , & \mbox{otherwise}.
\end{array}
\right. \nonumber
\end{equation}
Obviously $\underline{\underline D}^{\mathrm T}
= - \underline{\underline D}$.

The constant term is
\begin{eqnarray}
C & = &
\frac{-1}{8 \xi^2} \sum_{j=1}^l [ { q}_j - { q}_{j'} ]^2
\nonumber \\ & = &
\frac{-1}{8 \xi^2}
\left\{ ({ q}'_{l-1})^2 + ({ q}'_1)^2 +
\sum_{j=1}^{l-2} [ { q}'_j - { q}'_{j'} ]^2 \right\}
\nonumber \\ & = &
 \underline{\underline E} : \underline Q' \, \underline Q' ,
\end{eqnarray}
with
\begin{equation}
E_{jk}
=
\frac{-1}{4\xi^2}
\left\{ \begin{array}{ll}
1  , & k = j  ,  \\
\displaystyle
\frac{-1}{2} , & k = j \pm 1 ,  \\
0 , & \mbox{otherwise}.
\end{array}
\right.
\end{equation}
This is a tri-diagonal matrix.
(For $l=2$ this is the scalar
$\underline{\underline E}^{(2)} = {-2}/{4\xi^2}$.)

Now set the potential energy to zero,
$U({\bf q}^{n+l}) = 0$, and set $N=l$.
Also suppress the sum over the three components of direction.

Completing the squares gives
\begin{eqnarray}
\mbox{exp}
& = &
\frac{-1}{2} \underline{\underline A} :
\left[ \underline P - \underline{\underline A}^{-1} \underline B \right]
\,
\left[ \underline P - \underline{\underline A}^{-1} \underline B \right]
\nonumber \\ &  & \mbox{ }
+ C
+ \frac{1}{2} \underline B \cdot \underline{\underline A}^{-1} \underline B
\nonumber \\ & = &
\frac{-1}{2} \underline{\underline A} :
\left[ \underline P - \underline{\underline A}^{-1} \underline B \right]
\,
\left[ \underline P - \underline{\underline A}^{-1} \underline B \right]
\nonumber \\ &  & \mbox{ }
%- \beta U({\bf q}^l)
- \frac{1}{2} \underline{\underline F} : \underline Q' \, \underline Q' ,
\end{eqnarray}
where the $(l-1)\times (l-1)$ matrix,
$ \underline{\underline F}^{(l)} \equiv
-  2\underline{\underline E}
- \underline{\underline D}^\mathrm{T} \,\underline{\underline A}^{-1}
\underline{\underline D}  $,
is obviously real and positive definite.

It follows that the terms in the fugacity expansion of the grand potential
for the ideal gas are
\begin{eqnarray}
\lefteqn{
-\beta \Omega_l^\mathrm{id}
} \nonumber \\
& = &
\frac{z^l (l-1)!}{h^{3l} l!}
%\nonumber \\ && \mbox{ } \times
\int \mathrm{d}{\bm \Gamma}^{l} \;
e^{-\beta {\cal H}({\bm \Gamma}^{l})}
\chi^{(l)}({\bm \Gamma}^l)
\nonumber \\  & = &
(\pm 1)^{l-1} \frac{z^l (l-1)!}{h^{3l} l!}
\int \mathrm{d}{\bf p}^l \, \mathrm{d}{\bf q}^l \;
\nonumber \\ && \mbox{ }
\exp \sum_\alpha \left\{
\frac{-1}{2} \underline{\underline A}^{(l)} :
\left[ \underline P - \underline{\underline A}^{-1} \underline B \right]
\,
\left[ \underline P - \underline{\underline A}^{-1} \underline B \right]
\right. \nonumber \\ &  & \left. \mbox{ }
- \frac{1}{2} \underline{\underline F}^{(l)} : \underline Q' \, \underline Q'
\right\}_\alpha
\nonumber \\ & = &
 \frac{(\pm 1)^{l-1} z^l }{ l h^{3l}}
\frac{ (2\pi)^{3(2l-1)/2}
}{
 | \underline{\underline A}^{(l)} |^{3/2}
\; | \underline{\underline F}^{(l)} |^{3/2} } V .
\end{eqnarray}
Recall that $\underline{\underline A}$  is an $l\times l$ matrix
and that
$\underline{\underline F}^{(l)}$ is an $(l-1)\times(l-1)$ matrix,
which explains the factor of $(2\pi)^{3(2l-1)/2}$.
The factor of volume $V$ comes from the integral over ${\bf q}_l$,
which is decoupled from the $\underline Q'$.

It is straightforward to show that
\begin{equation}
\left| \underline{\underline A}^{(l)} \right|
=
a \left| \underline{\underline A}^{(l-1)}_\triangle \right|
- 2 b^2 \left| \underline{\underline A}^{(l-2)}_\triangle \right|
- 2 (-1)^{l} b^l .
\end{equation}
This holds for $l \ge 3$.
Here $\underline{\underline A}^{(l)}_\triangle$
is the strictly tri-diagonal entries in
$\underline{\underline A}^{(l)}$.
Its determinant obeys the recursion relation
\begin{equation}
\left| \underline{\underline A}_\triangle^{(l)} \right|
=
a \left| \underline{\underline A}^{(l-1)}_\triangle \right|
-  b^2 \left| \underline{\underline A}^{(l-2)}_\triangle \right| .
\end{equation}

One has for the first several results,
\begin{eqnarray}
| \underline{\underline A}^{(2)}_\triangle | & = & a^2 - b^2
\nonumber \\
| \underline{\underline A}^{(3)}_\triangle | & = & a^3 - 2 ab^2
\nonumber \\
| \underline{\underline A}^{(4)}_\triangle |  & = & a^4 - 3 a^2 b^2 + b^4 ,
%\nonumber \\
%| \underline{\underline A}^{(5)}_\triangle |  & = & a^5 - 4 a^3b^2 + 3 ab^4
%\nonumber \\
%| \underline{\underline A}^{(6)}_\triangle |
%& = & a^6 - 5 a^4b^2 + 6 a^2b^4 - b^6
%\nonumber \\
%| \underline{\underline A}^{(7)}_\triangle |
%& = & a^7 - 6 a^5b^2 + 10 a^3b^4 - 4ab^6.
\end{eqnarray}
and
\begin{eqnarray}
| \underline{\underline A}^{(2)} | & = & a^2 - 4 b^2
\nonumber \\
| \underline{\underline A}^{(3)} | & = & a^3 - 3 ab^2 + 2 b^3
\nonumber \\
| \underline{\underline A}^{(4)} |  & = & a^4 - 4 a^2 b^2 .
\end{eqnarray}

%%%%%%%%%%%%%%%%%%%%%%%%%%%%%%%%
\subsubsection{Large $\xi$ limit}

Here and below $a \equiv ({\beta}/{m}) + ({2 \xi^2}/{\hbar^2})$
and $b \equiv {-\xi^2}/{\hbar^2}$.
Hence take $a = -2 b + x$, with $| b | \gg x \equiv {\beta}/{m}$.
Retaining terms to linear order in $x$ is the large $\xi$ limit.

The general formula for
$| \underline{\underline A}^{(l)} | $
in the large $\xi$ limit is
\begin{equation}
| \underline{\underline A}^{(l)} |
=
% - 4 b x + {\cal O}(x^2) & l = 2, \\
l^2  (-b)^{l-1} x + {\cal O}(x^2) ,\; l \ge 2 .
\end{equation}

For use shortly,
recall that the thermal wave length is
$\Lambda \equiv \sqrt{2\pi \hbar^2/m k_\mathrm{B}T}$.

%%%%%%%%%%%%%%%%%%%%%%%%%%%%%%%%%%%%%%%%%%%%%%%%%%%%%%%%
\subsubsection{$l=2$}

Recall that  $\underline{\underline F}
\equiv - 2 \underline{\underline E} - \underline{\underline D}^\mathrm{T}
\, \underline{\underline A}^{-1} \, \underline{\underline D}$
is an $(l-1) \times (l-1)$ matrix.

For $l=2$ one has the scalar
$\underline{\underline E}^{(2)} = {-1}/{4\xi^2}$.
One can readily show that in this case the second term vanishes and hence
\begin{equation}
|\underline{\underline F}^{(2)} | =1/2\xi^{2} .
\end{equation}

Therefore one has for $l=2$
\begin{eqnarray}
%\lefteqn{
-\beta \Omega_2^\mathrm{id}
%} \nonumber \\
& = &
 \frac{\pm  \, z^2 }{ 2 h^{6}}
\frac{ (2\pi)^{9/2} V }{
 | \underline{\underline A}^{(2)} |^{3/2}
\; | \underline{\underline F}^{(2)} |^{3/2} }
\nonumber \\ & = &
 \frac{\pm  \, z^2 }{ 2 \hbar^{6}}
\frac{ (2\pi)^{-3/2} V
}{(a^2-4b^2)^{3/2} 2^{-3/2} \xi^{-3}} .
\end{eqnarray}

In the large $\xi$ limit one has
\begin{equation}
| \underline{\underline A}^{(2)} |
= a^2 - 4 b^2
= - 4 b x + {\cal O}(x^2)
= \frac{4 \beta \xi^2}{m \hbar^2}  .
\end{equation}
In this limit
\begin{eqnarray}
%\lefteqn{
-\beta \Omega_2^\mathrm{id}
%} \nonumber \\
& = &
 \frac{\pm  \, z^2 }{ 2 \hbar^{6}}
(2\pi)^{-3/2} V
\left(  \frac{4 \beta \xi^2}{m \hbar^2} \right)^{-3/2}
2^{3/2} \xi^{3}
\nonumber \\ & = &
\pm  \, z^2 2^{-5/2} \Lambda^{-3} V.
\end{eqnarray}
This is independent of the wave packet width.

%%%%%%%%%%%%%%%%%%%%%%%%%%%%%%%%%%%%%%%%%%%%%%%%%%%%%%%%
\subsubsection{$l=3$}

It is tedious but straightforward to show that
\begin{equation}
\left| \underline{\underline F}^{(3)}\right|
=
3 \frac{(1+c)^2}{4^2 \xi^4} ,
\end{equation}
where
\begin{equation}
c \equiv \frac{\xi^2}{\hbar^2} \frac{a^2 + ab - 2b^2 }{a^3-3ab^2 + 2b^3}.
\end{equation}

In the large $\xi$ limit,
\begin{eqnarray}
c
& = &
\frac{\xi^2}{\hbar^2}
\frac{4b^2 -4bx - 2b^2 + bx - 2b^2 }{-8b^3 + 12b^2x + 6b^3 - 3b^2x + 2b^3}
+{\cal O}(x^2)
\nonumber \\ & = &
\frac{\xi^2}{\hbar^2}
\frac{ -3bx  }{ 9b^2x }
\nonumber \\ & = &
 \frac{1}{3} .
\end{eqnarray}
Hence in this limit
\begin{equation}
\left| \underline{\underline F}^{(3)}\right|
=
\frac{3 }{4^2 \xi^4} \frac{4^2}{3^2}
= \frac{1 }{3 \xi^4} .
\end{equation}

Of course one also has
\begin{equation}
\left| \underline{\underline A}^{(3)}\right|
=
a^3 - 3 ab^2 + 2 b^3
=
 \frac{9 \xi^4}{\hbar^4} \frac{\beta}{m}
+{\cal O}(x^2) .
\end{equation}
Hence one has
\begin{eqnarray}
%\lefteqn{
-\beta \Omega_3^\mathrm{id}
%}  \nonumber \\
& = &
 \frac{(\pm 1)^{3-1} z^3 }{ 3 h^{9}}
\frac{ (2\pi)^{15/2}
}{
 | \underline{\underline A}^{(3)} |^{3/2}
\; | \underline{\underline F}^{(3)} |^{3/2} } V
\nonumber \\ & = &
\frac{z^3(2\pi)^{3/2}}{3 \hbar^9} V
\left( \frac{\hbar^4}{9 \xi^4} \frac{m}{\beta} \right)^{3/2}
\left( {3 \xi^4} \right)^{3/2}
\nonumber \\ & = &
\frac{z^3}{3^{5/2}} \Lambda^{-3} V .
\end{eqnarray}

%%%%%%%%%%%%%%%%%%%%%%%%%%%%%%%%%%%%%%%%%%%%%%%%%%%%%%%%
\subsubsection{Fugacity Expansion for the Quantum Ideal Gas}
\label{Sec:Ideal-z-Expn}

From the above, the grand potential for the quantum ideal gas is
\begin{eqnarray}
\lefteqn{
\Omega_\mathrm{id}(\mu,V,T)
} \nonumber \\
& = &
\Omega_\mathrm{cl,id}(\mu,V,T)
+ \sum_{l=2}^\infty \Omega_l^\mathrm{id}(\mu,V,T)
\nonumber \\  & = &
\Omega_\mathrm{cl,id}(\mu,V,T)
\nonumber \\ & & \mbox{ }
- k_\mathrm{B}T  \sum_{l=2}^\infty
 \frac{(\pm 1)^{l-1} z^l }{ l h^{3l}}
\frac{ (2\pi)^{3(2l-1)/2} V
}{
 | \underline{\underline A}^{(l)} |^{3/2}
\; | \underline{\underline F}^{(l)} |^{3/2} }
\end{eqnarray}
The classical ideal gas grand potential is
\begin{equation}
\Omega_\mathrm{cl,id}(\mu,V,T)
= - z k_\mathrm{B} T \Lambda^{-3} V .
\end{equation}
Since the pressure is $p = -\Omega /V$,
this and the explicit results given above for $l=2$ and $l=3$,
give the first three terms in the fugacity expansion
of the dimensionless pressure
for the quantum ideal gas,
\begin{eqnarray}
\beta p_\mathrm{id} \Lambda^3
& \equiv &
\frac{ - \beta \Omega_\mathrm{id}(\mu,V,T)  \Lambda^3 }{V}
\nonumber \\ & =&
z \pm \frac{z^2}{2^{5/2}} + \frac{z^3}{3^{5/2}} \pm \ldots
\end{eqnarray}
This holds in the large $\xi$ limit.

The known fugacity expansion for the quantum ideal gas is\cite{Pathria72}
\begin{equation} \label{Eq:p(z)}
\beta p_\mathrm{id} \Lambda^3
= \sum_{l=1}^\infty b_l z^l
, \;\; b_l = (\pm 1)^{l-1} l^{-5/2} .
\end{equation}
Hence the first three terms obtained explicitly here
for the ideal gas in the large $\xi$ limit
of the zeroth order entropy eigenfunctions
in the present theory
are in agreement with the known results.
It is nought but an exercise to evaluate
all terms in the present formal expression
for the quantum ideal gas.

%%%%%%%%%%%%%%%%%%%%%%%%%%%%%%%%%%%%%%%%%%%%%%%%%%%%%%%%%%%%%%%%%%%%%%%%%%
\subsection{Plane Wave Derivation} \label{Sec:Ideal-Plane-Fug}

Now a much shorter derivation of the fugacity expansion
will be given in which all terms are  readily evaluated.

For ideal particles,
the entropy eigenfunctions are the product of plane waves,
\begin{equation}
\zeta_{\bf  n}({\bf r}) =
\frac{1}{V^{N/2}} \prod_{j=1}^N e^{i {\bf k}_j \cdot {\bf r}_j}
, \;\; {\bf  k}_j = \frac{2\pi}{V^{1/3}} {\bf n}_j .
\end{equation}
This is just a wave packet of infinite width.
The energy eigenvalues are
\begin{equation}
{\cal H}_{\bf  n}  =
\frac{\hbar^2}{2m} \sum_{j=1}^N   k_j^2 .
\end{equation}
The overlap factor for an $l$-loop is
\begin{eqnarray}
\chi^{(l)}_{\bf n} & = &
(\pm 1 )^{l-1} \left\langle
\zeta_{\bf  n}({\bf r}_2,{\bf r}_3,\ldots,{\bf r}_l,{\bf r}_1)
|
\zeta_{\bf  n}({\bf r}_1,{\bf r}_2,\ldots,{\bf r}_l)
\right\rangle
\nonumber \\ & = &
(\pm 1 )^{l-1} \frac{1}{V^N} \int \mathrm{d}{\bf r} \;
\prod_{j=1}^N
e^{i {\bf k}_j \cdot [ {\bf r}_j - {\bf r}_{j-1}]} ,
\end{eqnarray}
where the indeces are counted mod $l$.
This reduces to a Kronecker-$\delta$,
but it is better to leave it as an integral as the continuum limit
for the entropy microstates will now be taken.

Since for ideal particles there are no interactions,
the monomers can be neglected, $N=l$,
and the average of the total weighted overlap factor
can be replaced by the total weighted overlap factor.
Hence the grand potential for an $l$-loop is
\begin{eqnarray}
\lefteqn{
- \beta \Omega_l^\mathrm{id}
} \nonumber \\
& = &
\frac{ z^l(l-1)!}{l!}
\sum_{\bf n} e^{-\beta {\cal H}_{\bf  n} }
\chi^{(l)}_{\bf n}
\nonumber \\ & = &
\frac{ z^l}{l}
\frac{V^{l}}{(2\pi)^{3l}} \int \mathrm{d}{\bf k} \;
e^{-\beta {\cal H}_{\bf  k} }
\chi^{(l)}_{\bf k}
\nonumber \\ & = &
\frac{ z^l}{l} \frac{(\pm 1 )^{l-1} }{(2\pi)^{3l}}
\int \mathrm{d}{\bf k} \;
\prod_{j=1}^l
e^{-(\beta \hbar^2/2m) k_j^2 }
\nonumber \\  && \mbox{ } \times
\int \mathrm{d}{\bf r} \;
\prod_{j=1}^l
e^{i {\bf k}_j \cdot [ {\bf r}_j - {\bf r}_{j-1}]}
\nonumber \\ & = &
\frac{ z^l}{l} \frac{(\pm 1 )^{l-1} }{(2\pi)^{3l}}
\int \mathrm{d}{\bf k} \; \int \mathrm{d}{\bf r} \;
\nonumber \\  && \mbox{ } \times
\prod_{j=1}^l
e^{-(\beta \hbar^2/2m)k_j^2  + i {\bf k}_j \cdot {\bf r}_j
 - i {\bf k}_j \cdot {\bf r}_{j-1} }
\nonumber \\ & = &
\frac{ z^l}{l} \frac{(\pm 1 )^{l-1} }{(2\pi)^{3l}}
\int \mathrm{d}{\bf k} \; \prod_{j=1}^l
\nonumber \\  && \mbox{ }
e^{-(\beta \hbar^2/2m)
\left[ {\bf k}_j + i (m/\beta \hbar^2) {\bf r}_j
 - i (m/\beta \hbar^2) {\bf r}_{j-1} \right]^2}
\nonumber \\  && \mbox{ } \times
 \int \mathrm{d}{\bf r} \; \prod_{j=1}^l
e^{-(m/2\beta \hbar^2) \left[ {\bf r}_j  - {\bf r}_{j-1} \right]^2}
\nonumber \\ & = &
\frac{ z^l}{l} \frac{(\pm 1 )^{l-1} }{(2\pi)^{3l}}
\left[ \frac{2\pi m}{\beta\hbar^2} \right]^{3l/2}
\nonumber \\  && \mbox{ } \times
 \int \mathrm{d}{\bf r}_1' %\, \mathrm{d}{\bf r}_2'
 \ldots \mathrm{d}{\bf r}_{l-1}' \, \mathrm{d}{\bf r}_{l} \;
e^{-(m/2\beta \hbar^2) \underline{\underline G}^{(l)}: {\bf r}' {\bf r}' }
\nonumber \\ & = &
\frac{(\pm 1 )^{l-1}z^l }{l\Lambda^{3l}}
\left( \frac{ 2 \pi m}{ \beta \hbar^2 } \right)^{3(l-1)/2}
\left| \underline{\underline G}^{(l)} \right|^{-3/2} V
\nonumber \\ & = &
\frac{(\pm 1 )^{l-1}z^l V }{l \Lambda^{3}}
\left| \underline{\underline G}^{(l)} \right|^{-3/2} .
\end{eqnarray}
Here ${\bf r}_j' \equiv  {\bf r}_j - {\bf r}_l$, $j=1,2,\ldots,l-1$,
and $\underline{\underline G}^{(l)}$
is an $(l-1)\times(l-1)$ tridiagonal matrix
with $2$ on the main diagonal
and $-1$ just above and just below the main diagonal,
with all other entries 0.
It is straightforward to show that the determinant is
\begin{eqnarray}
\left| \underline{\underline G}^{(l)} \right|
& = &
2 \left| \underline{\underline G}^{(l-1)} \right|
- \left| \underline{\underline G}^{(l-2)} \right|
\nonumber \\ & = &
l.
\end{eqnarray}
This gives the fugacity expansion of the quantum ideal gas as
\begin{eqnarray}
\beta p_\mathrm{id} \Lambda^3
& = &
\frac{- \beta \Omega^\mathrm{id} \Lambda^3}{V}
\nonumber \\ & = &
\sum_{l=1}^\infty \frac{- \beta \Omega^\mathrm{id}_l \Lambda^3}{V}
\nonumber \\ & = &
\sum_{l=1}^\infty
\frac{(\pm 1 )^{l-1}z^l  }{l^{5/2} } .
\end{eqnarray}
This is the known result, Eq.~(\ref{Eq:p(z)}).

%\newpage
%%%%%%%%%%%%%%%%%%%%%%%%%%%%%%%%%%%%%%%%%%%%%%%%%%%%%%%%%%%%%%%%%%%%%%%%%%
\section{Second Fugacity Coefficient} \label{Sec:b2}
\setcounter{equation}{0} \setcounter{subsubsection}{0}

The virial expansion for the pressure is
\begin{equation}
\beta p V
= -\beta \Omega
= \ln \Xi
= V \Lambda^{-3} \sum_{n=1}^\infty B_n \,\Lambda^{3n} \rho^n ,
\end{equation}
where $\rho = N/V$ is the number density
and $\Lambda = \sqrt{2\pi \hbar^2/m k_\mathrm{B}T}$ is the thermal wave length,.
Note that here the virial coefficients $B_n$ are dimensionless.
In classical statistical mechanics,
the virial coefficients are  defined in terms of Mayer cluster integrals,
$B_{n+1} = -\beta_n n/(n+1)$.\cite{TDSM,Pathria72}

The fugacity expansion for the pressure is
\begin{equation}
\beta p \Lambda^3 = \sum_{n=1}^\infty b_n z^n .
\end{equation}
Here $z= \exp \beta \mu$ is the fugacity.

The virial and fugacity coefficients
can be related to each other,
since from the general thermodynamic relationship,
$\overline \rho = {z}  {\partial \ln \Xi(\mu,V,T)}/V{\partial z}$,
one has
\begin{equation}
\overline \rho \Lambda^3
= \sum_{l=1}^\infty l b_l z^l .
\end{equation}
The first coefficients are $B_1 = b_1 = 1$,
and the second coefficients are related by
\begin{equation}
B_2 = - b_2.
\end{equation}

For a classical fluid,
the second fugacity coefficient is\cite{TDSM,Pathria72}
\begin{equation} \label{Eq:B2cl}
b_2^\mathrm{cl} = \frac{1}{2 \Lambda^3} \int \mathrm{d}{\bf r} \,
\left[ e^{-\beta u(r)} - 1 \right],
\end{equation}
where $u(r)$ is the central pair potential.

In the present formulation of quantum statistical mechanics,
the grand potential was given as a loop expansion.
Using Eq.~(\ref{Eq:Omega_l=<chi^l_tot>})
allows the pressure to be written
\begin{eqnarray}
\beta p \Lambda^3 & = &
\frac{- \beta\Omega\Lambda^3}{  V}
 \\ & = &
\frac{- \beta\Omega_1\Lambda^3}{V}
- \frac{ \beta\Lambda^3}{V} \sum_{l=2}^\infty \Omega_l
\nonumber \\ & = &
\sum_{n=1}^\infty b_{n}^\mathrm{mo} z^n
%\nonumber \\ && \mbox{ }
+ \frac{ \Lambda^3}{V} \sum_{l=2}^\infty
(l-1)! z^l
\left< \chi^{(l)}_\mathrm{tot}({\bf q}^{N-l})  \right>_\mu  .
\nonumber
\end{eqnarray}
Here has been written the original monomer notation
 $\Omega_1$ rather than  $\Omega_\mathrm{cl}$ or  $\Omega_\mu$.
 The reason for doing this
will be clarified below.

In general one cannot add to
the monomer fugacity coefficient $b_n^\mathrm{mo}$
the remaining quantum loop coefficient for $l=n$,
to get the fugacity coefficient $b_n$  of the quantum system
because the classical average itself depends upon the fugacity.
Instead one has to do a fugacity expansion of the average
and then collect coefficients.
However for the leading quantum correction,
%since the average is already multiplied by $z^2$,
one can write $ z^2 \left<  b_2^\mathrm{qu} \right>_\mathrm{cl}
= z^2 b_2^\mathrm{qu} + {\cal O}(z^3)$,
which gives
\begin{equation}
b_2 = b_{2}^\mathrm{mo}
+ b_2^\mathrm{qu} ,
\end{equation}
where
\begin{eqnarray}
b_2^\mathrm{qu}
& = &
\frac{ \Lambda^3}{V}
\chi^{(2)}_\mathrm{tot}({\bf q}^{N-2})
\nonumber \\ & = &
\frac{ \Lambda^3}{2V} \sum_{{\bf n}_{12}}
e^{-\beta {\cal H}^{(2)}_{{\bf n}_{12}}  }  \chi^{(2)}_{{\bf n}_{12}} .
\end{eqnarray}
The corrections due to the classical average of this
would contribute to the higher order quantum fugacity
coefficients.
Also, there is no monomer contribution to this,
because $N=2$;
the first monomer contribution to the dimer overlap factor has weight $z^3$.
The dimer overlap factor is
\begin{equation}
 \chi^{(2)}_{{\bf n}_{12}}
=
 \pm \langle
 \zeta_{{\bf n}_{12}}({\bf r}_2,{\bf r}_1)
 |  \zeta_{{\bf n}_{12}}({\bf r}_1,{\bf r}_2)
 \rangle ,
\end{equation}
with the upper sign for bosons
and the lower sign for fermions.

%%%%%%%%%%%%%%%%%%%%%%%%%%%%%%%%%%%%%%%%%%%%%%%%%%%%
\subsubsection{`Classical' Second Fugacity Coefficient}

One might think that the monomer second fugacity coefficient,
$b_{2}^\mathrm{mo}$,
which is derived from $\Omega_1 \equiv \Omega_\mathrm{cl}$,
is the classical coefficient given above, Eq.~(\ref{Eq:B2cl}).
However using this would give rise to an inconsistency.

The classical result is derived in classical phase space,
which in the present formalism arose
from invoking monomer wave packets of zero width.
This is exact in the thermodynamic limit,
$N\rightarrow \infty$, $V\rightarrow \infty$,
$N/V = \rho = \mbox{const}$.
However in evaluating the second fugacity coefficient
for the quantum system,
the exact entropy eigenfunction, $\zeta_{{\bf n}_{12}}({\bf r}_1,{\bf r}_2)$,
for $N=2$ will be used.
%(As well, these neglect the influence of any monomers
%because this only contributes to higher powers of the fugacity.)

It is necessary to evaluate the monomer coefficient, $b_2^\mathrm{mo}$,
under exactly the same conditions.
One requires  the exact rather than the approximate cancelation of the states
that are forbidden for fermions,
as well as the exact symmetric addition of the states for bosons.

For $N=2$, and neglecting the remaining monomers,
the required coefficient is
\begin{equation}
b_2^\mathrm{mo}
=
\frac{ \Lambda^3}{2V}
\sum_{{\bf n}_{12}}  \chi^{(1)}_{{\bf n}_{12}}
e^{-\beta {\cal H}^{(2)}_{{\bf n}_{12}}  } .
\end{equation}
The monomer loop overlap factor corresponds to the identity permutation,
\begin{equation}
 \chi^{(1)}_{{\bf n}_{12}}
=
\langle
 \zeta_{{\bf n}_{12}}({\bf r}_1,{\bf r}_2)
 |  \zeta_{{\bf n}_{12}}({\bf r}_1,{\bf r}_2)
 \rangle
 = 1 .
\end{equation}
Obviously since for $N=2$ one cannot use zero-width wave packets
and therefore the coefficient cannot be evaluated in classical phase space.
This is the reason for using the monomer superscript `mo'
rather than the superscript `cl', which would only be appropriate
in the limit $N \rightarrow \infty$.

%%%%%%%%%%%%%%%%%%%%%%%%%%%%%%%%%%%%%%%%%%%%%%%%%%%%
\subsubsection{Entropy Eigenfunction for a Collision}

In this sub-section the wave function for two particles is analyzed.
The material is standard
and the derivation follows closely that given by
Pathria, \S 9.5,\cite{Pathria72}
up until the point of application to the present formulation.
%The discussion is somewhat abbreviated
%until the results impact directly on the present formulation
%of quantum statistical mechanics.

The energy eigenvalues are
given by  the two-body stationary Schr\"odinger equation,
\begin{equation}
\hat{\cal H}^{(2)}({\bf r}_1 , {\bf r}_2)
\Psi_{{\bf n}_{12}}({\bf r}_1 , {\bf r}_2)
=
{\cal H}^{(2)}_{{\bf n}_{12}} \, \Psi_{{\bf n}_{12}}({\bf r}_1 , {\bf r}_2) .
\end{equation}
Here $\hat{\cal H}^{(2)}$ is taken to be the Hamiltonian operator
for the two particles alone.

For a central two-body potential $u(r)$,
the entropy eigenfunction is the product of functions
of the center of mass coordinate
${\bf R} \equiv [{\bf r}_1 + {\bf r}_2]/2$,
and of the relative coordinate $ {\bf r} \equiv {\bf r}_2 - {\bf r}_1$,
\begin{eqnarray}
\Psi_{{\bf n}_{12}}({\bf r}_1 , {\bf r}_2)
& = &
\psi_{\bf j}({\bf R}) \, \psi_{\bf n}({\bf r})
\nonumber \\ & = &
\left\{ \frac{1}{V^{1/2}}e^{- {\bf P}_{\bf j} \cdot {\bf R}/i\hbar} \right\}
\, \psi_{\bf n}({\bf r}) .
\end{eqnarray}
The energy eigenvalue is
\begin{equation}
{\cal H}^{(2)}_{{\bf n}_{12}} = \frac{ P_{\bf j}^2 }{4m} + \epsilon_{\bf n} ,
\end{equation}
and the wave equation for the relative motion is
\begin{equation}
\left[ \frac{-\hbar^2}{m } \nabla^2_r + u(r) \right]  \psi_{\bf n}({\bf r})
 =  \epsilon_{\bf n}  \psi_{\bf n}({\bf r}) ,
\end{equation}
with $m/2$ being the reduced mass of the pair.

The wave function for relative motion
can be written as the product of a radial function
and a spherical harmonic,
\begin{equation}
\psi_{klm}({\bf r}) = A_{klm} \frac{\chi_{kl}(r)}{r} Y_{l,m}(\theta,\phi) .
\end{equation}
(This $m$ is the magnetic quantum number, not the mass.)

The radial function must vanish at some large value,
$\chi_{kl}(R_0) = 0$.
Its asymptotic form is
\begin{equation}
\chi_{kl}(r) \propto
\sin \left( kr - \frac{l\pi}{2} + \eta_l(k) \right) .
\end{equation}
Here $\eta_l(k) $ is the scattering phase shift for the $l$th partial wave
of wave number $k$.
Hence
\begin{equation}
 kR_0  - \frac{l\pi}{2} + \eta_l(k) = o \pi , \;\; o=0,1,2, \ldots .
\end{equation}

The spherical harmonics have a definite symmetry:
\begin{equation}
\psi_{klm}(- {\bf r}) = (-1)^l \psi_{klm}({\bf r}) .
\end{equation}
Obviously negating the separation vector
corresponds to interchanging the two particles.
In conventional formulations,
this symmetry is exploited to yield the allowed states
for the particles:
 $l$ must be even for bosons and odd for fermions.
In the present formulation,
all states are allowed,
and it is the overlap factors
that take care of particle interchange symmetry.

For the present case of the second virial coefficient,
the dimer overlap factor is
\begin{eqnarray}
 \chi^{(2)}_{{\bf n}_{12}}
& = &
\pm \left\langle
\zeta_{{\bf n}_{12}}({\bf r}_2,{\bf r}_1)
\, \right| \left. \zeta_{{\bf n}_{12}}({\bf r}_1,{\bf r}_2)
\right \rangle
\nonumber \\ & = &
\pm \left\langle
\psi_{\bf j}({\bf R}) \psi_{klm}(-{\bf r})
\,  \right| \left.  \psi_{\bf j}({\bf R}) \psi_{klm}({\bf r})
\right \rangle
\nonumber \\ & = &
\pm (-1)^l .
\end{eqnarray}

Combining this result with that for the one-loop overlap factor,
$ \chi^{(1)}_{{\bf n}_{12}} = 1 $,
the second fugacity coefficient for a quantum system is
\begin{eqnarray}
b_2 & \equiv &
b_2^\mathrm{mo} + b_2^\mathrm{qu}
\nonumber \\ & = &
\frac{ \Lambda^3}{2V}
\sum_{{\bf n}_{12}}
e^{-\beta {\cal H}^{(2)}_{{\bf n}_{12}}  }
\left\{  \chi^{(1)}_{{\bf n}_{12}}
+  \chi^{(2)}_{{\bf n}_{12}} \right\}
\nonumber \\ & = &
\frac{ \Lambda^3}{2V}
\sum_{{\bf n}_{12}}
e^{-\beta {\cal H}^{(2)}_{{\bf n}_{12}}  }
\left\{  1 \pm  (-1)^l \right\} .
\end{eqnarray}
For bosons the even $l$ terms are non-zero,
and for fermions it is the odd $l$ terms that survive.

With this result the present  analysis joins exactly that
given by Pathria, \S 9.5.\cite{Pathria72}
Pathria shows how to derive the density of states,
how to convert the sum over states to an integral,
and how to subtract the ideal gas contribution
to remove the dependence on $R_0$.
Rather than repeat those standard results,
a discussion of the present manipulations is more useful.

In order to achieve consonance
between the present and standard formulations,
it was necessary to back-track somewhat
on the present classical contributions.
Rather than evaluating them in classical phase space,
it is necessary to evaluate them
on exactly the same basis as the quantum corrections.
That is, one has to use in both cases the same entropy eigenfunctions,
and to sum over the same entropy eigenstates
in order to get the exact cancelations
that are necessary for particle symmetry.

The mathematical justification for doing this in the present case
is straightforward.
The classical grand potential and classical phase space upon which
it is predicated only hold in the thermodynamic limit
where the entropy eigenfunctions for the monomers
become exactly wave packets of zero width.
Contrariwise, for $N=2$ in the present case,
one cannot set the monomer grand potential
equal to the classical grand potential,
$\Omega_1 \ne \Omega_\mathrm{cl}$,
$z \rightarrow 0$, $V$ fixed.
Instead one must use the exact entropy eigenfunction for $N=2$
with a monomer overlap factor $\chi^{(1)}_{{\bf n}_{12}} = 1$  to evaluate it,
as was done here.

But one should be a little wary of drawing a general conclusion
from this analysis of the second fugacity coefficient.
The limit $V \rightarrow \infty$, $z \rightarrow 0$,
(equivalently $N \rightarrow 0$), is of, well, limited interest.
For terrestrial condensed matter computations,
more relevant is the thermodynamic limit,
$V \rightarrow \infty$, $z $ fixed,
(equivalently $N \rightarrow \infty$, $\rho$ fixed).
In this case the monomer grand potential is dominated by terms
$N \approx \overline N(\mu,V)\rightarrow \infty$,
and is given by $\Omega_1 = \Omega_\mathrm{cl}$
with a relative error ${\cal O}(N^{-1/2})$.
Similar comments can be made for the quantum corrections to the grand potential,
the average total loop overlap factors. These are also dominated by monomers
in the thermodynamic limit, which is to say classical,
while the loop particles of the overlap factors
required characterization by an appropriate entropy eigenfunction
in the presence of the (fixed) monomers.

%\newpage $\;$ \newpage
%%%%%%%%%%%%%%%%%%%%%%%%%%%%%%%%%%%%%%%%%%%%%%%%%%%%%%%%%%%%%%%%%%%%%%%%%%
\section{Discursive Summary and Conclusion}
\setcounter{equation}{0} \setcounter{subsubsection}{0}

The three impediments to the computation of the properties
of a quantum many particle system listed in the introduction were
superposing states,
symmetrizing the wave function,
and finding  appropriate eigenfunctions.
The latter requires
choosing an appropriate operator or representation,
obtaining  eigenvalues,
choosing a reasonably complete set of approximate eigenfunctions,
and systematically improving and orthogonalizing them.

It was pointed out that the superposition of states
is suppressed in quantum statistical mechanics,
since entanglement of the sub-system with the reservoir
collapses the wave function into a mixture of pure states.
This solves the first problem.
Moreover, the mechanism of the collapse
reveals that the pure states are entropy eigenstates,
which in the case of the canonical equilibrium system
are just energy states.
This strongly suggests that for the third problem,
the appropriate eigenfunctions are entropy eigenfunctions.

Proceeding further down this path
it was observed that most large terrestrial systems
can be accurately characterized using classical statistical mechanics.
Classical mechanics  of course does not suffer from the problem
of superposition.
This suggests that it would be fecund to treat
quantum statistical mechanics as a perturbation expansion
about  classical statistical mechanics.
This in turn implies that the entropy microstates
that play a preferred role in quantum statistical mechanics
should be identified with the points in classical phase space,
which provides the microstates for classical statistical mechanics.
The logical consequence of this identification is that
the entropy eigenfunctions ought to bear a label
that is in one-to-one correspondence with the points
in classical phase space
(at least this ought to be the case for large systems).
Since wave packets bear just such a label,
there is strong motivation to use them as trial entropy eigenfunctions.

Before proceeding with this idea of the preferred role
of the entropy microstates and of wave packets as entropy eigenfunctions,
a general conceptual problem arises
from the symmetrization of the wave function.
The indistinguishability of particles
means that the set of distinct microstates
is smaller than the set of all microstates.
(In the case of one-particle states,
a permutation of the order of the one-particle labels
does not give a distinct microstate.)
The number of equivalent microstates varies with the particular microstate.
(For example, in the case of one-particle states,
if all states are different the correction factor is $N!$,
but if two or more states are the same,
in the case of bosons it is less than $N!$.)
The partition function, whose logarithm is the total entropy,
must be the weighted number of  distinct microstates.
This is consistent with the analysis of Messiah, Ch.~XIV, \S 6,\cite{Messiah61}
who says that microstates composed of one-particle states
in permuted order count as one and the same microstate.
The partition function can be written as the sum over all microstates provided
that a microstate-dependent permutation overlap factor is introduced
that corrects for the double counting of identical microstates.
This necessary correction factor does not usually appear explicitly
in conventional presentations of the formula
for the quantum partition function or for the quantum statistical average.

This permutation overlap factor is directly relevant to the use of
wave packets as entropy eigenfunctions,
or more generally, to the use of any non-orthogonal sets of eigenfunctions.
Wave packets in general have finite width.
Classical phase space is a continuum.
This means that wave packets necessarily overlap.
Since it would be practically impossible to orthogonalize
a continuous set of such finite-width wave packets,
the present approach is instead to re-formulate quantum  statistical mechanics
to cope with  non-orthogonal eigenfunctions.
A little thought shows that the overlap of  non-orthogonal eigenfunctions
has the same conceptual origin as the multiple occupancy
of single-particle states.
Accordingly both problems of over-counting
are solved by the microstate-dependent permutation overlap factor.

The introduction of the  permutation overlap factor
also resolves a second conceptual problem
that arises from wave function symmetrization
and the continuum limit.
The wave function must vanish if two fermions occupy the same state.
How is this to be interpreted when the states form a continuum?
For two particles
the permutation overlap factor turns out to be a Gaussian
in the entropy microstate labels, which correspond to classical phase space.
Hence it varies continuously between 2 and 1 for bosons
and between 0 and 1 for fermions,
depending on their separation in phase space.
This is the continuum analogue of the occupancy rules for quantum microstates.

Further, the overlap factor shows that the effects of symmetrization are local,
by which is meant that only particles in neighboring microstates
(such as phase space, which includes the projection onto configuration space)
are affected by interchange symmetry.
This is similar to a point made by  Messiah, Ch.~XIV, \S 8,\cite{Messiah61}
namely that if the particles are represented by wave packets,
then one does not have to symmetrize the wave function with respect to
interchange of particles in non-overlapping wave packets.
Localization resolves this third conceptual problem:
in calculating the properties of, say, selenium in a terrestrial laboratory,
one does not have to take into account  interchange symmetry with
particles on the moon.
More practically,
in a computer simulation of a quantum system,
one does not have to symmetrize the wave function
with respect to all particles,
but only within clusters of neighboring particles.

The overlap factor lends itself to a permutation expansion.
In general an arbitrary permutation of the elements of a finite set
can be decomposed into a product of closed permutation loops.
Hence the permutation overlap factor can be expanded
in terms of loops of increasing size and their products.

At this stage, on the basis of localization,
two approximations are invoked to simplify the results.
First, the expectation value of a product of overlap loops
is approximated as the product of the expectation values.
Second, the average (over the monomers) of the
product of overlap loop expectation values
is approximated as the product of the averages.
Because of localization,
these two approximations are expected to be valid
when the quantum loop overlap density is low.
%which is argued to be the case at terrestrial densities and temperatures.
It is argued that the corrections to these two factorizations
vanish in the thermodynamic limit.

The quantum correction for the properties of liquid water
at standard temperature and pressure
is on the order of one part in ten thousand.
(This is based on the value of the dimensionless parameter
$\rho \Lambda^3 $, which characterizes spatial symmetrization effects.)
Since this is so small, one would expect it to be dominated
by a single dimer overlap loop.
The contribution from two dimers loops
consists of their uncorrelated product,
which is retained in the present expansion,
and their direct correlation,  which has  been neglected here
because it is expected to be significantly smaller.
Accordingly one would expect that the neglected contribution from
the correlated part to be very much less than one part in one hundred million.
This is a persuasive argument
that the factorization approximations invoked in the present theory
are likely to be accurate for terrestrial condensed matter.

The factorization approximations
allow an infinite order re-summation of the permutation expansion
that results in the exponential of the sum of individual average
loop overlap factors.
Accordingly the grand potential of the quantum system
becomes the sum of monomer and individual loop grand potentials.

Crucially each term in the sum is extensive.
It may seem a little trivial to identify this as the fourth conceptual point,
but in fact extensivity is essential to the formulation of thermodynamics.
The only way to demonstrate this extensivity
is to perform an infinite resummation
of the wave function symmetrization expansion
using the factorization approximation.
Without the factorization and resummation,
the  grand potential is not extensive.

The monomer grand potential and averages over the monomers
deserve particular comment.
For the monomers,
the set of `bare' wave packet was analyzed with a view to optimizing
them as entropy eigenfunctions.
For the case of real interacting particles,
it was shown that in the thermodynamic limit $N \rightarrow \infty$,
the wave packet width vanished and the bare wave packet became an
exact entropy eigenfunction.
(For the case of an ideal particle
it was shown that the wave packet width became
infinite, which is the known, exact result.)
In a further calculation it was shown that wave packets
that were systematically modified to improve them as entropy eigenfunctions
tended to bare wave packets in the thermodynamic limit.

The significance of this result is that
in the thermodynamic limit
the entropy eigenfunctions are exactly given by infinitely sharp wave packets.
This means that each particle has a well-defined and simultaneous
position  and momentum, which is to say
that the entropy microstate of the system is a point in classical phase space.
This explains how classical mechanics arises from quantum mechanics
via  an open quantum system in the thermodynamic limit.

The last point,
which is the fifth conceptual point addressed by the present results,
requires clarification, as there appears to be two gaps in the chain of
logical reasoning that leads to the conclusion that quantum mechanics
implies classical mechanics.
It is certainly true
that quantum mechanics implies quantum statistical mechanics
for an open quantum system,\cite{QSM}
and, by the results in this paper,
quantum statistical mechanics implies classical statistical mechanics
(for a macroscopic system).
It is also true, from other arguments,\cite{TDSM}
that classical mechanics implies classical statistical mechanics.
But this information is not enough
to conclude that classical statistical mechanics implies classical mechanics.
Nor does it prove that classical phase space evolves
via the classical equations of motion.
One needs two additional links to close the chain.

The Ehrenfest theorem says in essence that
Hamilton's equations of motion in operator form hold
when considered as expectation values.\cite{Messiah61,Merzbacher70}
(The rate of change of the expectation value of position
is equal to the expectation value of the momentum derivative
of the Hamiltonian operator,
and the rate of change of the expectation value of momentum
is equal to the expectation value of the negative of the position derivative
of the Hamiltonian operator.)
The present results show that for a macroscopic open quantum system,
the entropy eigenfunctions are wave packets of zero width,
and therefore they are also exactly and simultaneously
position and momentum eigenfunctions.
From this it follows that the expectation value of a function of the
position and momentum operators is equal to the function of the
position and momentum eigenvalues.
Combined with the Ehrenfest theorem,
this proves that points in phase space evolve according to
the classical Hamilton's equations of motion.

One more case needs to be addressed,
namely that of macroscopic isolated objects.
A macroscopic object can be thought of as composed of sub-systems
that interact with each other across their boundaries.
Therefore each sub-system is  an open quantum system,
and therefore at any given instant the particles within each occupy a point in
classical phase space.
By the above,
they therefore move according to the classical equations of motion.
Therefore, even though the macroscopic object is isolated
and is itself not an open quantum system,
it nevertheless as a whole obeys the classical equations of motion.
The same conclusion holds for interacting but otherwise isolated
macroscopic objects.
This completes the chain of reasoning:
quantum mechanics implies classical mechanics for macroscopic objects.

Returning to the quantum overlap loops,
their evaluation requires
the entropy eigenvalues and eigenfunctions for the loop particles
in the presence of fixed monomers that are subsequently averaged over.
The loop particle entropy eigenfunctions
cannot be infinitely sharp bare wave packets,
since there is typically only a small number of particles in a loop,
but it is possible that they are well-approximated by modified wave packets
of optimized width.
Obtaining and averaging the total loop overlap factor is likely
to be computationally challenging,
although there appear to be several possibilities
for improving the efficiency of the process.

The present expansion for the quantum grand potential
was explicitly tested against the known quantum fugacity expansion
for two cases.
(The present expansion is in terms  of increasing quantum overlap loop size,
which, although it contains explicit powers of the fugacity,
is not a strict fugacity expansion.)
For the ideal gas,
using wave packets in which the width becomes infinite,
the first three terms of the expansion were calculated explicitly
(ie.\ the classical term and the first two quantum corrections),
\S \ref{Sec:Ideal-Packet-Fug},
and using plane waves, all terms in the full expansion
were obtained explicitly \S \ref{Sec:Ideal-Plane-Fug}.
Both cases agree with the known results for the quantum ideal gas.
The second fugacity coefficient for two interacting quantum particles
was also obtained explicitly from the expansion
and it was shown that this agreed with
the known second quantum fugacity coefficient.

Of course many of the fundamental issues raised here
have been previously addressed in the literature
and the present results are not completely orthogonal to earlier work,
with certain overlaps and parallels being evident.
Nevertheless one can identify several results
that arguably stand out from previous work.
First,
the restriction of the sum over states to distinct states,
and the transformation of this to the sum over all states
by the explicit inclusion of the overlap factor.
Second,
factorizing permutations into loops,
which gives the expansion of the overlap factor,
and which in turn allows the  re-summation of the grand partition function
to yield an expansion for the grand potential of a quantum system.
Third,
the argument that the entropy representation
is the most appropriate representation of a quantum many-particle system.
Fourth,
the demonstration that wave packets form suitable trial entropy eigenfunctions.
And fifth,
the proof that in the thermodynamic limit
the  wave packet width goes to zero,
and that this is the origin
of classical statistical mechanics and of classical mechanics.

In practical terms, for the application
of quantum statistical mechanics to condensed matter,
here it has been shown not only that the appropriate representation
is the entropy one,
but also that the appropriate approximate eigenfunctions are wave packets.
The expansion in terms of quantum overlap loops is a significant saving
because the loops are localized, they can be considered in isolation,
and they comprise few particles,
which reduce the space of trial eigenfunctions to be explored.
The formulation obviates the need to symmetrize or to orthogonalize
the eigenfunctions,
which eliminates these particular impediments
to a feasible computational approach to quantum many-particle problems.
Whether the tasks that remain are tractable
has yet to be demonstrated by explicit numerical computation.

%\section{References}
%\newpage $\;$ \newpage
%%%%%%%%%%%%%%%%%%%%%%%%%%%%%%%%%%%%%%%%%%%%%%%%%%%%%%%%%%%%%%%%%%%%%%%%%%

%%%%%%%%%%%%%%%%%%%%%%%%%%%%%%%%%%%%%%%%%%%%%%%%%%%%%%%%%%%%%%%%%%%%%%%%%%%

\appendix

\setcounter{equation}{0}
\renewcommand{\theequation}{\Alph{section}.\arabic{equation}}

%\newpage $\;$ \newpage
%%%%%%%%%%%%%%%%%%%%%%%%%%%%%%%%%%%%%%%%%%%%%%%%%%%%%%%%%%%%%%%%%%%%%%%%%
%
\section{Overlapping or Non-Orthogonal Eigenfunctions} \label{Sec:genoverlap}
%
%%%%%%%%%%%%%%%%%%%%%%%%%%%%%%%%%%%%%%%%%%%%%%%%%%%%%%%%%%%%%%%%%%%%%%%%%

This appendix analyzes the formulation
of quantum statistical mechanics
in the case that the basis functions are non-orthogonal or overlapping.
The analysis has wider application
than the minimal uncertainty wave packets that are used
to illustrate the results.

%%%%%%%%%%%%%%%%%%%%%%%%%%%%%%%%%%%%%%%%%%%%%%%%%%%%%%%%%%%%%%%%%%%%%%
\subsection{Gram-Schmidt Background}

Consider two normalized vectors
$|a_1\rangle$ and $|a_2\rangle$
that are not orthogonal, $\langle a_1|a_2\rangle  \ne 0$.
The Gram-Schmidt procedure may be used
\begin{equation}
|a_2'\rangle =
\frac{|a_2\rangle - \langle a_1|a_2\rangle \, |a_1\rangle
}{ \sqrt{1 - \langle a_1|a_2\rangle \langle a_2|a_1\rangle } },
\end{equation}
to create an orthonormal set, $\langle a_1|a_2'\rangle = 0$.
An arbitrary vector
(more precisely, its projection onto the sub-space spanned by the two vectors)
then has representation,
\begin{eqnarray}
|x\rangle & = &
\langle a_1|x\rangle \, |a_1\rangle
+ \langle a_2'|x\rangle \, |a_2'\rangle
 \\ & = &
\left[ 1 - |\langle a_1|a_2\rangle |^2 \right]^{-1}
\nonumber \\ && \mbox{ } \times
\left\{ \rule{0cm}{0.4cm}
\langle a_1|x\rangle \, |a_1\rangle
- |\langle a_1|a_2\rangle |^2 \langle a_1|x\rangle \, |a_1\rangle
\right. \nonumber \\ && \left. \mbox{ }
+
\left[
\langle a_2| x \rangle - \langle a_2|a_1\rangle \, \langle a_1| x \rangle
\right]
%\right. \nonumber \\ && \left. \mbox{ } \times
\left[ |a_2\rangle - \langle a_1|a_2\rangle \, |a_1\rangle \right]
\rule{0cm}{0.4cm} \right\}
%%%%%%%%%%%%%%%%%%%%%%%%%%%%%%%%%%%%%%%
\nonumber \\ & = &
\left[ 1 - |\langle a_1|a_2\rangle |^2 \right]^{-1}
%\nonumber \\ && \mbox{ } \times
\left\{ \rule{0cm}{0.4cm}
\langle a_1|x\rangle \, |a_1\rangle
+ \langle a_2|x\rangle \, |a_2\rangle
\right. \nonumber \\ && \left. \mbox{ }
- \langle a_1|a_2\rangle \, \langle a_2| x \rangle \, |a_1\rangle
- \langle a_2|a_1\rangle \, \langle a_1| x \rangle \, |a_2\rangle
\rule{0cm}{0.4cm} \right\} .
\nonumber
\end{eqnarray}
In the braces the first pair of terms comprises
the usual projections on the axes.
The second pair of terms removes the double counting due to the overlap of the
basis vectors.

The generalization of this to a complete, non-orthogonal set of basis vectors
is
\begin{eqnarray}
|x\rangle & = &
\frac{1}{\eta}
\sum_i \langle a_i|x\rangle \, |a_i\rangle
- \sum_{i \ne j} \langle a_i|a_j\rangle \,
\langle a_j|x\rangle \, |a_i\rangle
\nonumber \\  & = &
\frac{1}{\eta}
\sum_i
\left\{ 2 \langle a_i|x\rangle \,
- \sum_{j}  \langle a_i|a_j\rangle \, \langle a_j|x\rangle
\right\} |a_i\rangle
\nonumber \\ & \equiv &
\sum_i c_i(x;[a]) \, |a_i\rangle .
\end{eqnarray}
Here $\eta$ is a normalization  constant (see below).

%%%%%%%%%%%%%%%%%%%%%%%%%%%%%%%%%%%%%%%%%%%%%%%%%%%%%%%%%%%%%%%%%%%%%%%%%%%
\subsection{Wave Packets}

Consider now the quantum problem,
with the basis wave functions $\zeta_{\bf n}({\bf r})$
not necessarily orthogonal.
This includes wave packets, which are localized but not orthogonal.
It will be assumed below  that
the non-orthogonal basis functions are at least approximately
entropy eigenfunctions,
\begin{equation}
\hat{\cal H}({\bf r}) \zeta_{\bf n}({\bf r})
\approx {\cal H}_{\bf n} \zeta_{\bf n}({\bf r}) .
\end{equation}

In the non-orthogonal or overlapping case
one has a `soft' Kronecker-$\delta$,
\begin{eqnarray}
\langle \zeta_{\bf n'} |\zeta_{\bf n} \rangle
& \equiv &
\delta_\xi({\bf n'}-{\bf n})
\\ \nonumber & = &
e^{-({\bf q}_{\bf n'}-{\bf q}_{\bf n})^2/4\xi^2}
e^{-\xi^2({\bf p}_{\bf n'}-{\bf p}_{\bf n})^2/\hbar^2} .
\end{eqnarray}
Note that $ \delta_\xi({\bf 0}) = 1$.
The first definition holds in general
for any non-orthogonal set of basis functions.
The second equality holds only for the
minimum uncertainty wave packets,
which are the zeroth order entropy eigenfunctions.

For the unordered states the integral of this is
\begin{eqnarray} \label{Eq:lambda(xi)}
\lambda & \equiv &
\sum_{\bf n'} \delta_\xi({\bf n'}-{\bf n})
\nonumber \\  & = &
\frac{1}{[\Delta_{\bf q}\Delta_{\bf p}]^{3N}}
\int \mathrm{d}{\bf \Gamma}'\;
e^{-({\bf q}'-{\bf q})^2/4\xi^2}
e^{-\xi^2({\bf p}'-{\bf p}^2/\hbar^2}
\nonumber \\  & = &
\frac{1}{[\Delta_{\bf q}\Delta_{\bf p}]^{3N}}
[ 4 \pi \xi^2 ]^{3N/2}  [ \pi \hbar^2/\xi^2 ]^{3N/2}
\nonumber \\  & = &
\left[ \frac{h}{\Delta_{\bf q}\Delta_{\bf p}} \right]^{3N}.
\end{eqnarray}
Again the first definition holds in general
for any non-orthogonal set of basis functions.
It is important that $\lambda$ is independent of ${\bf n}$ in general.
It is interesting that in the case of the
present minimum uncertainty wave packets,
it is also independent of $\xi$.
This independence would hold even if
$\xi$ were replaced by a microstate-dependent
correlation matrix, $\xi^{-2} \Rightarrow {\bm \sigma}_{\bf n}^{-1}$.

Strictly speaking, the left hand side must be strictly greater than unity
(because all terms in the sum are non-negative,
and there is at least one term of unity, namely ${\bf n'}={\bf n}$).
However it is traditional to choose the quantized volume of phase space
as equal to Planck's constant, $\Delta_{\bf q}\Delta_{\bf p} = h$,
which choice makes the final right hand side unity.
The choice of this or any other value is immaterial
as it is just a constant factor
multiplying the partition function,
(ie.\ an additive constant for the free energy).

%This doesn't appear to work for ordered states:
%$\lambda$ becomes microstate-dependent.

Denote the `true' orthogonal, complete set
of non-symmetrized entropy eigenfunctions by $\phi_{\bf m}$.
Since for the present canonical equilibrium system,
the entropy operator is proportional to the energy operator,
$\hat S_\mathrm{r} = - \hat{\cal H}/T$,
one has
\begin{equation}
\hat{\cal H}({\bf r}) \phi_{\bf m}({\bf r})
= {\cal H}_{\bf m} \phi_{\bf m}({\bf r}) .
\end{equation}
Although these microstates are degenerate,
it is assumed that a Gram-Schmidt procedure has been used
to orthogonalize them,
\begin{equation} \label{Eq:<m'|m>}
\langle  \phi_{\bf m'} |  \phi_{\bf m} \rangle
 = \delta( {\bf m}'-{\bf m}) ,
\end{equation}
where a true Kronecker-$\delta$ appears.

In view of the analysis in the preceding sub-section,
the `true' entropy eigenfunctions can be expanded in the complete set
of non-orthogonal basis wave functions,
\begin{equation}
\phi_{\bf m}({\bf r})
=
\sum_{\bf n} c_{{\bf m} {\bf n}} \, \zeta_{\bf n}({\bf r}) .
\end{equation}
The transformation coefficient is
\begin{eqnarray}
 c_{{\bf m} {\bf n}}
 & \equiv &
 c_{\bf n}(\phi_{\bf m};[\zeta])
\nonumber \\ & = &
\frac{1}{\eta} \left\{
\langle \zeta_{\bf n}|\phi_{\bf m}\rangle \,
- \sum_{\bf n'}\!\!^{(\ne {\bf n})}
\langle \zeta_{\bf n}|\zeta_{\bf n'}\rangle \,
\langle \zeta_{\bf n'}|\phi_{\bf m}\rangle
\right\}
\nonumber \\ & = &
\frac{1}{\eta}
\left\{
2 \langle \zeta_{\bf n}|\phi_{\bf m}\rangle \,
- \sum_{\bf n'}
\delta_{\xi}({\bf n}-{\bf n'}) \,
\langle \zeta_{\bf n'}|\phi_{\bf m}\rangle \right\}
\nonumber \\ & \approx &
\frac{1}{\eta}
\left\{
2 \langle \zeta_{\bf n}|\phi_{\bf m}\rangle \,
- \langle \zeta_{\bf n}|\phi_{\bf m}\rangle
\sum_{\bf n'} \delta_{\xi}({\bf n}-{\bf n'}) \,
 \right\}
\nonumber \\ & = &
\frac{2-\lambda}{\eta}
\langle \zeta_{\bf n}|\phi_{\bf m}\rangle .
\end{eqnarray}
Setting $ \phi_{\bf m}({\bf r}) = \zeta_{{\bf n}'}({\bf r})$,
it is straightforward to show that the normalization constant is
$\eta = (2-\lambda) \lambda$,
which is again a constant independent of the microstate ${\bf n}$.
%(In this appendix  $\eta$  simply denotes a constant;
%it is unrelated to the chemical potential.)
The fact that the prefactor is independent of ${\bf n}$ and ${\bf m}$
means that the non-orthogonality only introduces a scale factor
into quantum statistical mechanics,
which is not a big issue.

Now $ c_{{\bf m} {\bf n}}^*  c_{{\bf m} {\bf n}}$
is the proportion of $\phi_{\bf m}$ uniquely in $\zeta_{\bf n}$.
But $\delta_\xi({\bf n}-{\bf n'})$ is spread among $\lambda$ states,
and so there must be $\lambda^{-1}$ of the total in each state.
Hence one must have
\begin{eqnarray}
\lambda^{-1}  & = &
  \sum_{\bf n} c_{{\bf m} {\bf n}}^* c_{{\bf m} {\bf n}}
\nonumber \\ & = &
\frac{(2-\lambda)^2}{ \eta^2}
\sum_{\bf n}
\langle \phi_{\bf m}|\zeta_{\bf n}\rangle \,
\langle \zeta_{\bf n}|\phi_{\bf m}\rangle .
\end{eqnarray}
%If this is summed over ${\bf m}$ then this gives
%$\sum_{\bf m} = [{\lambda(2-\lambda)^2}/{\eta^{2}}] \sum_{\bf n}$.

Since the set $\zeta_{\bf n}$ is complete,
this means that
\begin{equation}
\frac{\lambda(2-\lambda)^2}{\eta^2}
\sum_{\bf n}
|\zeta_{\bf n}\rangle \,
\langle \zeta_{\bf n}|
= {\bf I} ,
\end{equation}
or
\begin{equation}
\frac{\lambda(2-\lambda)^2}{\eta^2}
\sum_{\bf n}
 \zeta_{\bf n}({\bf r}')^* \,  \zeta_{\bf n}({\bf r})
= \delta( {\bf r}'-{\bf r}) .
\end{equation}
This result can also be obtained directly
by inserting the expansion for $\phi_{\bf m}({\bf r})$
into Eq.~(\ref{Eq:<m'|m>}).

This result is equivalent to
\begin{equation}
\sum_{{\bf n}} c_{{\bf m}' {\bf n}}^* c_{{\bf m} {\bf n}}
= \lambda^{-1} \delta({\bf m}' -{\bf m}).
\end{equation}
Conversely one has
\begin{eqnarray}
\sum_{{\bf m}} c_{{\bf m} {\bf n}}\, c_{{\bf m} {\bf n}'}^*
& = &
\frac{(2-\lambda)^2}{\eta^2}
\sum_{{\bf m}}
\langle \zeta_{\bf n}|\phi_{\bf m} \rangle \,
\langle \phi_{\bf m}  | \zeta_{\bf n'}\rangle
\nonumber \\ & = &
\frac{(2-\lambda)^2}{\eta^2}
\langle \zeta_{\bf n} | \zeta_{\bf n'}\rangle
\nonumber \\ & = &
\frac{(2-\lambda)^2}{\eta^2} \delta_\xi({\bf n}-{\bf n}') .
\end{eqnarray}
Note that the results from Eq.~(\ref{Eq:<m'|m>}) and following
hold for a general non-orthogonal set of basis wave functions,
not just for the set composed of the minimum uncertainty wave packets.

For wave packets and
the conventional volume element of classical statistical mechanics,
$\Delta_{\bf q}\Delta_{\bf p} = h$,
one has $\lambda = \left[{h}/{\Delta_{\bf q}\Delta_{\bf p}}\right]^{3N}
= 1$,
and  $\eta = (2-\lambda) \lambda = 1$.

%%%%%%%%%%%%%%%%%%%%%%%%%%%%%%%%%%%%%%%%%%%%%%%%%%%%%%%%%%%%%%%%%%%%%%%%%%%%%
\subsection{The Partition Function}

Summing over unique `true' entropy microstates
the partition function is
\begin{eqnarray}
Z(N,V,T)
& = &
\sum_{\hat O{\bf m}}  e^{-\beta {\cal H}_{\bf m}}
\nonumber \\ & = &
\frac{1}{N!} \sum_{\bf m} \chi(\phi_{\bf m}) e^{-\beta {\cal H}_{\bf m}} ,
\end{eqnarray}
where $\hat O{\bf m}$ is the unique ordered arrangement of
${\bf m}$ and its permutations,
and where the `true' entropy eigenfunction overlap factor is
\begin{equation}
\chi(\phi_{\bf m})
=
\sum_{\hat P} (\pm 1)^{p}
\langle \phi_{{\bf m}} | \phi_{\hat P{\bf m}} \rangle  .
\end{equation}
This overlap factor arises from the full symmetrization
of the quantum system, as discussed in the text.

In view of the fact that each wave packet state $\zeta_{\bf n}$
is `worth' $\eta^2/\lambda(2-\lambda)^2$ entropy microstates $\phi_{\bf m}$,
one expects that the partition function can also be written as
\begin{equation}
Z(N,V,T)
=
\frac{\lambda(2-\lambda)^2}{\eta^2}
\frac{1}{N!} \sum_{\bf n} \chi(\zeta_{\bf n}) e^{-\beta {\cal H}_{\bf n}} .
\end{equation}
The constant scale factor
is an additive constant for the total entropy or free energy,
which is of course immaterial and can be neglected.
(Recall that $\eta = (2-\lambda) \lambda$,
so that the prefactor reduces to
${\lambda(2-\lambda)^2}/{\eta^2} = \lambda^{-1}$.
For wave packets and
the conventional volume element of classical statistical mechanics,
$\Delta_{\bf q}\Delta_{\bf p} = h$,
one has $\lambda = \left[{h}/{\Delta_{\bf q}\Delta_{\bf p}}\right]^{3N}
= 1$,
and so the pre-factor is in fact unity.)
The wave packet overlap factor is
\begin{eqnarray}
\chi(\zeta_{\bf n})
& = &
\sum_{\hat P} (\pm 1)^{p}
\langle \zeta_{{\bf n}} | \zeta_{\hat P{\bf n}} \rangle
\nonumber \\ & = &
\sum_{\hat P} (\pm 1)^{p}
\delta_{\xi}({\bf n}-\hat P{\bf n})
\nonumber \\ & = &
\frac{\eta^2}{(2-\lambda)^2}
\sum_{\hat P} (\pm 1)^{p} \sum_{\bf m}
c_{\bf mn}\, c_{{\bf m}\hat P{\bf n}}^*
\nonumber \\ & = &
\sum_{\hat P} (\pm 1)^{p} \sum_{\bf m}
\langle \zeta_{\bf n} | \phi_{\bf m} \rangle \,
\langle \phi_{\bf m} | \zeta_{\hat P{\bf n}} \rangle .
\end{eqnarray}
The fourth equality in fact follows directly from the first
using the complete nature of the entropy microstates.
It appears that this overlap factor depends upon the width of the wave packet.

These two expressions for the partition function must be equal.
One has
\begin{eqnarray}
\lefteqn{
\frac{\lambda(2-\lambda)^2}{\eta^2 N!}
\sum_{\bf n} \chi(\zeta_{\bf n}) e^{-\beta {\cal H}_{\bf n}}
}\nonumber \\
& = &
\frac{\lambda(2-\lambda)^2}{\eta^2 N!} \sum_{\bf n}
\sum_{\hat P} (\pm 1)^{p} \sum_{\bf m}
\langle \zeta_{\bf n} | \phi_{\bf m} \rangle \,
\langle \phi_{\bf m} | \zeta_{\hat P{\bf n}} \rangle
e^{-\beta {\cal H}_{\bf n}}
\nonumber \\ & \approx &
\frac{\lambda(2-\lambda)^2}{\eta^2 N!} \sum_{\bf n}
\sum_{\hat P} (\pm 1)^{p} \sum_{\bf m}
\langle \phi_{\bf m} | \zeta_{\hat P{\bf n}} \rangle \,
\langle e^{-\beta \hat {\cal H}} \zeta_{\bf n} | \phi_{\bf m} \rangle
\nonumber \\ & = &
\frac{\lambda(2-\lambda)^2}{\eta^2 N!} \sum_{\bf n}
\sum_{\hat P} (\pm 1)^{p} \sum_{\bf m}
\langle \phi_{\hat P{\bf m}} | \zeta_{\bf n} \rangle \,
\langle \zeta_{\bf n} |  e^{-\beta \hat {\cal H}}\phi_{\bf m} \rangle
\nonumber \\ & = &
\frac{1}{N!} \sum_{\bf m}
\sum_{\hat P} (\pm 1)^{p}
\langle \phi_{\hat P{\bf m}} |  e^{-\beta \hat {\cal H}}\phi_{\bf m} \rangle
\nonumber \\ & = &
\frac{1}{N!} \sum_{\bf m}
\sum_{\hat P} (\pm 1)^{p}
\langle \phi_{\hat P{\bf m}} | \phi_{\bf m} \rangle
e^{-\beta {\cal H}_{\bf m}}
\nonumber \\ & = &
\frac{1}{N!} \sum_{\bf m} \chi(\phi_{\bf m}) e^{-\beta {\cal H}_{\bf m}},
\;\mbox{ QED}.
\end{eqnarray}
%\emph{quod erat demonstrandum}.
The third equality follows from the fact that the Hamiltonian operator
is Hermitian.
The approximation embodied in the second equality is that
the non-orthogonal basis functions are approximately entropy eigenfunctions,
$ \hat{\cal H}({\bf r}) \zeta_{\bf n}({\bf r})
\approx {\cal H}_{\bf n} \zeta_{\bf n}({\bf r}) $.
The analysis in this subsection holds for a general set of non-orthogonal
entropy eigenfunctions, not just the set of entropy eigenfunctions
based on wave packets.

%\newpage %$\;$ \newpage
%%%%%%%%%%%%%%%%%%%%%%%%%%%%%%%%%%%%%%%%%%%%%%%%%%%%%%%%%%%%%%%%%%%%%%%%%
%
\section{Position Localization and Partition Function Expansion}
\label{Sec:Local-Posn}
%
%%%%%%%%%%%%%%%%%%%%%%%%%%%%%%%%%%%%%%%%%%%%%%%%%%%%%%%%%%%%%%%%%%%%%%%%%

One can look at localization
in the position representation of the entropy eigenfunction,
largely following Pathria, \S 5.5.\cite{Pathria72}
This reveals certain physical properties
for wave function symmetrization,
which are used in \S \ref{Sec:Local-n-r}.
Of interest is the problem raised in  \S \ref{Sec:Zb},
with the naive expansion of the partition function
based upon symmetrization.
Actually resolving this problem was the original motivation for this paper,
and it lead directly to the re-summation in \S \ref{Sec:Zcl+Z1-0}.
The localization of the eigenfunction
segues into the formulation of the configuration probability density
in terms of clusters,
and the virial expansion based on cluster integrals, \S \ref{Sec:Zb}.

%%%%%%%%%%%%%%%%%%%%%%%%%%%%%%%%%%%%%%%%%%%%%%%%%%%%%%%%%%%%%%%
\subsection{Localization}

The un-normalized probability density in the position representation
for a system of $N$ particles
may be written as
\begin{eqnarray}
W_N({\bf r};{\bf r}')
& = &
N! \Lambda^{3N}
\langle {\bf r} | e^{- \beta \hat{\cal H}} | {\bf r}' \rangle
 \\  & = &
N! \Lambda^{3N}
\langle {\bf r} | e^{- \beta \hat{\cal H}}
\sum_{\bf n}\!'\, | \zeta_{\bf n}^\mathrm{S/A}   \rangle\,
\langle \zeta_{\bf n}^\mathrm{S/A}
|  {\bf r}' \rangle
\nonumber \\ & = &
N! \Lambda^{3N}
\sum_{\bf n}\!'\, e^{- \beta {\cal H}_{\bf n}}
\langle {\bf r} | \zeta_{\bf n}^\mathrm{S/A}   \rangle\,
\langle \zeta_{\bf n}^\mathrm{S/A} |  {\bf r}' \rangle
\nonumber \\ & = &
N! \Lambda^{3N}
\sum_{\bf n}\!'\, e^{- \beta {\cal H}_{\bf n}}
\zeta_{\bf n}^\mathrm{S/A}({\bf r}) \, \zeta_{\bf n}^\mathrm{S/A}({\bf r}')^* .
\nonumber
\end{eqnarray}
The position ket $|{\bf r}\rangle$
can be thought of as a Dirac $\delta$-function, $\delta({\bf r}-{\bf s})$,
where ${\bf s}$ is the inner product position index,
for example
$\langle \zeta_{\bf n}^\mathrm{S/A} |  {\bf r}' \rangle
= \int \mathrm{d}{\bf s} \, \zeta_{\bf n}^\mathrm{S/A}({\bf s})^*
\delta({\bf r}'-{\bf s}) = \zeta_{\bf n}^\mathrm{S/A}({\bf r}')^*$.
Here it is assumed that the symmetrized entropy eigenfunctions
are complete, $\sum_{\bf n}'  | \zeta_{\bf n}^\mathrm{S/A}   \rangle\,
\langle \zeta_{\bf n}^\mathrm{S/A} | = \hat{\mathrm I}$,
where the sum over distinct states is
$\sum_{\bf n}' = \sum_{\bf n} \chi_{\bf n}/N!$.
(This restriction to distinct states differentiates
the present treatment from that of Pathria, \S 9.6.)\cite{Pathria72}
The prefactor $N! \Lambda^{3N}$ is used to ensure certain
scaling and asymptotic factorization behavior that is discussed below.
The off-diagonal elements ${\bf r}' \ne {\bf r}$
have no meaning in classical probability.
The diagonal elements,
$W_N({\bf r}) \equiv W_N({\bf r};{\bf r})$,
are proportional to the probability of the $N$ particles being at ${\bf r}$.

It is important to note that this is not a reduced probability density,
which is common in classical statistical mechanics,
but rather the configuration probability density
for all the particles in the system.

For an ideal gas
the  energy eigenvalues are
\begin{equation}
E_K \equiv \frac{\hbar^2 K^2}{2m}
= \frac{\hbar^2 }{2m} \sum_{j=1}^N k_j^2.
\end{equation}
The unsymmetrized entropy (energy) eigenfunctions are the product
of single particle functions,
\begin{equation}
\phi_{{\bf n}}({\bf r}) =
\prod_{j=1}^N
\phi_{{\bf n}_j}({\bf r}_j) ,
\end{equation}
with
\begin{equation}
\phi_{{\bf n}_j}({\bf r}_j) =
\frac{1}{V^{1/2}} e^{i {\bf k}_j\cdot{\bf r}_j },
\;\; {\bf k}_j =  \frac{2 \pi}{V^{1/3}}  {\bf n}_j ,
\end{equation}
with $ n_{j\alpha} = 0, \pm 1, \pm 2 , \ldots$, and $\alpha = x,y,z$.
The symmetrized wave function is
\begin{equation}
\phi_{\bf n}^\mathrm{S/A}({\bf r})
= \frac{1}{\sqrt{N!\chi_{\bf n}}} \sum_{\hat{\mathrm P}} (\pm 1)^p
\phi_{{\bf n}}(\hat{\mathrm P}{\bf r}) .
\end{equation}
Here has been included the overlap factor $\chi_{\bf n}$,
which is unity unless one or more of the single particle states
in the microstate ${\bf n}$
are multiply occupied,
(ie.\ any of the ${\bf n}_j$ are equal).
%In the continuum limit, which is next invoked,
%multiply occupied states form a set of measure zero,
%which can be neglected by setting the overlap factor to unity.

Modifying slightly Pathria, Eq.~(5.5.12),\cite{Pathria72}
the configuration probability density
for the ideal gas may be written as
\begin{eqnarray}
\lefteqn{
W_N^\mathrm{id}({\bf r};{\bf r}')
} \nonumber \\
& = &
%N! \Lambda^{3N}
%\langle {\bf r} | e^{- \beta \hat{\cal H}^\mathrm{id}}  | {\bf r}' \rangle
%\nonumber \\ & = &
N! \Lambda^{3N} \sum_{{\bf n}}\!'
e^{-\beta \hbar^2 K^2/2m}
\phi_{\bf n}^\mathrm{S/A}({\bf r})
\phi_{\bf n}^\mathrm{S/A}({\bf r}')^*
\nonumber \\ & = &
\Lambda^{3N} \sum_{{\bf n}}\!' \frac{1}{\chi_{\bf n}}
e^{-\beta \hbar^2 K^2/2m}
\nonumber \\ && \mbox{ } \times
\sum_{\hat{\mathrm P},\hat{\mathrm P}'} (\pm 1)^{p+p'}
\phi_{{\bf n}}(\hat{\mathrm P}{\bf r})
\phi_{{\bf n}}(\hat{\mathrm P}'{\bf r}')^*
\nonumber \\ & = &
\Lambda^{3N}  \sum_{\bf n} e^{-\beta \hbar^2 K^2/2m}
\sum_{\hat{\mathrm P}} (\pm 1)^{p}
\phi_{{\bf n}}(\hat{\mathrm P}{\bf r})
\phi_{{\bf n}}({\bf r}')^*
\nonumber \\ & = &
\frac{ \Lambda^{3N} V^N}{(2\pi)^{3N}}
\sum_{\hat{\mathrm P}} (\pm 1)^{p}
%\nonumber \\ && \mbox{ } \times
\int \mathrm{d}{\bf k}
 e^{-\beta \hbar^2 K^2/2m}
\phi_{{\bf n}}(\hat{\mathrm P}{\bf r})
\phi_{{\bf n}}({\bf r}')^*
\nonumber \\ & = &
\frac{\Lambda^{3N} V^N}{(2\pi)^{3N}}
\sum_{\hat{\mathrm P}} (\pm 1)^{p}
\prod_{j=1}^N
\int \mathrm{d}{\bf k}_j
 e^{-\beta \hbar^2 k_j^2/2m}
 e^{i {\bf k}_j \cdot  [{\bf r}_{\hat{\mathrm P}j}-{\bf r}_j'] }
\nonumber \\ & = &
V^N
\sum_{\hat{\mathrm P}} (\pm 1)^{p}
\prod_{j=1}^N f({\bf r}_{\hat{\mathrm P}j}-{\bf r}_j').
\end{eqnarray}
The thermal wave length is $\Lambda = \sqrt{ 2 \pi \hbar^2/m k_\mathrm{B}T }$.
In the final equality appears the Gaussian
\begin{equation}
f({\bf r}_j) = e^{ -\pi r_j^2/\Lambda^2} .
\end{equation}

The diagonal element of the probability density is
\begin{eqnarray} \label{Eq:Wid-diag}
%\lefteqn{
W_N^\mathrm{id}({\bf r})
%} \nonumber \\
& = &
V^N
\sum_{\hat{\mathrm P}} (\pm 1)^{p}
\prod_{j=1}^N f({\bf r}_{\hat{\mathrm P}j}-{\bf r}_j).
\end{eqnarray}
For the sum over permutations,
the leading term comes from the identity permutation, $\hat{\mathrm I}$,
in which case $f({\bf 0}) = 1$.
The next term comes from a single transposition
of particles $j$ and $k$, $\hat{\mathrm P}_{jk}$,
which gives $\pm f({\bf r}_{jk}) f({\bf r}_{kj})$, and so on.
Hence
\begin{eqnarray}
\sum_{\hat{\mathrm P}} (\pm 1)^{p}
\lefteqn{
\prod_{j=1}^N f({\bf r}_{\hat{\mathrm P}j}-{\bf r}_j)
} \nonumber \\
& = &
1 \pm \sum_{j<k}^N f({\bf r}_{jk}) f({\bf r}_{kj}) + \ldots
\end{eqnarray}
The function $f(r_{jk})$ vanishes when $r_{kj} \gg \Lambda$.
Hence in the low density, high temperature limit,  $\rho \Lambda^3 \ll 1$,
the correction due to quantum symmetrization is negligible.
This decay of symmetrization effects with distance
is used to illustrate the localization argument in  \S \ref{Sec:Local-n-r}.

%%%%%%%%%%%%%%%%%%%%%%%%%%%%%%%%%%%%%%%%%%%%%%%%%%%%%%%%%%%%%%%
\subsection{Cluster Expansion of the Partition Function} \label{Sec:Zb}

%%%%%%%%%%%%%%%%%%%%%%%%%%%%%%%%%%%
\subsubsection{Partition Function}

One might naively attempt to use these results to evaluate
the partition function
by writing
(cf.\ Eqs~(5.5.17) and (5.5.19) of Pathria)\cite{Pathria72}
\begin{eqnarray}
Z_N & = &
\mbox{TR } e^{-\beta \hat{\cal H}}
\nonumber \\ & = &
\frac{1}{V^N} \int {\mathrm d}{\bf r} \;
\langle {\bf r} | e^{- \beta \hat{\cal H}} | {\bf r} \rangle
\nonumber \\ & = &
\frac{V^N \Lambda^{-3N}}{N!}
\pm
\frac{\Lambda^{-3N}}{N!} \int {\mathrm d}{\bf r} \;
\sum_{j<k}^N f({\bf r}_{jk}) f({\bf r}_{kj})
+ \ldots
\nonumber \\ & = &
\frac{V^N \Lambda^{-3N}}{N!}
\pm
\frac{\Lambda^{-3N}}{N!} \frac{N(N-1)}{2} V^{N-2}
\nonumber \\ & & \mbox{ } \times
 \int {\mathrm d}{\bf r}_1 \, {\mathrm d}{\bf r}_2 \;
 e^{ -2 \pi r_{12}^2/\Lambda^2}
+ \ldots
\nonumber \\ & = &
\frac{V^N \Lambda^{-3N}}{N!}
\pm
\frac{\Lambda^{-3N}}{N!} \frac{N^2V^{N-1}}{2}
\left(\frac{2\pi\Lambda^2}{4\pi}\right)^{3/2}
\nonumber \\ & & \mbox{ }
+ \ldots
\end{eqnarray}
(The partition function denoted $Z$ here and throughout
is denoted $Q$ by Pathria.
The configuration integral denoted $Q$ by the present author
is denoted $Z$ by Pathria.)\cite{Pathria72}
The problem with this is the way that
successive terms scale with volume.
It is not possible to continue this  form
to obtain a meaningful expansion of the partition function
that can be terminated after a finite number of terms.
An infinite re-summation has to be carried out in order to secure
the correct extensivity of the Helmholtz free energy
or of the grand potential.
This is done in \S \ref{Sec:Zcl+Z1-0}.

An alternative to the loop expansion of \S \ref{Sec:Zcl+Z1-0}
is the virial expansion,
which takes into account the clustering properties of the probability density.
Following Pathria, \S 9.6\cite{Pathria72}
(except that here only distinct states are included),
the canonical partition function can be written in terms
of the configuration probability density as
\begin{eqnarray}
%\lefteqn{
Z_N
%(N,V,T)
% } \nonumber \\
& = &
\mbox{TR } e^{-\beta \hat{\cal H}}
\nonumber \\ & = &
\sum_{\bf n}\!'\, e^{- \beta {\cal H}_{\bf n}}
\nonumber \\ & = &
\sum_{\bf n} \frac{\chi_{\bf n}}{N!}
 e^{- \beta {\cal H}_{\bf n}}
\int_V \mathrm{d}{\bf r} \;
\zeta_{\bf n}^\mathrm{S/A}({\bf r})^* \, \zeta_{\bf n}^\mathrm{S/A}({\bf r})
\nonumber \\ & = &
\frac{1}{N! \Lambda^{3N}}
\int_V \mathrm{d}{\bf r} \;
W_N({\bf r}) .
\end{eqnarray}
Hence $\wp({\bf r}) \equiv W_N({\bf r})/N! \Lambda^{3N}Z_N$
is the configuration probability density that is normalized to unity.

For a system consisting of a single particle, $N=1$,
which is therefore ideal,
Eq.~(\ref{Eq:Wid-diag}) gives the configuration probability density
matrix as
\begin{equation}
W_1({\bf r}'_1 ;{\bf r}_1 )
=
e^{-\pi({\bf r}_1'-{\bf r}_1)^2/\Lambda^2} .
\end{equation}
One sees now that the scale factor $N! \Lambda^{3N}$
ensures that the diagonal element is unity,
$ W_1({\bf r}_1 )=1$.
The partition function for a single particle is
\begin{equation}
Z_1 =
\frac{1}{\Lambda^{3}} \int_V \mathrm{d}{\bf r} \; W_1({\bf r})
=  V \Lambda^{-3}.
\end{equation}

%%%%%%%%%%%%%%%%%%%%%%%%%%%%%%%%%%%
\subsubsection{Cluster Localization}

Now comes the very important clustering property.
In classical statistical mechanics,
groups of particles that are far-separated have no influence
on each other and they are therefore uncorrelated.
One expects the same situation to hold in  classical statistical mechanics,
particularly since the effects of wave function
symmetrization with respect to particle interchange
are limited to a range on the order of the thermal wave length $\Lambda$,
Eq.~(\ref{Eq:Wid-diag}).

For the general quantum case of $N$ interacting particles,
suppose that the configuration divides into two far separated groups,
${\bf r}^N = \{ {\bf r}^A , {\bf r}^B \}$,
with $r_{jk} \gg \Lambda'$ if ${\bf r}_j \in A$ and ${\bf r}_k \in B$.
Here $\Lambda'$ is the larger of the thermal wave length
and the range of the inter-particle potential.
In this case one expects that
\begin{equation}
W_N({\bf r}^N) = W_A({\bf r}^A)\,W_B({\bf r}^B)  .
\end{equation}
Hence for two particles
\begin{equation}
W_2({\bf r}_1,{\bf r}_2) \sim  W_1({\bf r}_1 )\,W_1({\bf r}_2)
= 1
, \;\; r_{12} \rightarrow \infty.
\end{equation}
(Note the notational difference between these diagonal elements
and the off-diagonal elements,
which would be written as $W_2({\bf r}_1',{\bf r}_2'; {\bf r}_1,{\bf r}_2)$,
or more simply as $W_2(1',2'; 1,2)$.)

%%%%%%%%%%%%%%%%%%%%%%%%%%%%%%%%%%%%%%%%%%%%%%%%%%%
\subsubsection{Two Ideal Particles}

One can illustrate this factorization property explicitly
for two ideal particles.
Since there are no inter-particle interactions,
the factorization should be exact  in this case
for separations much greater than the thermal wave length.

For two non-interacting particles,
the un-symmetrized entropy eigenfunction is
\begin{equation}
\zeta_{\bf n}({\bf r})
=
\frac{1}{V} e^{i {\bf k}_1 \cdot {\bf r}_1} e^{i {\bf k}_2 \cdot {\bf r}_2}  ,
\end{equation}
with $ k_{j\alpha} = 2\pi n_{j\alpha} /V^{1/3}$,
$j=1,2$, $\alpha = x,y,z$, and $n_{j\alpha} = 0, \pm 1, \pm 2,\ldots$.
The symmetrized  entropy eigenfunction is
\begin{equation}
\zeta_{\bf n}^\mathrm{S/A}({\bf r})
=
\frac{1}{V\sqrt{2! \chi_{\bf n}}}
\left[
 e^{i {\bf k}_1 \cdot {\bf r}_1} e^{i {\bf k}_2 \cdot {\bf r}_2}
 \pm
 e^{i {\bf k}_1 \cdot {\bf r}_2} e^{i {\bf k}_2 \cdot {\bf r}_1}
\right] .
\end{equation}
The overlap factor is
\begin{eqnarray}
 \chi_{\bf n} & = &
 1 \pm \langle \zeta_{\bf n}({\bf r}_2,{\bf r}_1)|
 \zeta_{\bf n}({\bf r}_1,{\bf r}_2) \rangle
 \nonumber \\ & = &
 1 \pm \frac{1}{V^2} \int {\mathrm d}{\bf r}_1 \, {\mathrm d}{\bf r}_2 \;
e^{-i {\bf k}_1 \cdot {\bf r}_2} e^{-i {\bf k}_2 \cdot {\bf r}_1}
e^{i {\bf k}_1 \cdot {\bf r}_1} e^{i {\bf k}_2 \cdot {\bf r}_2}
 \nonumber \\ & = &
 1 \pm \delta_{{\bf n}_{12}} ,
\end{eqnarray}
where a Kronecker-$\delta$ appears.

With these the canonical partition function is
\begin{eqnarray}
Z_2^\mathrm{id}  & = &
\sum_{{\bf n}_2 \ge {\bf n}_1}
e^{-\beta {\cal H}_{\bf n}}
 \nonumber \\ & = &
\frac{1}{2!} \sum_{\bf n} \chi_{\bf n}
e^{- \beta\hbar^2 (k_1^2 + k_2^2)/2m}
 \nonumber \\ & = &
\frac{1}{2!} \sum_{{\bf n}_1,  {\bf n}_2}
e^{- \beta\hbar^2 (k_1^2 + k_2^2)/2m}
\pm \frac{1}{2!} \sum_{{\bf n}_1}
e^{- 2\beta\hbar^2 k_1^2 /2m}
\nonumber \\ & = &
\frac{1}{2!} \frac{V^2}{(2\pi)^6}
\int  {\mathrm d}{\bf k}_1 \, {\mathrm d}{\bf k}_2 \;
e^{- \beta\hbar^2 (k_1^2 + k_2^2)/2m}
\nonumber \\ &  & \mbox{ }
\pm \frac{1}{2!}  \frac{V}{(2\pi)^3}
\int  {\mathrm d}{\bf k}_1 \;
e^{- 2\beta\hbar^2 k_1^2 /2m}
\nonumber \\ & = &
\frac{1}{2!} \frac{V^2}{(2\pi)^6}
\left(\frac{2 \pi m  }{\beta \hbar^2}\right)^{3}
\pm \frac{1}{2!}  \frac{V}{(2\pi)^3}
\left(\frac{2 \pi m  }{\beta \hbar^2}\right)^{3/2}
\nonumber \\ & = &
\frac{V^2}{2\Lambda^6}
\pm \frac{V}{2\Lambda^3} .
\end{eqnarray}

The two-particle ideal configuration probability density is,
with ${\bf r} \equiv \{ {\bf r}_1, {\bf r}_2\}$,
\begin{eqnarray} \label{Eq:W2id}
\lefteqn{
W_2^\mathrm{id}({\bf r},{\bf r}')
} \nonumber \\ & = &
2! \Lambda^{6}
\sum_{\bf n}\!'\, e^{- \beta {\cal H}_{\bf n}^\mathrm{id}}
\zeta_{\bf n}^\mathrm{S/A}({\bf r}) \, \zeta_{\bf n}^\mathrm{S/A}({\bf r}')^*
\nonumber \\ & = &
2! \Lambda^{6} \sum_{\bf n} \frac{\chi_{\bf n}}{2!}
 e^{- \beta {\cal H}_{\bf n}}
\zeta_{\bf n}^\mathrm{S/A}({\bf r}) \, \zeta_{\bf n}^\mathrm{S/A}({\bf r}')^*
\nonumber \\ & = &
\frac{2 \Lambda^{6} }{2} \frac{1}{2V^2} \sum_{\bf n}
e^{-\beta {\cal H}_{\bf n}}
\left[
 e^{i {\bf k}_1 \cdot {\bf r}_1} e^{i {\bf k}_2 \cdot {\bf r}_2}
 \pm
 e^{i {\bf k}_1 \cdot {\bf r}_2} e^{i {\bf k}_2 \cdot {\bf r}_1}
\right]
\nonumber \\ && \mbox{ } \times
\left[
 e^{-i {\bf k}_1 \cdot {\bf r}_1'} e^{-i {\bf k}_2 \cdot {\bf r}_2'}
 \pm
 e^{-i {\bf k}_1 \cdot {\bf r}_2'} e^{-i {\bf k}_2 \cdot {\bf r}_1'}
\right]
\nonumber \\ & = &
\frac{\Lambda^{6}}{2V^2} \sum_{{\bf n}_1,{\bf n}_2}
e^{-\beta {\cal H}_{\bf n}}
\nonumber \\ && \mbox{ } \times
\left[
 e^{i {\bf k}_1 \cdot ( {\bf r}_1-{\bf r}_1') }
 e^{i {\bf k}_2 \cdot ({\bf r}_2-{\bf r}_2') }
%\right. \nonumber \\ && \left.  \mbox{ }
+
 e^{i {\bf k}_1 \cdot ( {\bf r}_2-{\bf r}_2') }
 e^{i {\bf k}_2 \cdot ({\bf r}_1-{\bf r}_1') }
\right. \nonumber \\ && \left.  \mbox{ }
\pm
 e^{i {\bf k}_1 \cdot ( {\bf r}_1-{\bf r}_2') }
 e^{i {\bf k}_2 \cdot ({\bf r}_2-{\bf r}_1') }
%\right. \nonumber \\ && \left.  \mbox{ }
 \pm
 e^{i {\bf k}_1 \cdot ( {\bf r}_2-{\bf r}_1') }
 e^{i {\bf k}_2 \cdot ({\bf r}_1-{\bf r}_2') }
\right]
\nonumber \\ & = &
\frac{\Lambda^{6}}{2 \Lambda^6}
\left\{
2 e^{- \pi\left( {\bf r}_1-{\bf r}_1' \right)^2/\Lambda^2 }
e^{- \pi\left( {\bf r}_2-{\bf r}_2' \right)^2/\Lambda^2 }
\right. \nonumber \\ && \left. \mbox{ }
\pm
2 e^{- \pi\left( {\bf r}_1-{\bf r}_2' \right)^2/\Lambda^2 }
e^{- \pi\left( {\bf r}_2-{\bf r}_1' \right)^2/\Lambda^2 }
\right\} .
\end{eqnarray}
The first term in the penultimate equality is
\begin{eqnarray}
\lefteqn{
\frac{ \Lambda^{6} }{2V^2} \sum_{{\bf n}_1,{\bf n}_2}
e^{-\beta {\cal H}_{\bf n}}
 e^{i {\bf k}_1 \cdot ( {\bf r}_1-{\bf r}_1') }
 e^{i {\bf k}_2 \cdot ({\bf r}_2-{\bf r}_2') }
} \nonumber \\
& = &
\frac{ \Lambda^{6} }{2V^2} \frac{V^2}{(2\pi)^6}
\int  {\mathrm d}{\bf k}_1 \, {\mathrm d}{\bf k}_2 \;
e^{- \beta\hbar^2 (k_1^2 + k_2^2)/2m}
\nonumber \\ && \mbox{ } \times
 e^{i {\bf k}_1 \cdot ( {\bf r}_1-{\bf r}_1') }
 e^{i {\bf k}_2 \cdot ({\bf r}_2-{\bf r}_2') }
 \nonumber \\ & = &
\frac{\Lambda^{6}}{2(2\pi)^6}
\int  {\mathrm d}{\bf k}_1  \;
e^{- \beta\hbar^2 \left[{\bf k}_1
+ i m( {\bf r}_1-{\bf r}_1')/\beta\hbar^2 \right]^2/2m}
\nonumber \\ && \mbox{ } \times
\int  {\mathrm d}{\bf k}_2  \;
e^{- \beta\hbar^2 \left[{\bf k}_2
+ i m( {\bf r}_2-{\bf r}_2')/\beta\hbar^2 \right]^2/2m}
\nonumber \\ && \mbox{ } \times
e^{- m\left( {\bf r}_1-{\bf r}_1' \right)^2/2\beta\hbar^2 }
e^{- m\left( {\bf r}_2-{\bf r}_2' \right)^2/2\beta\hbar^2 }
 \nonumber \\ & = &
\frac{\Lambda^{6}}{2(2\pi)^6}
\frac{(2\pi)^3 m^3 }{(\beta\hbar^2)^3}
e^{- \pi\left( {\bf r}_1-{\bf r}_1' \right)^2/\Lambda^2 }
e^{- \pi\left( {\bf r}_2-{\bf r}_2' \right)^2/\Lambda^2 }
 \nonumber \\ & = &
 \frac{1}{2}
e^{- \pi\left( {\bf r}_1-{\bf r}_1' \right)^2/\Lambda^2 }
e^{- \pi\left( {\bf r}_2-{\bf r}_2' \right)^2/\Lambda^2 } .
\end{eqnarray}
The remaining three terms give a similar result
with the subscripts interchanged.
Adding them together gives the final equality.

For the near-diagonal terms,
$ {\bf r}_1 \approx {\bf r}_1'$ and $ {\bf r}_2 \approx {\bf r}_2'$,
in the asymptotic limit,
$r_{12} \gg \Lambda$,
the second term in the final equality in Eq.~(\ref{Eq:W2id}) goes to zero.
This leaves the configuration probability density as
\begin{eqnarray}
W_2^\mathrm{id}({\bf r};{\bf r}')
& \sim &
e^{- \pi\left( {\bf r}_1-{\bf r}_1' \right)^2/\Lambda^2 }
e^{- \pi\left( {\bf r}_2-{\bf r}_2' \right)^2/\Lambda^2 }
, \; \; r_{12} \gg \Lambda
\nonumber \\ &=&
W_1({\bf r}_1,{\bf r}_1' )W_1({\bf r}_2,{\bf r}_2') .
\end{eqnarray}
One sees that for these ideal particles the factorization is exact.
One notes that this exact factorization
depends upon including $N!$ in the scale factor
in the definition of the configuration probability density.

This result for the ideal gas holds when $r_{12} \gg \Lambda$.
For interacting particles one expects the same result
when the particles are beyond the range of the inter-particle potential.

%%%%%%%%%%%%%%%%%%%%%%%%%%%%%%%%%%%%%%%%%%%%%%%%%%%
\subsubsection{Ursell Cluster Functions}

Evidently then one can define the Ursell or cluster functions
\cite{Ursell27,Kahn38}
that asymptote to zero as any one of the particles
become far-separated from the rest
(see Pathria, \S 9.6).\cite{Pathria72}
These are the analogue of the total correlation function
that occurs in classical statistical mechanics.\cite{TDSM}
The idea is that by exhibiting the asymptote explicitly,
what remains must be short-ranged.
Writing $j \equiv {\bf r}_j$, one has
\begin{eqnarray}
W_1(1';1) & = & U_1(1';1) ,
 \\
W_2(1',2';1,2) & = &
U_2(1',2';1,2)
+ U_1(1';1) U_1(2';2),
\nonumber \\
W_3(1',2',3';1,2,3) & = &
U_3(1',2',3';1,2,3)
\nonumber \\ && \mbox{ }
+ U_2(2',3';2,3) U_1(1';1)
\nonumber \\ && \mbox{ }
+ U_2(3',1';3,1) U_1(2';2)
\nonumber \\ && \mbox{ }
+ U_2(1',2';1,2) U_1(3';3)
\nonumber \\ && \mbox{ }
+ U_1(1';1) U_1(2';2) U_1(3';3) .
\nonumber
\end{eqnarray}
The general formula for $N$ particles
(for the diagonal elements) is
\begin{eqnarray}
\lefteqn{
W_N(1,\ldots,N)
} \nonumber \\
& = &
\sum_{\{m_l\}}\!'
\sum_{\hat P}
U_1(j_1) \ldots U_1(j_{m_1})
\nonumber \\ && \mbox{ } \times
U_2(j_{m_1+1},j_{m_1+2}) \ldots U_2(j_{m_1+2m_2-1},j_{m_1+2m_2})
\nonumber \\ && \mbox{ } \times
\ldots
\nonumber \\ & = &
\sum_{\{m_l\}}\!'
\sum_{\hat P}
U_1(\cdot)^{m_1} U_2(\cdot,\cdot)^{m_2} \ldots
\end{eqnarray}
The sum is over all sets $\sum_{l=1}^N l m_l = N$, $m_l = 0, 1,2,\ldots$.
The permutation sum is over the
$N!/\prod_l (l!)^{m_l} m_l!$
distinct arrangements of the particles for each set.
Explicitly the Ursell functions are
\begin{eqnarray}
U_1(1';1) & = & W_1(1';1) ,
 \\
U_2(1',2';1,2) & = &
W_2(1',2';1,2)
- W_1(1';1) W_1(2';2),
\nonumber \\
U_3(1',2',3';1,2,3) & = &
W_3(1',2',3';1,2,3)
\nonumber \\ && \mbox{ }
- W_2(2',3';2,3) W_1(1';1)
\nonumber \\ && \mbox{ }
- W_2(3',1';3,1) W_1(2';2)
\nonumber \\ && \mbox{ }
- W_2(1',2';1,2) W_1(3';3)
\nonumber \\ && \mbox{ }
+ 2 W_1(1';1) W_1(2';2) W_1(3';3) .
\nonumber
\end{eqnarray}
Since the left hand side must be short-ranged,
the coefficients on the right hand side must sum to zero in each case.

Using the diagonal elements of the Ursell cluster functions,
the cluster integrals are defined as
\begin{equation}
b_l \equiv \frac{1}{l! \Lambda^{3(l-1)}V}
\int \mathrm{d}{\bf r}^l\;
U_l({\bf r}^l) .
\end{equation}
Because the argument is short-ranged,
this is independent of volume in the limit $V \rightarrow \infty$.

In view of the formulation of the configuration probability density
in  terms of cluster functions,
the canonical partition function can be re-written as
\begin{eqnarray}
\lefteqn{
Z(N,V,T)
} \nonumber \\
& = &
\frac{1}{N! \Lambda^{3N}}
\int_V \mathrm{d}{\bf r} \;
W_N({\bf r})
\nonumber \\ & = &
\frac{1}{N! \Lambda^{3N}}
\int {\mathrm d}{\bf r}^N \;
\sum_{\{m_l\}}\!'
\sum_{\hat P}
U_1(\cdot)^{m_1} U_2(\cdot,\cdot)^{m_2} \ldots
\nonumber \\ & = &
\frac{1}{N! \Lambda^{3N}}
\sum_{\{m_l\}}\!'
\frac{N!}{\prod_l (l!)^{m_l} m_l!}
\int {\mathrm d}{\bf r}^N \;
U_1(1)\ldots U_1(m_1)
\nonumber \\ && \mbox{ } \times
U_2(m_1+1,m_1+2)  \ldots  U_2(m_1+m_2-1,m_1+m_2)
\nonumber \\ && \mbox{ } \times
U_3(m_1+m_2+1,m_1+m_2+2,m_1+m_2+3)
\nonumber \\ && \mbox{ } \times
\ldots
\nonumber \\ & = &
\frac{1}{N! \Lambda^{3N}}
\sum_{\{m_l\}}\!'
\frac{N!}{\prod_l (l!)^{m_l} m_l!}
\prod_{l=1}^N \left[ l! \Lambda^{3(l-1)} V b_l \right]^{m_l}
\nonumber \\ & = &
\sum_{\{m_l\}}\!' \prod_{l=1}^N
\frac{1}{m_l!}
 \left[ \Lambda^{-3} V b_l \right]^{m_l} .
\end{eqnarray}
This is now in the form of classical cluster theory,
as originally enunciated by Mayer.\cite{Mayer37}
In particular,
the grand partition function is
\begin{eqnarray}
%\lefteqn{
\Xi(\mu,V,T)
%} \nonumber \\
& = &
\sum_{N=0}^\infty  z^N Z(N,V,T)
\nonumber \\ & = &
\prod_{l=1}^\infty \sum_{m_l=0}^\infty
\frac{1}{m_l!}
 \left[ \Lambda^{-3} V z^l b_l \right]^{m_l}
\nonumber \\ & = &
\prod_{l=1}^\infty \exp  \left[ \Lambda^{-3} V z^l b_l \right] .
\end{eqnarray}
Hence the grand potential is
\begin{equation}
\Omega(\mu,V,T)
=
- k_\mathrm{B}T \sum_{l=1}^\infty \Lambda^{-3} V z^l b_l ,
\end{equation}
and the dimensionless pressure is
\begin{equation}
\beta p \Lambda^{3}
= \frac{ - \beta \Omega \Lambda^{3}}{V}
=
\sum_{l=1}^\infty   z^l b_l .
\end{equation}

Lee and Yang\cite{Lee59} developed a binary collision method
for evaluating the cluster integrals.
Each is expressed as a sum of multi-dimensional temperature integrals,
with the integrands  being an infinite sum of products
of derivative operators and binary kernels
(see Pathria, \S 9.7).\cite{Pathria72}
The convergence properties of the expansions are not well understood.
The method is far more complicated than
is required to evaluate the corresponding classical cluster integrals.
\cite{TDSM}
This is not encouraging, given the fact that
the classical virial expansion itself is not a feasible approach
to evaluating the properties of classical condensed matter systems.
For these three reasons
there appears to be little motivation to attempt
to implement the method of Lee and Yang for quantum condensed matter systems.
Superficially at least, the evaluation of the permutation loop grand potentials
given in \S  \ref{Sec:Zcl+Z1-0} appears more straightforward
than the  Lee-Yang method for evaluating the quantum cluster integrals.
It also appears that the permutation loop expansion itself
will be more rapidly converging
for terrestrial quantum condensed matter than the quantum virial expansion.
Absent an explicit demonstration,
such speculation remains just that.

%\newpage
%%%%%%%%%%%%%%%%%%%%%%%%%%%%%%%%%%%%%%%%%%%%%%%%%%%%%%%%%%%%%%%%%%%%%%%%%
%
\section{Wave Packet Expectation of the Square of the Energy Operator}
\label{Sec:<H2>n}
%
%%%%%%%%%%%%%%%%%%%%%%%%%%%%%%%%%%%%%%%%%%%%%%%%%%%%%%%%%%%%%%%%%%%%%%%%%

In this appendix
the expectation value of the square of the Hamiltonian operator
for a wave packet is derived in detail.

For this one requires the general formula
$\nabla^2 BC = C \nabla^2 B + B \nabla^2 C
+ 2 \nabla B \cdot \nabla C$.
Writing the exponent of the wave packet as $A$,
with $\nabla^3 A  = 0$, one has
\begin{eqnarray} \label{Eq:H^2e^A}
\lefteqn{
[ \nabla^2 + U] \,
[ \nabla^2 + U] \, e^A
}  \nonumber \\
& = &
[ \nabla \cdot \nabla  + U]
[ \nabla^2 A + \nabla A \cdot \nabla A  + U ]e^A
\nonumber  \\ & = &
[ U \nabla^2 A  + U \nabla A \cdot \nabla A + U^2 ] e^A
\nonumber  \\ &  & \mbox{ }
+ \nabla \cdot \left\{
[ 2 \nabla A \cdot \nabla \nabla A  + \nabla U ]
 e^A \right\}
\nonumber  \\ &  & \mbox{ }
+ \nabla \cdot \left\{
[  \nabla A \nabla^2 A + \nabla A \nabla A \cdot \nabla A  + U \nabla A  ]
 e^A \right\}
\nonumber  \\ & = &   %%%%%%%%%%%%%%%%%%%%%%%%%%%%%%%%%%%%%%%%
[ U \nabla^2 A  + U \nabla A \cdot \nabla A + U^2 ] e^A
\nonumber  \\ &  & \mbox{ }
+ [ 2 ( \nabla \nabla A ) :( \nabla \nabla A )
+ \nabla^2 U ] e^A
\nonumber  \\ &  & \mbox{ }
+ [ 2 \nabla A \cdot (\nabla \nabla A ) \cdot \nabla A
+ \nabla A \cdot \nabla U ] e^A
\nonumber  \\ &  & \mbox{ }
+ [  \nabla^2  A \nabla^2 A
+ \nabla^2 A \nabla A \cdot \nabla A
+ 2 \nabla A \cdot \nabla \nabla A \cdot \nabla A
\nonumber  \\ &  & \mbox{ }
+ \nabla A \cdot \nabla U
+ U \nabla^2 A ]
e^A
\nonumber  \\ &  & \mbox{ }
+ [ \nabla A \cdot \nabla A \nabla^2 A
+ (\nabla A \cdot \nabla A)^2
+  U \nabla A \cdot \nabla A ]
e^A
\nonumber  \\ & = &   %%%%%%%%%%%%%%%%%%%%%%%%%%%%%%%%%%%%%%%%
\left\{ \rule{0cm}{0.4cm}
U \nabla^2 A  + U \nabla A \cdot \nabla A + U^2
\right. \nonumber  \\ &  & \left. \mbox{ }
+  2 ( \nabla \nabla A ) :( \nabla \nabla A )
+ \nabla^2 U
\right. \nonumber  \\ &  & \left. \mbox{ }
+ 4 \nabla A \cdot (\nabla \nabla A ) \cdot \nabla A
+ 2 \nabla A \cdot \nabla U
\right. \nonumber  \\ &  & \left. \mbox{ }
+  \nabla^2  A \nabla^2 A
+ 2\nabla^2 A \nabla A \cdot \nabla A
+ U \nabla^2 A
\right. \nonumber  \\ &  & \left. \mbox{ }
+ (\nabla A \cdot \nabla A)^2
+  U \nabla A \cdot \nabla A
\rule{0cm}{0.4cm} \right\} e^A .
\end{eqnarray}
Obviously one has to multiply $\nabla^2$ by $-\hbar^2/2m$.
Since
$A \equiv -{\bm \varepsilon}_{{\bf n}}({\bf r})^2/ 4 \xi^2
 - {\bf p}_{{\bf n}} \cdot {\bm \varepsilon}_{{\bf n}}({\bf r})/{i\hbar} $,
one has
\begin{eqnarray}
\nabla A & = &
\frac{ -{\bm \varepsilon}_{{\bf n}}({\bf r}) }{ 2 \xi^2 }
 - \frac{ {\bf p}_{{\bf n}} }{i\hbar}
\end{eqnarray}
and
\begin{eqnarray}
\nabla \nabla A & = &
\frac{ -1 }{ 2 \xi^2 } {\bf I} .
\end{eqnarray}
One will also require the expectation value of the tetradic,
\begin{eqnarray}
\langle \varepsilon_a \varepsilon_b \varepsilon_c \varepsilon_d \rangle_{\bf n}
& = &
\xi^4 \left[ \delta_{ab}\delta_{cd}  + \delta_{ac}\delta_{bd}
+ \delta_{ad}\delta_{bc} \right]
\nonumber \\ && \mbox{ }
+ 3(3-1) \xi^4  \delta_{ab}\delta_{ac}\delta_{ad} .
\end{eqnarray}

Now the expectation values of the twelve terms
on the right hand side of the final equality in Eq.~(\ref{Eq:H^2e^A})
are obtained.
These will turn out to be of ${\cal O}(N^2)$ and ${\cal O}(N)$.
(The orders cited below do not include any $N$-dependence
of the wave packet width unless specifically stated.)
No terms are neglected in the following results.
It will prove necessary to retain the ${\cal O}(N)$  term
for the final result.

The expectation value of the first term  in Eq.~(\ref{Eq:H^2e^A}) gives
\begin{eqnarray}
\lefteqn{
\frac{-\hbar^2}{2m} \left\langle U \nabla^2 A \right\rangle_{\bf n}
}  \nonumber \\
& = &
\frac{-\hbar^2}{2m}
\left\langle
\left[ U_{\bf n} + {\bf U}_{\bf n}' \cdot {\bm \varepsilon}_{\bf n}
+ \frac{1}{2}{\bf U}_{\bf n}'' :
{\bm \varepsilon}_{\bf n} {\bm \varepsilon}_{\bf n}
\right]
\frac{-3N}{2\xi^2}
\right\rangle_{\bf n}
\nonumber \\ & = &
\frac{3N\hbar^2}{4m\xi^2}  U_{\bf n}
+ \frac{3N\hbar^2}{8m} \mbox{TR }{\bf U}_{\bf n}'' .
\end{eqnarray}
Note that here and below the potential is not expanded beyond second order.
These are  ${\cal O}(N^2)$.
The second term gives
\begin{eqnarray}
\lefteqn{
\frac{-\hbar^2}{2m}
\left\langle U \nabla A \cdot \nabla A \right\rangle_{\bf n}
}  \nonumber \\
& = &
\frac{-\hbar^2}{2m}
\left\langle
\left[ U_{\bf n} + {\bf U}_{\bf n}' \cdot {\bm \varepsilon}_{\bf n}
+ \frac{1}{2}{\bf U}_{\bf n}'' :
{\bm \varepsilon}_{\bf n} {\bm \varepsilon}_{\bf n}
\right]
\right. \nonumber \\ && \left. \mbox{ }
\left( \frac{ -{\bm \varepsilon}_{\bf n} }{ 2 \xi^2 }
 - \frac{ {\bf p}_{\bf n} }{i\hbar} \right)
 \cdot
 \left( \frac{ -{\bm \varepsilon}_{\bf n} }{ 2 \xi^2 }
 - \frac{ {\bf p}_{\bf n} }{i\hbar} \right)
\right\rangle_{\bf n}
\nonumber \\ & = &
\frac{-\hbar^2}{2m} U_{\bf n}
\left[ \frac{3N\xi^2}{4\xi^4} - \frac{2m {\cal K}_{\bf n}}{\hbar^2}
\right]
\nonumber \\ &&
+ \frac{-\hbar^2}{2m} \frac{2}{2\xi^2i\hbar} \xi^2
{\bf U}_{\bf n}' \cdot {\bf p}_{\bf n}
\nonumber \\ &&
+ \frac{-\hbar^2}{2m}
\frac{1}{2}{\bf U}_{\bf n}'' :
\left\langle {\bm \varepsilon}_{\bf n}{\bm \varepsilon}_{\bf n}
\left[
\frac{1}{4\xi^4} {\bm \varepsilon}_{\bf n} \cdot {\bm \varepsilon}_{\bf n}
- \frac{2m {\cal K}_{\bf n}}{\hbar^2} \right]
 \right\rangle_{\bf n}
 \nonumber \\ & = &
\frac{-3N\hbar^2}{8m\xi^2} U_{\bf n}
+  U_{\bf n} {\cal K}_{\bf n}
+ \frac{i\hbar}{2m} {\bf U}_{\bf n}' \cdot {\bf p}_{\bf n}
\nonumber \\ && \mbox{ }
- \frac{\hbar^2}{4m}
\left[ \frac{\xi^4  (3N+8) }{4\xi^4}
- \frac{2m \xi^2{\cal K}_{\bf n}}{\hbar^2} \right]
\mbox{TR }{\bf U}_{\bf n}''
 \nonumber \\ & = &
\frac{-3N\hbar^2}{8m\xi^2} U_{\bf n}
+  U_{\bf n} {\cal K}_{\bf n}
%\nonumber \\ && \mbox{ }
- \left[  \frac{  3N\hbar^2}{16m} - \frac{\xi^2}{2}  {\cal K}_{\bf n} \right]
\mbox{TR }{\bf U}_{\bf n}''
\nonumber \\ && \mbox{ }
+ \frac{i\hbar}{2m} {\bf U}_{\bf n}' \cdot {\bf p}_{\bf n}
- \frac{  8\hbar^2}{16m} \mbox{TR }{\bf U}_{\bf n}'' .
\end{eqnarray}
These are  ${\cal O}(N^2)$,
except for the final two terms here, which are  ${\cal O}(N)$.
Here and below the kinetic energy is
${\cal K}_{\bf n} \equiv {\bf p}_{\bf n} \cdot {\bf p}_{\bf n}/2m$.
The third term gives
\begin{eqnarray}
\lefteqn{
\left\langle U^2 \right\rangle_{\bf n}
}  \nonumber \\
& = &
\left\langle
\left[ U_{\bf n} + {\bf U}_{\bf n}' \cdot {\bm \varepsilon}_{\bf n}
+ \frac{1}{2}{\bf U}_{\bf n}'' :
{\bm \varepsilon}_{\bf n} {\bm \varepsilon}_{\bf n}
\right]^2 \right\rangle_{\bf n}
 \nonumber \\ & = &
U_{\bf n}^2
+ \xi^2  U_{\bf n} \mbox{TR }{\bf U}_{\bf n}''
%\nonumber \\ && \mbox{ }
+ \frac{\xi^4}{4} ( \mbox{TR }{\bf U}_{\bf n}'')^2
+ \frac{2 \xi^4}{4} {\bf U}_{\bf n}'' : {\bf U}_{\bf n}''
\nonumber \\ && \mbox{ }
+  \frac{6 \xi^4}{4} \sum_{j,\alpha} ( U''_{j\alpha,j\alpha})^2
+ \xi^2 {\bf U}_{\bf n}' \cdot {\bf U}_{\bf n}' .
\end{eqnarray}
These are  ${\cal O}(N^2)$,
except for the final two terms here, which are  ${\cal O}(N)$.

The fourth  term gives
\begin{eqnarray}
\frac{\hbar^4}{4m^2}
\left\langle  2 ( \nabla \nabla A ) :( \nabla \nabla A ) \right\rangle_{\bf n}
& = &
\frac{2\hbar^4}{4m^2}
\left\langle \frac{1}{4\xi^4} {\bf I} : {\bf I} \right\rangle_{\bf n}
 \nonumber \\ & = &
\frac{3N\hbar^4}{8 m^2 \xi^4}  .
\end{eqnarray}
This is ${\cal O}(N)$.
The fifth  term gives
\begin{eqnarray}
\frac{\hbar^2}{2m} \left\langle  \nabla^2 U \right\rangle_{\bf n}
& = &
\frac{\hbar^2}{2m}
\left\langle  \mbox{TR }{\bf U}_{\bf n}''  \right\rangle_{\bf n}
 \nonumber \\ & = &
\frac{\hbar^2}{2m}  \mbox{TR }{\bf U}_{\bf n}''  .
\end{eqnarray}
This is ${\cal O}(N)$.

The sixth  term gives
\begin{eqnarray}
\lefteqn{
\frac{\hbar^4}{4m^2}
\left\langle
 4 \nabla A \cdot (\nabla \nabla A ) \cdot \nabla A
 \right\rangle_{\bf n}
 }  \nonumber \\
& = &
\frac{4\hbar^4}{4m^2}\frac{-1}{2\xi^2}
\left\langle
\left( \frac{ -{\bm \varepsilon}_{\bf n} }{ 2 \xi^2 }
 - \frac{ {\bf p}_{\bf n} }{i\hbar} \right)
 \cdot
 \left( \frac{ -{\bm \varepsilon}_{\bf n} }{ 2 \xi^2 }
 - \frac{ {\bf p}_{\bf n} }{i\hbar} \right)
 \right\rangle_{\bf n}
 \nonumber \\ & = &
\frac{-\hbar^4}{2m^2\xi^2} \frac{3N \xi^2}{4\xi^4}
+
\frac{\hbar^4}{2m^2\xi^2} \frac{2m{\cal K}_{\bf n}}{\hbar^2} .
\end{eqnarray}
This is ${\cal O}(N)$.
The seventh  term gives
\begin{eqnarray}
\lefteqn{
\frac{-\hbar^2}{2m}
\left\langle
2 \nabla A \cdot \nabla U
 \right\rangle_{\bf n}
 }  \nonumber \\
& = &
\frac{-\hbar^2}{m}
\left\langle
\left( \frac{ -{\bm \varepsilon}_{\bf n} }{ 2 \xi^2 }
 - \frac{ {\bf p}_{\bf n} }{i\hbar} \right)
 \cdot
 \left( {\bf U}_{\bf n}' + {\bf U}_{\bf n}''
 \cdot {\bm \varepsilon}_{\bf n}   \right)
 \right\rangle_{\bf n}
 \nonumber \\ & = &
\frac{\hbar^2\xi^2}{2m\xi^2}  \mbox{TR }{\bf U}_{\bf n}''
- \frac{i\hbar}{m} {\bf U}_{\bf n}' \cdot {\bf p}_{\bf n} .
\end{eqnarray}
These are ${\cal O}(N)$.

The eighth term gives
\begin{eqnarray}
\frac{\hbar^4}{4m^2} \left\langle  \nabla^2 A \, \nabla^2 A
\right\rangle_{\bf n}
& = &
\frac{9N^2\hbar^4}{16m^2\xi^4}   .
\end{eqnarray}
This is ${\cal O}(N^2)$.
The ninth term gives
\begin{eqnarray}
\lefteqn{
\frac{\hbar^4}{4m^2}
\left\langle
2 \nabla^2 A \, \nabla A \cdot \nabla A
 \right\rangle_{\bf n}
 }  \nonumber \\
& = &
\frac{\hbar^4}{2m^2}
\frac{-3N}{2\xi^2}
\left\langle
\left( \frac{ -{\bm \varepsilon}_{\bf n} }{ 2 \xi^2 }
 - \frac{ {\bf p}_{\bf n} }{i\hbar} \right)
 \cdot
 \left( \frac{ -{\bm \varepsilon}_{\bf n} }{ 2 \xi^2 }
 - \frac{ {\bf p}_{\bf n} }{i\hbar} \right)
  \right\rangle_{\bf n}
 \nonumber \\ & = &
\frac{-9N^2\hbar^4}{16m^2\xi^4}
+
\frac{3N\hbar^2}{4m^2\xi^2} 2m {\cal K}_{\bf n} .
\end{eqnarray}
These are ${\cal O}(N^2)$.
The tenth term gives
\begin{eqnarray}
\lefteqn{
\frac{-\hbar^2}{2m}
\left\langle U \nabla^2 A  \right\rangle_{\bf n}
 }  \nonumber \\
& = &
\frac{-\hbar^2}{2m}
\frac{-3N}{2\xi^2}
\left\langle
\left[ U_{\bf n} + {\bf U}_{\bf n}' \cdot {\bm \varepsilon}_{\bf n}
+ \frac{1}{2}{\bf U}_{\bf n}'' :
{\bm \varepsilon}_{\bf n} {\bm \varepsilon}_{\bf n}
\right]  \right\rangle_{\bf n}
 \nonumber \\ & = &
\frac{3N\hbar^2}{4m\xi^2} U_{\bf n}
+
\frac{3N\hbar^2}{8m}  \mbox{TR }{\bf U}_{\bf n}'' .
\end{eqnarray}
These are ${\cal O}(N^2)$.

The eleventh term gives
\begin{eqnarray}
\lefteqn{
\frac{\hbar^4}{4m^2}
\left\langle (\nabla A \cdot \nabla A)^2 \right\rangle_{\bf n}
 }  \nonumber \\
& = &
\frac{\hbar^4}{4m^2}
\left\langle
\left[ \left( \frac{ -{\bm \varepsilon}_{\bf n} }{ 2 \xi^2 }
 - \frac{ {\bf p}_{\bf n} }{i\hbar} \right)
 \cdot
 \left( \frac{ -{\bm \varepsilon}_{\bf n} }{ 2 \xi^2 }
 - \frac{ {\bf p}_{\bf n} }{i\hbar} \right)
 \right]^2
 \right\rangle_{\bf n}
 \nonumber \\ & = &
\frac{\hbar^4}{4m^2}
\left\langle
 \left(
\frac{ {\bm \varepsilon}_{\bf n} \cdot {\bm \varepsilon}_{\bf n} }{ 4 \xi^4 }
 +
 \frac{ {\bm \varepsilon}_{\bf n} \cdot {\bf p}_{\bf n}  }{ i\hbar \xi^2 }
 - \frac{ {\bf p}_{\bf n} \cdot {\bf p}_{\bf n} }{\hbar^2} \right)^2
 \right\rangle_{\bf n}
  \nonumber \\ & = &
\left\langle
\frac{\hbar^4}{4m^2}
\frac{ ({\bm \varepsilon}_{\bf n} \cdot {\bm \varepsilon}_{\bf n})^2
}{16 \xi^8 }
- \frac{\hbar^4}{4m^2}
\frac{ ({\bm \varepsilon}_{\bf n} \cdot {\bf p}_{\bf n})^2 }{\hbar^2 \xi^4 }
\right. \nonumber  \\ &  & \left. \mbox{ }
+ \frac{\hbar^4}{4m^2}
\frac{ ({\bf p}_{\bf n} \cdot {\bf p}_{\bf n})^2 }{\hbar^4 }
- 2 \frac{\hbar^4}{4m^2}
\frac{ {\bf p}_{\bf n} \cdot {\bf p}_{\bf n}  }{\hbar^2 }
\frac{ {\bm \varepsilon}_{\bf n} \cdot {\bm \varepsilon}_{\bf n} }{ 4 \xi^4 }
 \right\rangle_{\bf n}
\nonumber \\ & = &
\frac{\hbar^4}{4m^2} \frac{ 9N^2+8N}{16 \xi^4 }
- \frac{\hbar^4}{4m^2} \frac{ 2m {\cal K}_{\bf n} }{\hbar^2 \xi^2 }
% \nonumber  \\ &  &  \mbox{ }
+   {\cal K}_{\bf n}^2
- \frac{\hbar^2}{m} \frac{ 3N  }{ 4 \xi^2 }{\cal K}_{\bf n}
\nonumber \\ & = &
\frac{ 9N^2 \hbar^4}{2^6m^2\xi^4 }
+   {\cal K}_{\bf n}^2
- \frac{3N\hbar^2}{4m\xi^2} {\cal K}_{\bf n}
+ \frac{ 8N \hbar^4}{2^6m^2\xi^4 } .
\end{eqnarray}
These are ${\cal O}(N^2)$,
except for the final term here, which is ${\cal O}(N)$.
The twelfth term is identical to the second term.

With these the expectation value of the square of the energy operator is
(grouping first the terms that are ${\cal O}(N^2)$,
and then the terms that are ${\cal O}(N)$)
\begin{eqnarray}
\lefteqn{
\langle \zeta_{\bf n} | \hat{\cal H}^2 | \zeta_{\bf n}\rangle
} \nonumber  \\
& = &
\frac{3N\hbar^2}{4m\xi^2}  U_{\bf n}
+ \frac{3N\hbar^2}{8m} \mbox{TR }{\bf U}_{\bf n}''
%\nonumber  \\ &  &  \mbox{ }                                  % 2 + 12
-\frac{3N\hbar^2}{4m\xi^2} U_{\bf n}
+ 2  U_{\bf n} {\cal K}_{\bf n}
\nonumber \\ && \mbox{ }
- \frac{\hbar^2}{2m}
\left[ \frac{ 3N }{4}
- \frac{2m \xi^2{\cal K}_{\bf n}}{\hbar^2} \right]
\mbox{TR }{\bf U}_{\bf n}''
\nonumber  \\ &  &  \mbox{ }                                  % 3
+ U_{\bf n}^2
+ \xi^2  U_{\bf n} \mbox{TR }{\bf U}_{\bf n}''
%\nonumber \\ && \mbox{ }
+ \frac{\xi^4}{4} ( \mbox{TR }{\bf U}_{\bf n}'')^2
+ \frac{ \xi^4}{2} {\bf U}_{\bf n}'' : {\bf U}_{\bf n}''
\nonumber  \\ &  &  \mbox{ }                                 % 8, 9
+ \frac{9N^2\hbar^4}{16m^2\xi^4}
%\nonumber  \\ &  &  \mbox{ }                                 % 9
-\frac{9N^2\hbar^4}{16m^2\xi^4}
+
\frac{3N\hbar^2}{2m\xi^2} {\cal K}_{\bf n}
+ \frac{3N\hbar^2}{4m\xi^2} U_{\bf n}
\nonumber  \\ &  &  \mbox{ }                                 % 10
+
\frac{3N\hbar^2}{8m}  \mbox{TR }{\bf U}_{\bf n}''
%\nonumber  \\ &  &  \mbox{ }                                 % 11
+\frac{ 9N^2 \hbar^4}{2^6m^2\xi^4 }
+   {\cal K}_{\bf n}^2
%\nonumber  \\ &  &  \mbox{ }
- \frac{3N\hbar^2}{4m\xi^2} {\cal K}_{\bf n}
%%%%%%%%%%%%%%%%%%%%%%%%%%%%%%%%%%%%%%%%%%%%%%%%%%%%%%%%%%%%%%%%%%%%%%%
\nonumber  \\ &  &  \mbox{ }                                 % 2  O(N)
+ \frac{i\hbar}{2m} {\bf U}_{\bf n}' \cdot {\bf p}_{\bf n}
- \frac{  8\hbar^2}{16m} \mbox{TR }{\bf U}_{\bf n}''
\nonumber  \\ &  &  \mbox{ }                                 % 3
+  \frac{6 \xi^4}{4} \sum_{j,\alpha} ( U''_{j\alpha,j\alpha})^2
+ \xi^2 {\bf U}_{\bf n}' \cdot {\bf U}_{\bf n}'
\nonumber  \\ &  &  \mbox{ }                                 % 4,5,6
+\frac{3N\hbar^4}{8 m^2 \xi^4}
+\frac{\hbar^2}{2m}  \mbox{TR }{\bf U}_{\bf n}''
-\frac{3N \hbar^4}{8m^2 \xi^4}
+ \frac{\hbar^2}{m\xi^2}  {\cal K}_{\bf n}
\nonumber  \\ &  &  \mbox{ }                                 % 7,11
+ \frac{\hbar^2}{2m}  \mbox{TR }{\bf U}_{\bf n}''
- \frac{i\hbar}{m} {\bf U}_{\bf n}' \cdot {\bf p}_{\bf n}
+ \frac{ 8N \hbar^4}{2^6m^2\xi^4 }
\nonumber  \\ & = &  %%%%%%%%%%%%%%%%%%%%%%%%%%%%%%%%%%%%%%%%%
\left(  U_{\bf n} + {\cal K}_{\bf n} \right)^2
+
\frac{3N\hbar^2}{4m\xi^2 }
\left(  U_{\bf n} + {\cal K}_{\bf n} \right)
+ \frac{ 9N^2 \hbar^4}{2^6 m^2   \xi^4 }
\nonumber  \\ &  &  \mbox{ }
+ \left\{
\xi^2{\cal K}_{\bf n}
+ \xi^2  U_{\bf n}
+ \frac{3N\hbar^2}{8m}
\right\}
\mbox{TR }{\bf U}_{\bf n}''
\nonumber \\ && \mbox{ }
+ \frac{\xi^4}{4} ( \mbox{TR }{\bf U}_{\bf n}'')^2
+ \frac{ \xi^4}{2} {\bf U}_{\bf n}'' : {\bf U}_{\bf n}''
\nonumber \\ && \mbox{ }   %%%%%%%%%%%%%%%%%%%%%%  O(N):
- \frac{i\hbar}{2m} {\bf U}_{\bf n}' \cdot {\bf p}_{\bf n}
+ \frac{  \hbar^2}{2m} \mbox{TR }{\bf U}_{\bf n}''
+  \frac{6 \xi^4}{4} \sum_{j,\alpha} ( U''_{j\alpha,j\alpha})^2
\nonumber  \\ &  &  \mbox{ }
+ \xi^2 {\bf U}_{\bf n}' \cdot {\bf U}_{\bf n}'
%\nonumber  \\ &  &  \mbox{ }
+ \frac{\hbar^2}{m\xi^2}  {\cal K}_{\bf n}
+ \frac{ N \hbar^4}{8m^2\xi^4 } .
\end{eqnarray}
All orders have been kept here.
This result is used in \S \ref{Sec:olxi}.

%\newpage $\;$ \newpage
%%%%%%%%%%%%%%%%%%%%%%%%%%%%%%%%%%%%%%%%%%%%%%%%%%%%%%%%%%%%%%%%%%%%%%%%%
%
\section{Eigenvalue Equation to Quadratic Order} \label{Sec:2Oeigen}
%
%%%%%%%%%%%%%%%%%%%%%%%%%%%%%%%%%%%%%%%%%%%%%%%%%%%%%%%%%%%%%%%%%%%%%%%%%

The eigenvalue equation is here expanded and solved to quadratic order,
which means that the eigenfunction is required to fourth order.
Besides revealing further generic properties of the modifying function,
namely that to fourth order
the departure from unity vanishes in the thermodynamic limit,
the explicit expressions for the coefficients
may be of use in computational simulations of quantum condensed matter.

The modifying function is
\begin{eqnarray}
f_{\bf n}({\bf r})
& = &
1
+ {\bf f}^{\bf  n} \cdot {\bm \varepsilon}_{{\bf n}}({\bf r})
+ \frac{1}{2} {\bf f}^{\bf  n  n}
: {\bm \varepsilon}_{{\bf n}}({\bf r})
{\bm \varepsilon}_{{\bf n}}({\bf r})
\nonumber \\ && \mbox{ }
+ \frac{1}{3!} {\bf f}^{\bf  n  n  n}
\; \raisebox{-1mm}{\vdots}\;
{\bm \varepsilon}_{{\bf n}}({\bf r})^3
%\nonumber \\ && \mbox{ }
+ \frac{1}{4!}
{\bf f}^{\bf nnnn} \, \raisebox{-1mm}{$\stackrel{:}{:}$}\,
{\bm \varepsilon}_{{\bf n}}({\bf r})^4 .
\end{eqnarray}
The coefficients are symmetric in all indeces
and so it is not necessary below to write out all the combinations
or to specify what indeces are summed over for the trace,
or which are involved in the various scalar products.

It will be assumed that all the
${\bf f}^{\bf n}$, ${\bf f}^{\bf nn}$, \ldots,
are  ${\cal O}(N^{-1})$.
The validity of this assumption can be judged by the final results.
An expansion of the eigenvalue equation will be made
to  ${\cal O}(N^{0})$,
which is the  leading order.
Scalar products contributes a factor of $N$
(not counting scalar products with ${\bm \varepsilon}_{{\bf n}}$).

The gradient to quadratic order in ${\bm \varepsilon}$ is
\begin{eqnarray}
\nabla_r f_{\bf n}({\bf r})
& = &
{\bf f}^{\bf n}
+ {\bf f}^{\bf n  n}
\cdot {\bm \varepsilon}_{{\bf n}}({\bf r})
+ \frac{1}{2!}
{\bf f}^{\bf nnn}
:
{\bm \varepsilon}_{{\bf n}}({\bf r})
{\bm \varepsilon}_{{\bf n}}({\bf r})
\nonumber \\ && \mbox{ }
+ \frac{1}{3!} {\bf f}^{\bf nnnn} \; \raisebox{-1mm}{\vdots}\;
{\bm \varepsilon}_{{\bf n}}({\bf r})^3
\nonumber \\ & = &
{\bf f}^{\bf n}
+ {\bf f}^{\bf n  n}
\cdot {\bm \varepsilon}_{{\bf n}}({\bf r})
+ \frac{1}{2!}
{\bf f}^{\bf nnn}
:
{\bm \varepsilon}_{{\bf n}}({\bf r})
{\bm \varepsilon}_{{\bf n}}({\bf r})
\nonumber \\ & &  \mbox{ }
+ {\cal O}({\bm \varepsilon}_{{\bf n}}^{3}).
\end{eqnarray}
%Because of the symmetry of the coefficients,
%one does not have to indicate explicitly which indeces are summed over.
The Laplacian is
\begin{eqnarray}
\nabla_r^2 f_{\bf n}({\bf r})
& = &
\mbox{TR }  {\bf f}^{\bf n n}
%\nonumber \\ && \mbox{ }
+ \mbox{TR}^{(1)} \{ {\bf f}^{\bf nnn} \}
\cdot
{\bm \varepsilon}_{{\bf n}}({\bf r})
\nonumber \\ && \mbox{ }
+ \frac{1}{2} \mbox{TR}^{(1)} \{ {\bf f}^{\bf nnnn} \}
: {\bm \varepsilon}_{{\bf n}}({\bf r}) {\bm \varepsilon}_{{\bf n}}({\bf r})
\nonumber \\ & = &
\mbox{TR } {\bf f}^{\bf n n}
+ \mbox{TR}^{(1)} \{ {\bf f}^{\bf nnn} \} \cdot
{\bm \varepsilon}_{{\bf n}}({\bf r})
\nonumber \\ && \mbox{ }
+ \mbox{TR}^{(1)} \{ \tilde{\bf f}^{\bf nnnn} \} :
{\bm \varepsilon}_{{\bf n}}({\bf r})^2
+{\cal O}({\bm \varepsilon}_{{\bf n}}^{3}).
\nonumber \\
\end{eqnarray}
Since $f_{\bf n}({\bf r})^{-1} = 1 + {\cal O}(N^{-1})$,
the expansions for
$\nabla_r f_{\bf n}({\bf r})/f_{\bf n}({\bf r})$ and
$\nabla_r^2 f_{\bf n}({\bf r})/f_{\bf n}({\bf r})$
are unchanged from these.

The gradient of the reference wave function is
\begin{equation}
\nabla_r \tilde \zeta_{\bf n}({\bf r})
=
\left[
\frac{-1}{2\xi^2} {\bm \varepsilon}_{{\bf n}}({\bf r})
- \frac{1}{i\hbar} {\bf p}_{{\bf n}}
\right]
\tilde \zeta_{\bf n}({\bf r}) .
\end{equation}

Hence to ${\cal O}({\bm \varepsilon}^2)$
and to ${\cal O}(N,N^0)$ one has
\begin{eqnarray}
\lefteqn{
\frac{2}{f_{\bf n}({\bf r})}
\left[ \nabla_r f_{\bf n}({\bf r}) \right]
\cdot
\left[
\frac{-1}{2\xi^2} {\bm \varepsilon}_{{\bf n}}({\bf r})
- \frac{1}{i\hbar} {\bf p}_{{\bf n}}
\right]
} \nonumber \\
& = &
2 \left[
{\bf f}^{\bf n}
+  {\bf f}^{\bf n n} \cdot {\bm \varepsilon}_{{\bf n}}({\bf r})
+ \frac{1}{2!}
{\bf f}^{\bf nnn}
:
{\bm \varepsilon}_{{\bf n}}({\bf r})
{\bm \varepsilon}_{{\bf n}}({\bf r})
\right]
\nonumber \\ & &  \mbox{ } \cdot
\left[
\frac{-1}{2\xi^2} {\bm \varepsilon}_{{\bf n}}({\bf r})
- \frac{1}{i\hbar} {\bf p}_{{\bf n}}
\right]
\nonumber \\ & = &   %%%%%%%%%%%%%%%%%%%
\frac{-2}{i\hbar} {\bf p}_{{\bf n}} \cdot  {\bf f}^{\bf n}
- \frac{2}{i\hbar}  {\bf f}^{\bf n n} :
{\bf p}_{{\bf n}} {\bm \varepsilon}_{{\bf n}}({\bf r})
\nonumber \\ && \mbox{ }
- \frac{1}{i\hbar} {\bf p}_{{\bf n}} \cdot
{\bf f}^{\bf nnn} :
{\bm \varepsilon}_{{\bf n}}({\bf r})
{\bm \varepsilon}_{{\bf n}}({\bf r})
\nonumber \\ & & \mbox{ }
- \frac{1}{\xi^2}
{\bf f}^{\bf  n} \cdot  {\bm \varepsilon}_{{\bf n}}({\bf r})
%\nonumber \\ & & \mbox{ }
- \frac{1}{\xi^2}
{\bf f}^{\bf n n}
: {\bm \varepsilon}_{{\bf n}}({\bf r})
 {\bm \varepsilon}_{{\bf n}}({\bf r}) .
\end{eqnarray}
All the terms here are ${\cal O}(N^0)$,
except for the final two,
which are ${\cal O}(N^{-1})$ and may therefore be neglected.

%The various terms from the kinetic energy
%have to be multiplied by $-\hbar^2/2m$ in the eigenvalue equation.

The eigenvalue equation is
\begin{eqnarray}
\lefteqn{
\hat{\cal H}({\bf r}) \zeta_{\bf n}({\bf r})
} \nonumber  \\
& = &
U({\bf r}) \zeta_{\bf n}({\bf r})
- \frac{\hbar^2}{2m}
\frac{\tilde C_{\bf n}}{C_{\bf n}}
\left\{
 \tilde \zeta_{\bf n}({\bf r}) \nabla_{\bf r}^2 f_{\bf n}({\bf r})
\right. \nonumber \\ && \left. \mbox{ }
+ 2
\nabla_{\bf r} f_{\bf n}({\bf r}) \cdot
\nabla_{\bf r}\tilde \zeta_{\bf n}({\bf r})
+
f_{\bf n}({\bf r}) \nabla_{\bf r}^2\tilde \zeta_{\bf n}({\bf r})
\right\}
\nonumber  \\ & = &
U({\bf r}) \zeta_{\bf n}({\bf r})
- \frac{\hbar^2}{2m}
\left\{
\frac{1}{f_{\bf n}({\bf r})} \nabla_{\bf r}^2 f_{\bf n}({\bf r})
\right. \nonumber \\ && \left. \mbox{ }
+
\frac{2}{f_{\bf n}({\bf r})}
\nabla_{\bf r} f_{\bf n}({\bf r}) \cdot
\left[
\frac{-1}{2\xi^2} {\bm \varepsilon}_{{\bf n}}({\bf r})
- \frac{1}{i\hbar} {\bf p}_{{\bf n}}
\right]
\right. \nonumber \\ && \left. \mbox{ }
+
\left[ \frac{-1}{2\xi^2} {\bm \varepsilon}_{{\bf n}}({\bf r})
- \frac{1}{i\hbar} {\bf p}_{{\bf n}} \right]^2
-\frac{3N}{2\xi^2}
\right\} \zeta_{\bf n}({\bf r}) .
\end{eqnarray}
The above results can be inserted into this
and the terms grouped according to powers of
${\bm \varepsilon}_{{\bf n}}({\bf r})$.
The potential, assumed continuous, is
\begin{eqnarray}
U({\bf r}) & = &
U({\bf q}_{\bf n})
+ {\bf U}'_{\bf n}
\cdot {\bm \varepsilon}_{{\bf n}}({\bf r})
\nonumber \\ && \mbox{ }
+ \frac{1}{2} {\bf U}''_{\bf nn}
: {\bm \varepsilon}_{{\bf n}}({\bf r}) {\bm \varepsilon}_{{\bf n}}({\bf r}).
\end{eqnarray}

The zeroth order or scalar equation is the eigenvalue
\begin{eqnarray}
{\cal H}_{\bf n} & \equiv &
U({\bf q}_{\bf n})
+ \frac{1}{2 m } {\bf p}_{\bf n} \cdot {\bf p}_{\bf n}
+ \frac{3 N \hbar^2}{4 m \xi^2}
\nonumber \\ && \mbox{ }
- \frac{ \hbar^2}{2m} \mbox{TR } {\bf f}^{\bf nn}
- \frac{i\hbar}{m} {\bf p}_{\bf n} \cdot {\bf f}^{\bf n} .
\end{eqnarray}
The first three terms are  ${\cal O}(N)$ and give the energy eigenvalue.
The final two terms are ${\cal O}(N^0)$ and
the neglected terms are  ${\cal O}(N^{-1})$.
The energy eigenvalue equals the classical Hamiltonian
function of the nominal positions and momenta, plus an immaterial constant,
\begin{eqnarray}
{\cal H}_{\bf n}
& = &
{\cal H}({\bf q}_{\bf n}, {\bf p}_{\bf n})
+ \frac{3 N \hbar^2}{4 m \xi^2}
\nonumber \\ & = &
U({\bf q}_{\bf n})
+ \frac{1}{2 m } {\bf p}_{\bf n} \cdot {\bf p}_{\bf n}
+ \frac{3 N \hbar^2}{4 m \xi^2}.
\end{eqnarray}

The final constant has no physical effect because only differences
in energy can be measured.
The  dependence of the constant on number does not appear significant
because again only the difference between chemical potentials
has a physical effect.
One could instead incorporate the constant into ${\bf f}^{\bf nn}$
by adding a term ${\bf I}/2\xi^2$,
but this is ${\cal O}(N^0)$
and similar terms have to added to all the even coefficients,
which is problematic.

The remainder of the zeroth order equation must vanish.
This gives the scalar condition on the coefficients of the modifying function,
\begin{equation}
0 =
\frac{ \hbar^2}{2m} \mbox{TR } {\bf f}^{\bf nn}
+ \frac{i\hbar}{m} {\bf p}_{\bf n} \cdot {\bf f}^{\bf n} .
\end{equation}

The term linear in
$  {\bm \varepsilon}_{{\bf n}}$
in the eigenvalue equation must vanish.
Hence the coefficient of this must be zero,
which gives the vector condition
\begin{eqnarray}
{\bf 0} & = &
{\bf U}'_{\bf n}
+ \frac{i\hbar}{m\xi^2 }  {\bf p}_{{\bf n}}
- \frac{\hbar^2}{2m} \mbox{TR}^{(1)} \{ {\bf f}^{\bf nnn} \}
\nonumber \\ && \mbox{ }
+ \frac{\hbar^2}{2m}
%\left\{
\frac{2}{i\hbar} {\bf f}^{\bf nn} \cdot
{\bf p}_{{\bf n}} .
%+ \frac{1}{\xi^2} {\bf f}^{{\bf n}} \right\}.
\end{eqnarray}
All the terms here are ${\cal O}(N^0)$.

The term quadratic in
$  {\bm \varepsilon}_{{\bf n}}$
in the eigenvalue equation must vanish.
Hence the  coefficient of this must be zero,
which gives the matrix condition
\begin{eqnarray}
{\bf 0} & = &
{\bf U}''_{\bf n}
%\nonumber \\ && \mbox{ }
- \frac{\hbar^2}{2m}
 \mbox{TR}^{(1)} \{ {\bf f}^{\bf nnnn} \}
%\nonumber \\ && \mbox{ }
- \frac{i\hbar}{2m} {\bf p}_{{\bf n}} \cdot {\bf f}^{\bf nnn} .
\end{eqnarray}

One now has a scalar, vector, and matrix equation for four unknown functions:
a vector $ {\bf f}^{\bf  n} $,
a matrix $ {\bf f}^{\bf nn}$,
a third order quantity ${\bf f}^{\bf nnn}$,
and  a fourth order quantity ${\bf f}^{\bf nnnn}$.
It is obviously an under-determined system.

For simplicity, terminate the expansion of the modifier function
at the third order by imposing the condition
\begin{equation}
{\bf f}^{\bf nnnn} = {\bf 0}.
\end{equation}
The equations to be solved then are
\begin{equation}
0 =
\frac{ \hbar^2}{2m} \mbox{TR } {\bf f}^{\bf nn}
+ \frac{i\hbar}{m} {\bf p}_{\bf n} \cdot {\bf f}^{\bf n} ,
\end{equation}
\begin{eqnarray}
{\bf 0} & = &
{\bf U}'_{\bf n}
+ \frac{i\hbar}{m\xi^2 }  {\bf p}_{{\bf n}}
- \frac{\hbar^2}{2m} \mbox{TR}^{(1)} \{ {\bf f}^{\bf nnn} \}
\nonumber \\ && \mbox{ }
- \frac{i\hbar}{m} {\bf f}^{\bf nn} \cdot {\bf p}_{{\bf n}}  ,
\end{eqnarray}
and
\begin{eqnarray}
{\bf 0} & = &
{\bf U}''_{\bf n}
- \frac{i\hbar}{2m} {\bf p}_{{\bf n}} \cdot {\bf f}^{\bf nnn}.
%\nonumber \\
\end{eqnarray}
This is still an under-determined system of equations,
since the matrix condition is insufficient to determine the third order quantity
${\bf f}^{\bf nnn}$.
Obviously, one cannot set ${\bf f}^{\bf nnn} = {\bf 0}$.

Take as an ansatz
\begin{equation}
{\bf f}^{\bf nnn} = \mbox{sym}\{{\bf A}_{\bf n} {\bf p}_{\bf n} \},
\end{equation}
where ${\bf A}_{\bf n}$ is a symmetric matrix to be determined.
Hence the matrix condition becomes
\begin{eqnarray}
\frac{2m}{i\hbar} {\bf U}''_{\bf n}
& = &
{\bf p}_{{\bf n}} \cdot {\bf f}^{\bf nnn}
 \\ \nonumber & = &
{\bf p}_{{\bf n}} \cdot {\bf p}_{{\bf n}} {\bf A}_{\bf n}
+
{\bf p}_{{\bf n}} {\bf A}_{\bf n} \cdot {\bf p}_{{\bf n}}
+
{\bf A}_{\bf n} \cdot {\bf p}_{{\bf n}} {\bf p}_{{\bf n}}.
\end{eqnarray}
Taking the scalar product of this with the momentum vector gives
\begin{eqnarray}
\frac{2m}{i\hbar} {\bf U}''_{\bf n}  \cdot {\bf p}_{\bf n}
& = &
{\bf p}_{{\bf n}} \cdot {\bf p}_{\bf n} {\bf A}_{\bf n} \cdot {\bf p}_{\bf n}
+
{\bf p}_{\bf n} {\bf p}_{\bf n} \cdot {\bf A}_{\bf n} \cdot {\bf p}_{{\bf n}}
\nonumber \\ && \mbox{ }
+
{\bf A}_{\bf n} \cdot {\bf p}_{{\bf n}} {\bf p}_{{\bf n}}\cdot{\bf p}_{\bf n}.
\end{eqnarray}
Finally taking another scalar product gives
\begin{eqnarray}
\frac{2m}{i\hbar}
 {\bf U}''_{\bf n} :  {\bf p}_{\bf n} {\bf p}_{\bf n}
& = &
3 {\bf p}_{{\bf n}} \cdot {\bf p}_{\bf n} \,
{\bf A}_{\bf n} : {\bf p}_{\bf n} {\bf p}_{\bf n} .
\end{eqnarray}
These successively give the scalar,
\begin{equation}
A_{\bf n}^{pp} \equiv
{\bf A}_{\bf n} : {\bf p}_{\bf n} {\bf p}_{\bf n}
=
\frac{1}{3i\hbar K_{\bf n}}
 {\bf U}''_{\bf n}  : {\bf p}_{\bf n} {\bf p}_{\bf n} ,
\end{equation}
the vector,
\begin{eqnarray}
{\bf A}_{\bf n}^{p}
& \equiv &
{\bf A}_{\bf n} \cdot {\bf p}_{\bf n}
\nonumber \\ & = &
\frac{1}{4mK_{\bf n}}
\left[
\frac{2m}{i\hbar} {\bf U}''_{\bf n}  \cdot {\bf p}_{\bf n}
- A_{\bf n}^{pp} {\bf p}_{\bf n}
\right] ,
\end{eqnarray}
and finally the matrix itself,
\begin{eqnarray}
{\bf A}_{\bf n}
& = &
\frac{1}{2mK_{\bf n}}
\left[
\frac{2m}{i\hbar} {\bf U}''_{\bf n}
- {\bf p}_{\bf n} {\bf A}_{\bf n}^{p}
-  {\bf A}_{\bf n}^{p}{\bf p}_{\bf n}
\right] .
\end{eqnarray}
This is ${\cal O}(N^{-1})$, as promised.
Everything on the right hand side is known.
The matrix condition is now satisfied by this ansatz.

Inserting the ansatz for ${\bf f}^{\bf nnn}$ into the vector condition gives
\begin{eqnarray}
{\bf 0} & = &
{\bf U}'_{\bf n}
+ \frac{i\hbar}{m\xi^2 }  {\bf p}_{{\bf n}}
- \frac{i\hbar}{m} {\bf f}^{\bf nn} \cdot {\bf p}_{{\bf n}}
\nonumber \\ && \mbox{ }
- \frac{\hbar^2}{2m}
\left[ {\bf p}_{\bf n} \mbox{TR } {\bf A}_{\bf n}
+ 2 {\bf A}_{\bf n}^{p} \right]
\nonumber \\ & = &
{\bf U}'_{\bf n}
+ \frac{i\hbar}{m\xi^2 }  {\bf p}_{{\bf n}}
- \frac{i\hbar}{m} {\bf f}^{\bf nn} \cdot {\bf p}_{{\bf n}}
\nonumber \\ && \mbox{ }
- \frac{\hbar^2}{2m}
\left\{
\frac{1}{i\hbar K_{\bf n}} {\bf p}_{\bf n}
\mbox{TR } {\bf U}''_{\bf n}
\right. \nonumber \\ && \left. \mbox{ }
- \frac{{A}_{\bf n}^{pp}}{mK_{\bf n}} {\bf p}_{\bf n}
+ 2 {\bf A}_{\bf n}^{p} \right\} .
\end{eqnarray}
Clearly this can be solved for ${\bf f}^{\bf nn}$
by making the latter a sum of the three dyadics
formed from  ${\bf p}_{\bf n}$ and
the three vectors ${\bf p}_{\bf n}$, ${\bf U}'_{\bf n}$,
and ${\bf U}''_{\bf n}\cdot {\bf p}_{\bf n}$.
Instead of the last of these, it is slightly simpler to use
the vector ${\bf A}_{\bf n}^{p}$,
which is a linear combination of ${\bf U}''_{\bf n}\cdot {\bf p}_{\bf n}$
and ${\bf p}_{\bf n}$.
Since ${\bf f}^{\bf nn}$ is a symmetric matrix the ansatz is
\begin{eqnarray}
{\bf f}^{\bf nn} & = &
a_{\bf n} {\bf p}_{\bf n} {\bf p}_{\bf n}
+ b_{\bf n}
[ {\bf p}_{\bf n} {\bf U}'_{\bf n} +  {\bf U}'_{\bf n} {\bf p}_{\bf n} ]
\nonumber \\ && \mbox{ }
+ c_{\bf n}
[ {\bf p}_{\bf n} {\bf A}_{\bf n}^{p} +  {\bf A}_{\bf n}^{p}{\bf p}_{\bf n} ] .
\end{eqnarray}
Rearranging the vector condition,
writing $u_{\bf n}'' \equiv \mbox{TR } {\bf U}''_{\bf n}$,
and  using this gives
\begin{eqnarray}
\lefteqn{
\frac{i\hbar}{m} {\bf f}^{\bf nn} \cdot {\bf p}_{{\bf n}}
} \nonumber \\
& = &
{\bf U}'_{\bf n}
+ \left[
\frac{i\hbar}{m\xi^2 }
+ \frac{i\hbar u_{\bf n}''}{2mK_{\bf n}}
+ \frac{\hbar^2{A}_{\bf n}^{pp}}{2m^2K_{\bf n}}
\right] {\bf p}_{{\bf n}}
- \frac{\hbar^2}{m} {\bf A}_{\bf n}^{p}
\nonumber \\ & = &
\frac{i\hbar}{m} \left[
2 m K_{\bf n} a_{\bf n} {\bf p}_{\bf n}
+  2 m K_{\bf n}  b_{\bf n} {\bf U}'_{\bf n}
+  b_{\bf n} {\bf p}_{\bf n}\cdot {\bf U}'_{\bf n} {\bf p}_{\bf n}
\right. \nonumber \\ && \left. \mbox{ }
+ 2 m K_{\bf n} c_{\bf n} {\bf A}_{\bf n}^{p}
+  c_{\bf n} {A}_{\bf n}^{pp} {\bf p}_{\bf n} \right].
\end{eqnarray}
Now equate  the coefficients of the individual vectors.
From ${\bf U}'_{\bf n}$ one obtains
\begin{equation}
b_{\bf n} = \frac{1}{2 i\hbar K_{\bf n} } .
\end{equation}
From ${\bf A}_{\bf n}^{p}$  one obtains
\begin{equation}
c_{\bf n} = \frac{i\hbar}{2 m K_{\bf n} } ,
\end{equation}
And from ${\bf p}_{\bf n}$  one obtains
\begin{eqnarray}
a_{\bf n} & = & \frac{1}{2 i\hbar K_{\bf n} }
\left[ \frac{i\hbar}{m\xi^2 }
+ \frac{i\hbar u_{\bf n}''}{2mK_{\bf n}}
+ \frac{\hbar^2{A}_{\bf n}^{pp}}{2m^2K_{\bf n}}
\right. \nonumber \\ & & \left. \mbox{ }
-  \frac{ i\hbar b_{\bf n}}{m} {\bf p}_{\bf n}\cdot {\bf U}'_{\bf n}
- \frac{ i\hbar c_{\bf n}}{m} {A}_{\bf n}^{pp} \right].
\end{eqnarray}
The three scalars $a_{\bf n}$,  $b_{\bf n}$, , and $c_{\bf n}$,
are all ${\cal O}(N^{-1})$, and hence so is ${\bf f}^{\bf nn}$.

The coefficients ${\bf f}^{\bf nnn}$ and ${\bf f}^{\bf nn}$
have now been determined.
The vector ${\bf f}^{\bf n}$ can now be determined from the scalar condition
\begin{eqnarray}
\frac{i\hbar}{m} {\bf p}_{\bf n} \cdot {\bf f}^{\bf n}
& = &
\frac{ -\hbar^2}{2m} \mbox{TR } {\bf f}^{\bf nn}
\nonumber \\ & = &
\frac{ -\hbar^2}{2m}
\left[
 a_{\bf n} {\bf p}_{\bf n} \cdot {\bf p}_{\bf n}
+ 2 b_{\bf n} {\bf U}'_{\bf n} \cdot {\bf p}_{\bf n}
\right. \nonumber \\ && \left. \mbox{ }
+ 2 c_{\bf n} {\bf A}^p_{\bf n} \cdot {\bf p}_{\bf n}
\right] .
\end{eqnarray}
Clearly this is satisfied by
\begin{eqnarray}
 {\bf f}^{\bf n} & = &
\frac{ i\hbar}{2}
\left[
 a_{\bf n} {\bf p}_{\bf n}
+ 2 b_{\bf n} {\bf U}'_{\bf n}
%\right. \nonumber \\ && \left. \mbox{ }
+ 2 c_{\bf n} {\bf A}^p_{\bf n}
\right] .
\end{eqnarray}

With these results the eigenvalue equation has been solved
to quadratic order in $\varepsilon_{\bf n}$.
Three non-zero coefficients of the eigenfunction have been determined.
This ansatz for $f({\bf r})$ is not unique.
Other solutions are possible for the present three coefficients,
and the values of these three coefficients will change
if the eigenvalue equation is solved to higher order.

The merits or otherwise of this or other ansatz compared with that given in
\S \ref{Sec:ctsU} remain to be seen.

Maybe the most important point is that all the coefficients
in the expansion of $f({\bf r})$ are ${\cal O}(N^{-1})$.
Hence in the thermodynamic limit they can all be neglected,
which leaves only the reference wave packet as the entropy eigenfunction.

%%%%%%%%%%%%%%%%%%%%%%%%%%%%%%%%%%%%%%%%%%%%%%%%%%%%%%%%%%%%%%%%%%%%%%%%%%

\begin{thebibliography}{99}




\bibitem{Messiah61}
Messiah, A. (1961),
\emph{Quantum Mechanics},
(North-Holland, Amsterdam, Vols I and II).


\bibitem{Merzbacher70}
Merzbacher, E. (1970),
\emph{Quantum Mechanics},
(Wiley, New York, 2nd edn).


\bibitem{QSM1}
Attard, P. (2013),
`Quantum Statistical Mechanics. I.
Decoherence, Wave function Collapse, and the von Neumann Density Matrix',
arXiv:1401.1786v1. % [cond-mat.stat-mech].

\bibitem{QSM}
 Attard, P. (2015),
\emph{Quantum Statistical Mechanics:
Equilibrium and Non-Equilibrium Theory from First Principles},
(IOP Publishing, Bristol).
%ISBN 978-0-7503-1188-5 (ebook)
%ISBN 978-0-7503-1189-2 (print)
%ISBN 978-0-7503-1190-8 (mobi)

\bibitem{Neumann27}
von Neumann, J. (1927),
%%"Wahrscheinlichkeitstheoretischer Aufbau der Quantenmechanik",
G\"ottinger Nachrichten {\bf 1}, 245. %–272.




\bibitem{Davies76}
Davies, E. B. (1976),
\emph{Quantum Theory of Open Systems},
(Academic Press, London).

\bibitem{Breuer02}
Breuer, H.-P. and Petruccione, F. (2002),
\emph{The Theory of Open Quantum Systems},
(Oxford University Press, Oxford).

\bibitem{Weiss08}
Weiss, U. (2008),
\emph{Quantum Dissipative Systems},
(World Scientific, Singapore, 3rd Ed.).

\bibitem{Zeh01}
Zeh, H. D. (2001),
\emph{The Physical Basis of the Direction of Time},
(Springer, Berlin, 4th ed.).

\bibitem{Zurek03}
Zurek, W. H. (2003),
`Decoherence, Einselection,
and the Quantum Origins of the Classical',
arXiv:quant-ph/0105127v3.




\bibitem{TDSM}
Attard, P. (2002),
\emph{Thermodynamics and Statistical Mechanics:
Equilibrium by Entropy Maximisation},
(Academic Press, London).
%%ISBN  0-12-066321-X.  Library of Congress 2002103392


\bibitem{Wigner32}
Wigner, E. (1932),
%`On the Quantum Correction For Thermodynamic Equilibrium'
Phys.\ Rev.\ {\bf 40}, 749.

\bibitem{Kirkwood33}
Kirkwood, J. (1933),
% `Quantum Statistics of alomost Calssical Particles',
Phys.\ Rev.\ {\bf 44}, 31.

\bibitem{Allen87}
Allen, M. P. and Tildesley, D. J. (1987),
\emph{Computer Simulations of Liquids},
(Oxford University Press, Oxford).

\bibitem{Kahn38}
Kahn, B. and Uhlenbeck, G. E. (1938),
Physica {\bf 5}, 399.

\bibitem{Lee59}
Lee, T. D. and Yang, C. N. (1959),
Phys.\ Rev.\ {\bf 113}, 1165; {\bf 116}, 25;
\emph{ibid} (1960), {\bf 117}, 12, 22, 897.

%\bibitem{Lee60}
%Lee, T. D. and Yang, C. N. (1960),
%Phys.\ Rev.\ {\bf 117}, 12, 22, 897.


\bibitem{Uhlenbeck36}
Uhlenbeck, G. E. and  Bethe, H. A. (1936),
Physica {\bf 3}, 729.

\bibitem{Uhlenbeck37}
Uhlenbeck, G. E. and  Bethe, H. A. (1937),
Physica {\bf 4}, 915.

\bibitem{Pathria72}
Pathria, R. K. (1972),
\emph{Statistical Mechanics},
(Pergamon Press, Oxford).


\bibitem{Ursell27}
Ursell, H. D. (1927),
Proc.\ Camb.\ Phil.\ Soc.\ {\bf 23}, 685.

\bibitem{Mayer37}
Mayer, J. E. \emph{et al.} (1937),
J. Chem.\ Phys.\ {\bf 5}, 67, 74.
%(1938), {\bf 6}; 87, 101;
%(1942), {\bf 10}, 629;
%(1951), {\bf 19} 1024.

%\bibitem{Moody04}
%Moody, M. P. and Attard, P. (2004),
%``Monte Carlo Simulation Methodology
%of the Ghost Interface Theory for the Planar Surface Tension'',
%J.\ Chem.\ Phys., {\bf 120}, 1892. %--1904


%\bibitem{Widom63}
%Widom, B. (1964)
%``Some Topics in the Theory of Fluids '',
%J.\ Chem.\ Phys., {\bf 39}, 2808.


%\bibitem{Attard99}
%Attard, P.
%``Markov Superposition Expansion for the Entropy
%and Correlation Functions in Two and Three Dimensions'',
%p.~189,
%% Date: 10 April, 1998
%% For James McGuire's Festschrift
%in \emph{Statistical Physics on the Eve of the Twenty-First Century},
%Batchelor, M. T. and Wille, L. T. (eds),
%(World Scientific, Singapore, 1999).



%%%%%%%%%%%%%%%%



%\bibitem{Kubo62}
%Kubo, R. (1962),
%`Generalized Cumulant Expansion Method',
%J. Phys.\ Soc.\ Japan {\bf 17}, 1100. %--1120 .





%\bibitem{Schiff68}
%Schiff, L. I. (1968),
%\emph{Quantum Mechanics},
%(Mcgraw-Hill, New York, 3rd ed.).

%\bibitem{NETDSM}
%Attard, P. (2012),
%\emph{Non-Equilibrium Thermodynamics and Statistical Mechanics:
%Foundations and Applications}, (Oxford University Press, Oxford).
%%ISBN  978-0-19-966276-0

%\bibitem{Risken84}
%Risken, H. (1984),
%\emph{The Fokker-Planck Equation},
%(Springer-Verlag, Berlin).




%\bibitem{AttardV}
%Attard, P. (2006),
%%``Statistical Mechanical Theory for Steady State Systems.
%%V. Non-equilibrium Probability Density.'',
%J. Chem.\ Phys.\ {\bf 124}, 224103.

%\bibitem{AttardIX}
%Attard, P. (2009),
%%``Statistical Mechanical Theory for Non-equilibrium Systems. IX.
%%Stochastic Molecular Dynamics.''
%J. Chem.\ Phys.\ {\bf 130}, 194113.



%\bibitem{Attard14}
%Attard, P. (2014),
%%'Simplified Derivation of the Non-Equilibrium Probability Distribution',
%arXiv:1405.1469 [cond-mat.stat-mech].

%\bibitem{Attard00}
%Attard, P. (2000),
%%``The Explicit Density Functional and its Connection with
%%Entropy Maximisation'',
%J. Stat.\ Phys.\ {\bf 100}, 445. %--473




%\bibitem{QSM2}
%Attard, P. (2013),
%%`Quantum Statistical Mechanics. II.
%%Stochastic Schr\"odinger Equation',
%arXiv:1401.1787v3 [cond-mat.stat-mech].

%\bibitem{QSM3}
%Attard, P. (2014),
%%'Quantum Statistical Mechanics. III. Equilibrium Probability',
%arXiv:1404.2683 [cond-mat.stat-mech].
%In this paper the fluctuation expansions
%were performed on the expectation entropy,
%which is a linearization of the correct actual entropy.




\end{thebibliography}
\end{document}